\documentclass[sigconf,10pt]{acmart}
\def\BibTeX{{\rm B\kern-.05em{\sc i\kern-.025em b}\kern-.08emT\kern-.1667em\lower.7ex\hbox{E}\kern-.125emX}}

\copyrightyear{2020} 
\acmYear{2020} 
\setcopyright{acmcopyright}
\acmConference[SIGMOD '20] {2020 ACM SIGMOD International Conference on Management of Data}{June 14--19, 2020}{Portland, OR, USA}
\acmBooktitle{2020 ACM SIGMOD International Conference on Management of Data (SIGMOD'20), June 14--19, 2020, Portland, OR, USA}
\acmPrice{15.00}
\acmDOI{10.1145/3318464.3380596}
\acmISBN{978-1-4503-6735-6/20/06}

\usepackage{enumitem}
\usepackage{amsmath,amssymb,amsfonts}
\usepackage{algorithm}
\usepackage{algcompatible}
\usepackage{graphicx}
\usepackage{comment}
\usepackage{multirow} 
\usepackage{amsthm}
\usepackage{textcomp}
\usepackage{xcolor}
\def\BibTeX{{\rm B\kern-.05em{\sc i\kern-.025em b}\kern-.08em
    T\kern-.1667em\lower.7ex\hbox{E}\kern-.125emX}}
\usepackage{amsmath}
\usepackage{rotating}
\usepackage{xspace}
\usepackage{caption}
\usepackage{subcaption}
\newtheorem{theorem}{Theorem}
\newtheorem{definition}{Definition}

\usepackage{mathtools}

\usepackage{xspace}

\newcommand{\eat}[1]{}

\newcommand{\stitle}[1]{\vspace{.33em}\noindent\textbf{#1}}

\newcommand{\system}{Crypt$\epsilon$\xspace}

\newcommand{\encD}{\boldsymbol{\tilde{\mathcal{D}}}}
\newcommand{\crossproduct}{\times}
\newcommand{\project}{\pi}
\newcommand{\filter}{\sigma}
\newcommand{\countagg}{count}
\newcommand{\groupbystar}{\gamma^{count}}
\newcommand{\groupby}{\tilde{\gamma}^{count}}
\newcommand{\countdistinct}{countD}
\newcommand{\encT}{\boldsymbol{\tilde{T}}}
\newcommand{\encB}{\boldsymbol{B}}
\newcommand{\encC}{\boldsymbol{c}}
\newcommand{\encV}{\boldsymbol{V}}
\newcommand{\lap}{Lap}
\newcommand{\noisymax}{NoisyMax}

\newcommand{\ldp}{\textsf{LDP}\xspace}
\newcommand{\cdp}{\textsf{CDP}\xspace}

\newcommand{\squishlist}{
	\begin{list}{$\bullet$}
		{
			\setlength{\itemsep}{0pt}
			\setlength{\parsep}{3pt}
			\setlength{\topsep}{3pt}
			\setlength{\partopsep}{0pt}
			\setlength{\leftmargin}{1em}
			\setlength{\labelwidth}{0.8em}
			\setlength{\labelsep}{0.3em} } }
	
\newcommand{\squishend}{
\end{list}  }\newcommand{\squishlistnum}{
	\begin{enumerate}
		{
			\setlength{\itemsep}{0pt}
			\setlength{\parsep}{3pt}
			\setlength{\topsep}{3pt}
			\setlength{\partopsep}{0pt}
			\setlength{\leftmargin}{1.5em}
			\setlength{\labelwidth}{1em}
			\setlength{\labelsep}{0.3em} } }
	
\newcommand{\squishendnum}{
\end{enumerate}  }

\newcommand{\AS}{\textsf{AS}\xspace}
\newcommand{\CSP}{\textsf{CSP}\xspace}

\newcommand{\CPS}{\textsf{CSP}\xspace}

\newif\ifpaper
\paperfalse   %%FULL VERSION
\newtheorem{innercustomgeneric}{\customgenericname}
\providecommand{\customgenericname}{}
\newcommand{\newcustomtheorem}[2]{%
  \newenvironment{#1}[1]
  {%
   \renewcommand\customgenericname{#2}%
   \renewcommand\theinnercustomgeneric{##1}%
   \innercustomgeneric
  }
  {\endinnercustomgeneric}
}
\usepackage{balance}
\newcustomtheorem{customthm}{Theorem}
\newcustomtheorem{customlemma}{Lemma}
\begin{document}
\title{\system: Crypto-Assisted Differential Privacy on Untrusted Servers}
\author{Amrita Roy Chowdhury}
  \affiliation{\institution{University of Wisconsin-Madison}}
    \email{amrita@cs.wisc.edu}
  \author{Chenghong Wang} 
  \affiliation{\institution{Duke University}}
  \email{cw374@duke.edu} 
  \author{Xi He }
  \affiliation{\institution{University of Waterloo}}
  \email{xihe@uwaterloo.ca}
  \author{Ashwin Machanavajjhala}
  \affiliation{\institution{Duke University}}
  \email{ashwin@cs.duke.edu} 
  \author{Somesh Jha} 
  \affiliation{\institution{University of Wisconsin-Madison}}
  \email{jha@cs.wisc.edu}
\begin{abstract}
Differential privacy (DP) is currently the de-facto standard for achieving privacy in data analysis, which is typically implemented either in the ``central'' or ``local'' model. 
The local model has been more popular for commercial deployments as it does not require a trusted data collector. This increased privacy, however, comes at the cost of utility and algorithmic expressibility as compared to the central model. 

In this work, we propose, \system, a  system and programming framework that (1) achieves the accuracy guarantees and algorithmic expressibility of the central model (2) without any trusted data collector like in the local model. \system achieves the ``best of both worlds'' by employing two non-colluding untrusted servers that run DP programs on encrypted data from the data owners. In theory, straightforward implementations of DP programs using off-the-shelf secure multi-party computation tools can achieve the above goal. However, in practice, they are beset with many challenges like poor performance and tricky security proofs. To this end, \system allows data analysts to author logical DP programs that are automatically translated to secure protocols that work on encrypted data. These protocols ensure that the untrusted servers learn nothing more than the noisy outputs, thereby guaranteeing DP (for computationally bounded adversaries) for all \system programs. \system supports a rich class of DP programs that can be expressed via a small set of transformation and measurement operators followed by arbitrary post-processing. Further, we propose performance optimizations leveraging the fact that the output is noisy. We demonstrate \system's practical feasibility with extensive empirical evaluations on real world datasets.
\end{abstract}
\eat{
}
%\begin{IEEEkeywords}
%\keywords{Differential Privacy, Homomorphic Encryption}
\settopmatter{printfolios=true}
\maketitle
\thispagestyle{empty}
%\end{IEEEkeywords}
%\vspace{-4mm}
\section{Introduction}
%There is a growing need for releasing aggregate properties from sensitive datasets in several domains  including social science, healthcare, and advertising. %Differentially private algorithms \cite{dwork}, whose outputs are insensitive to adding or removing a single row in the input dataset, have become the gold standard for these situations. Differential privacy provides a provable and persuasive guarantee of privacy to individuals in a dataset, and has seen adoption by government \cite{machanavajjhala08onthemap,Vilhuber17Proceedings} and commercial organizations \cite{Rappor1,Apple, Samsung}. %It is defined with respect to a privacy parameter $\epsilon > 0$ where lower the value of $\epsilon$, greater is the privacy guaranteed.
Differential privacy (DP) is a rigorous privacy definition that is currently the gold standard for data privacy. %and has enjoyed adoption by both government \cite{machanavajjhala08onthemap,Vilhuber17Proceedings} and commercial organizations \cite{Rappor1,Apple, Samsung}.
It is typically implemented in one of two models -- \textit{centralized differential privacy} (\cdp) and \textit{local differential privacy} (\ldp). In \cdp, data from individuals are collected and stored \textit{in the clear} in a \textit{trusted} centralized data curator which then executes DP programs on the sensitive data  and releases outputs to an untrusted data analyst. %A canonical algorithm in \cdp is the Laplace mechanism where the curator releases the output of a function $f$ by adding noise drawn from the Laplace distribution to hide the presence or absence of one row in the input database. %Depending on infrastructural constraints, like the viable trust model in a given setting, differential private algorithms in practice can have two different styles of implementation.  The more common and historically precedent implementation is the central differential privacy (\textsf{CDP}) model where a trusted data curator collates data from all individuals into a centrally held dataset and processes it in a privacy preserving way. %For example, the data curator could publish differentially private statistics of the data, that allows analysis on the data, without revealing individual information. 
%The curator is trusted to store the data \emph{in the clear} and mediates upon queries posed by a mistrustful analyst, interested in learning some synopsis of this dataset. Privacy is enforced by the curator by adding uncertainty to the answers for analyst's queries before releasing them. The other competing model of implementation, is the local differential privacy model (\textsf{LDP}) where the central data aggregating server is untrusted. Thus every data owner has to individually randomize his/her input before communicating it to the central aggregator. Hence the private data is concealed from the untrusted aggregator who attempts to infer statistics about the true population from the perturbed data instead. 
In \ldp, there is no trusted data curator. Rather, each individual perturbs his/her own data using a (local) DP algorithm. The data analyst uses these noisy data to infer aggregate statistics of the datasets. In practice, \cdp's assumption of a trusted server is ill-suited for many applications as it constitutes a single point of failure for data breaches, and saddles the trusted curator with legal and ethical obligations to uphold data privacy. %For instance, Google Chrome uses the \ldp model rather than \cdp to detect changes in browser properties of its userbase as it does not want the legal burden of storing highly sensitive browser fingerprints in the clear on its servers \cite{Rappor1}. 
Hence, recent commercial deployments of DP  \cite{Rappor1, Apple} have preferred  \ldp over \cdp. However, \ldp's attractive privacy properties comes at a cost. %All DP algorithms ensure privacy by introducing noise into the computation. 
Under the \cdp model, the expected additive error for a aggregate count over a dataset of size $n$ is at most $\Theta(1/\epsilon)$ to achieve $\epsilon$-DP. %(e.g., using the Laplace mechanism \cite{dwork})%. 
In contrast, under the \ldp model, at least $\Omega(\sqrt{n}/\epsilon)$ additive expected error must be incurred by any $\epsilon$-DP program \cite{error1,error2,error3}, owing to the randomness of each data owner. %\cite{Prochlo,Rappor1,Rappor2,LDP1}. 
The \ldp model in fact imposes additional penalties on the algorithmic expressibility;  the power of \ldp is equivalent to that of the statistical query model \cite{SQ1} and there exists an exponential separation between the accuracy and sample complexity of \ldp and \cdp algorithms \cite{Kasivi}. 
% As a consequence, \ldp requires enormous amounts of data to obtain reliable population statistics. Unsurprisingly, only large corporations  like Google \cite{Rappor1,Rappor2,Prochlo} and Apple \cite{Apple} have attempted deploying \ldp.
 
\eat{  Specifically, computation in \textsf{LDP} results in an additional error of $\Omega(\sqrt{n})$ where $n$ is the total number of data owners contributing to the noisy estimate \cite{error1,MPCtools,error3}. In contrast, we get a constant error bound for \textsf{CDP}.
For e.g., for a single counting query, in the \textsf{CDP} model, the trusted data curator first computes the true count in the clear. Next, adding a single instance of noise from the distribution $Lap(\frac{1}{\epsilon})$ to this true count guarantees $\epsilon$-differential privacy. Since the s.t.d of the distribution $Lap(\frac{1}{\epsilon})$ is given by $\frac{1}{\epsilon}$, we get an expected error of $O(\frac{1}{\epsilon})$. On the other hand, owing to the independent coin flips of each reporting data owner, the resulting noise in \textsf{LDP} induces a binomial distribution. This binomial distribution can be approximated by a normal distribution; the magnitude of this random Gaussian
noise however can be very large, its
standard deviation grows in proportion to the square root of
the report count $ \Omega\big(\frac{\sqrt{n}}{\epsilon}\big)$, and the noise is in practice higher by an
order of magnitude \cite{Prochlo,Rappor1,Rappor2,LDP1}. %often growing with the data dimension

 Thus, if a billion individuals'
reports are analyzed, then a common signal from even
up to a million reports may be missed.

The construct of the \textsf{LDP} model in fact imposes additional penalties in terms of the  algorithmic expressibility.  Kasiviswanathan et al. in \cite{Kasivi} showed that the power of \textsf{LDP} is equivalent to that of the statistical query model \cite{SQ1} from learning theory and there exists an exponential separation between the accuracy and sample complexity of local and central algorithms.  As a consequence, \textsf{LDP} requires enormous amounts of data \cite{Kasivi}
to obtain reliable population statistics. Unsurprisingly, only large corporations  like Google \cite{Rappor1,Rappor2,Prochlo} and Apple \cite{Apple}, with  billions of user base have had successful commercial deployment of \textsf{LDP}.
} %More abstractly, the power of the local model is equivalent tothe statistical query model from learning theory and t.

In this paper, we strive to bridge the gap between \textsf{LDP} and \textsf{CDP}. We propose, \system, a system and a programming framework for executing DP programs that: 
\squishlist
\item never stores or computes on sensitive data in the clear
\item achieves the accuracy guarantees and algorithmic expressibility of the \cdp model \squishend 
\system employs a pair of untrusted but non-colluding servers -- Analytics Server (\textsf{AS}) and  Cryptographic Service Provider (\textsf{CSP}). The \textsf{AS} executes DP programs (like the data curator in \cdp) but on \textit{encrypted} data records. The \textsf{CSP} initializes and manages the cryptographic primitives, and collaborates with the \textsf{AS} to generate the program outputs. Under the assumption that the \textsf{AS} and the \textsf{CSP} are semi-honest and do not collude (a common assumption in cryptographic systems \cite{Boneh1,Boneh2,Ridge2,Matrix2,secureML,LReg,Ver}), \system ensures $\epsilon$-DP guarantee for its programs via two cryptographic primitives -- linear homomorphic encryption (LHE) and garbled circuits. 
One caveat here is that due to the usage of cryptographic primitives, the DP guarantee obtained in \system is that of computational differential privacy or \textsf{SIM-CDP} \cite{CDP} (details in Section \ref{sec:security}).

 \system provides a data analyst with a programming framework to author logical DP programs just like in \cdp.  Like in prior work \cite{PINQ, FWPINQ, ektelo}, access to the sensitive data is restricted via a set of predefined transformations operators (inspired by relational algebra) and DP measurement operators (Laplace mechanism and Noisy-Max \cite{Dork}). Thus, any program that can be expressed as a composition of the above operators automatically satisfies $\epsilon$-DP (in the \cdp model) giving the analyst a proof of privacy for free. 
 \system programs support constructs like looping, conditionals, and can arbitrarily post-process outputs of measurement operators.  \\The main contributions of this work are:
\squishlist
\item \textbf{New Approach}: We present the design and implementation of \system, a novel system and programming framework for executing DP programs over encrypted data on two non-colluding and untrusted servers. %\system integrates the constant error bounds of \textsf{CDP} with the low trust assumption of \textsf{LDP}.
\item \textbf{Algorithm Expressibility}: \system supports a rich class of state-of -the-art DP programs expressed in terms of a small set of transformation and measurement operators. Thus, \system achieves the accuracy guarantees of the \cdp model without the need for a trusted data curator.  
\item \textbf{Ease Of Use:} \system allows the data analyst to express the DP program logic using high-level operators. \system automatically translates this to the underlying implementation specific secure protocols that work on encrypted data and provides a DP guarantee (in the \cdp model) for free. Thus, the data analyst is relieved of all concerns regarding secure computation protocol implementation.
%abstracts out all the low-level implementation details %like the choice of input data representation, SMC protocol, key management scheme etc 
%from the data analyst thereby reducing his/her burden of complex decision making. The data analysts need to only encode the DP program logic in terms of the \system operators. 
\item \textbf{Performance Optimizations}: %Existing techniques to compile logical \system programs into cryptographic protocols often result in inefficient programs.
We propose optimizations that speed up computation on encrypted data by at least an order of magnitude. A novel contribution of this work is a DP indexing optimization that leverages the fact that noisy intermediate statistics about the data can be revealed.% as long as DP is satisfied. 
\item \textbf{Practical for Real World Usage}: %We demonstrate the accuracy and efficiency of \system via extensive  evaluation. 
For the same tasks, \system programs achieve accuracy comparable to \textsf{CDP} and $50\times$ more than \textsf{LDP} for a dataset of size $\approx 30K$. \system runs within $3.6$ hours for a large class of programs on a dataset with $1$ million rows and $4$ attributes.
\item \textbf{Generalized Multiplication Using \textsf{LHE}}: Our implementation uses an efficient way for performing $n$-way multiplications using \textsf{LHE} which maybe of independent interest.
\squishend
\ifpaper The full version of the paper is available in \cite{anom}.\fi 
 
%\noindent\textbf{Organization}: Section \ref{sec:overview} presents an overview of \system and Section~\ref{sec:background} introduces the necessary background. We present the system description in Section~\ref{sec:sysdesc}. \system operators and their implementation are outlined in Sections~\ref{sec:operators} and \ref{sec:implementation}, respectively. Section~\ref{sec:optimization} describes \system optimizations and Section~\ref{sec:security} outlines the security sketch for \system. We empirically evaluate \system in Section~\ref{sec:experiments}. Related work and conclusions are discussed in Sections~\ref{sec:related-short} and \ref{sec:conclusions}, respectively.

\vspace{-1mm}
\section{\system Overview} \label{sec:overview}
\begin{figure}[t]
\centering

	\includegraphics[width=1\linewidth,height=3.9cm]{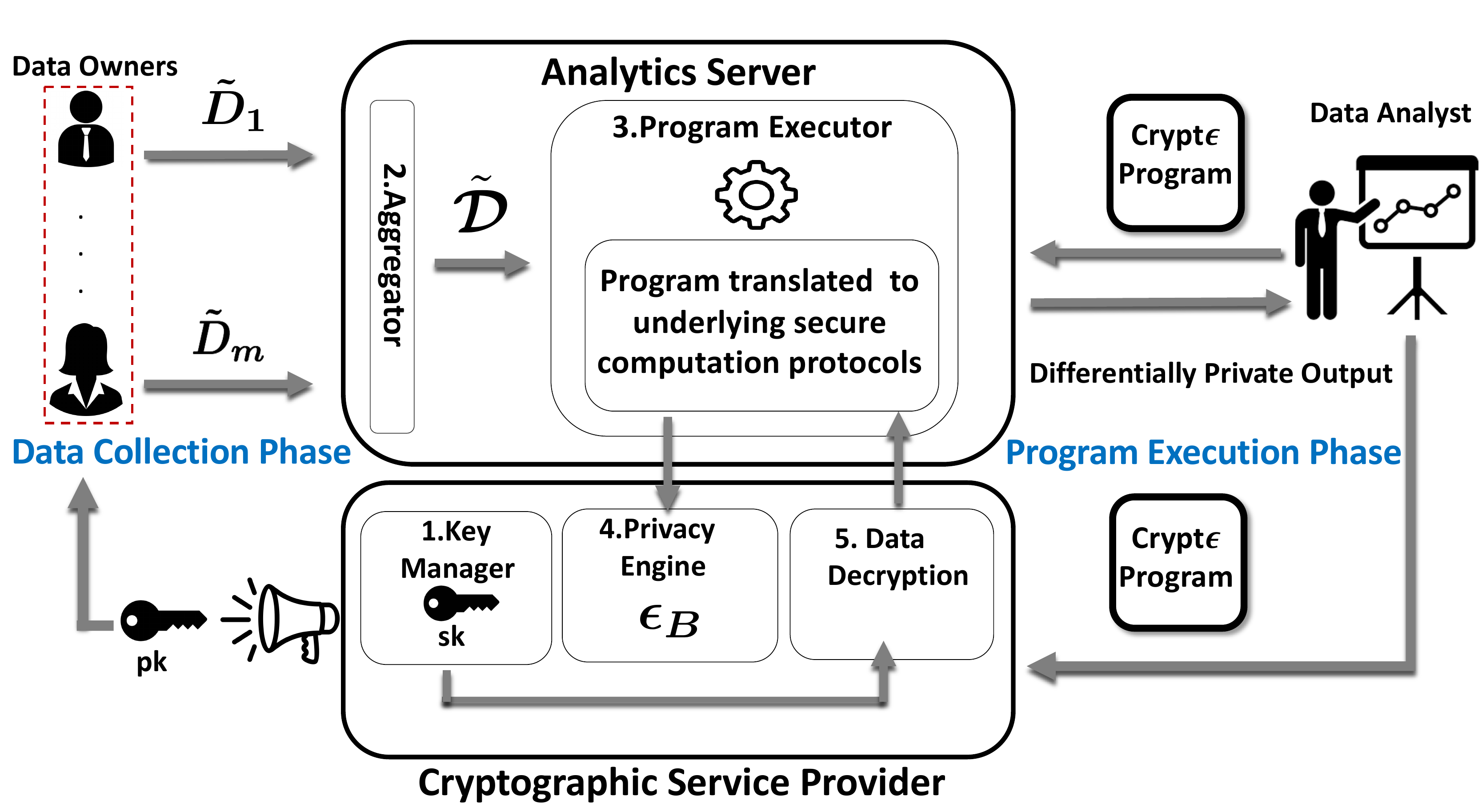}
		\vspace{-6mm}
	\caption{\label{fig:system} Crypt$\epsilon$ System %The  \textsf{AS} executes the Crypt$\epsilon$ programs. The \textsf{CSP} manages the cryptographic primitives.  
	}
	% Setting: The  \textsf{AS} runs the Crypt$\epsilon$ programs. The \textsf{CSP} manages the cryptographic primitves. } 
	\vspace{-0.5cm}
\end{figure}
%This section gives an overview of \system's architecture and design principles. \am{Can drop if you need space}
\subsection{System Architecture}
Figure ~\ref{fig:system} shows \system's system architecture. \system has two servers: Analytics server (\AS) and Cryptographic Service Provider (\CSP). %A set of \textit{data owners} ${\textsf{DO}_i, i\in [m]}$ have private data records ${D_i, i \in [m]}$. \system permits data analysts to author and execute programs $P$ that satisfy DP under the \cdp model without storing or computing on the private data records in the clear. 
At the very outset, the \CSP records the total privacy budget, $\epsilon^B$ (provided by the data owners), and generates the key pair, $\langle sk,pk \rangle $ (details in Section \ref{sec:background}), %\textit{Cryptographic Service Provider},
 for the encryption scheme. The data owners, ${\textsf{DO}_i, i\in [m]}$ ($m$ = number of data owners),  encrypt their data records, ${D_i}$, in the appropriate format %(per attribute one-hot-encoding, details in Section \ref{sec:overview})
 with the public key, $pk$, and send the encrypted records, $\boldsymbol{\tilde{D_i}}$, to the \AS
which aggregates them into a single encrypted database, $\encD$. Next, the \AS inputs logical programs from the data analyst and translates them to \system's implementation specific secure protocols that work on $\encD$.  A \system program typically consists of a sequence of transformation operators followed by a measurement operator. The \AS can execute most of the transformations on its own. %The transformation operators are designed in such a way that the \AS can perform majority of the associated computation on its own. 
 However, each measurement operator requires an interaction with the \textsf{CSP} for (a) decrypting the answer, and (b) checking that %the noise is scaled by the correct sensitivity (verified from the \system program) and that 
the total privacy budget, $\epsilon^B$, is not exceeded. In this way, the \AS and the \CSP compute the output of a \system program with the data owners being offline.
	\vspace{-2mm}\subsection{\system Design Principles}\label{sec:design}
\stitle{Minimal Trust Assumptions}: As mentioned above, the overarching goal of \system is to mimic the \cdp model but without a trusted server. A natural solution for dispensing with the trust assumption of the \cdp model is using cryptographic primitives \cite{Prochlo,mixnets,amplification,Shi,Shi2,kamara,Rastogi,DworkOurData,BeimelSFE+DP,Shrinkwrap}. Hence, to accommodate the use of cryptographic primitives, we assume a computationally bounded adversary in \system. However, a generic $m$-party SMC would be computationally expensive. This necessitates a third-party entity that can capture the requisite secure computation functionality in a 2-party protocol instead. %Moreover, since the data owners are no longer in the loop to monitor every query answering, the aforementioned entity should also watchdog the overall privacy budget expenditure.
This role is fulfilled by the \CSP in \system. For this two-server model, we assume semi-honest behaviour and non-collusion. This is a very common assumption in the two-server model \cite{Boneh1,Boneh2,Ridge2,Matrix2,secureML,LReg,Ver}.

\stitle{Programming Framework}:
Conceptually, the aforementioned goal of achieving the \textit{best of both worlds} can be obtained by implementing the required DP program using off-the-self secure multi-party computation (SMC) tools like \cite{EMP,MPCtools,ScaleMAMBA,ABY}. %This notion is similar in spirit with the works in \cite{DworkOurData,BeimelSFE+DP}. 
However, when it comes to real world usage, \system outperforms such approaches due to the following reasons.
 
 First, without the support of a programming framework like that of \system,  every DP program must be implemented from scratch. This requires the data analyst to be well versed in both DP and SMC techniques; he/she must know how to implement SMC protocols, estimate sensitivity of transformations and track privacy budget across programs. %Moreover, our setting has additional challenges owing to DP program execution on untrusted servers like tracking the \textit{sensitivity}, monitoring the privacy budget expenditure across all programs. %For example, revisiting our above mentioned database schema, let us assume that the data analyst is additionally interested in learning the output of a second query "Count the number of age values having at least 100 records". Typically, both of these queries will require two separate garbled circuits. 
%Thus, in a practical set-up, this is feasible only if %to be able to answer multiple queries on-the-fly 
%the data analyst is well-versed in both DP and SMC techniques. 
In contrast, \system allows the data analyst to write the DP program using a high-level and expressive programming framework. \system abstracts out all the low-level implementation details like the choice of input data format, translation of queries to that format, choice of SMC primitives and privacy budget monitoring
from the analyst thereby reducing his/her burden of complex decision making. Thus, every \system program is automatically translated to protocols corresponding to the underlying implementation. %In principle, any function can be computed on encrypted data via secure computation, and thus \system can support any differentially private algorithm. However, we currently limit the expressibility of the programs supported in \system to those that operate on the sensitive data with efficiently implementable operators, and whose privacy can be easily tracked by the \textsf{CSP}. Nevertheless, as shown later in the paper, \system can already support a rich class of state-of-the-art DP programs.%In fact, owing to this separation of the logical and physical layers, \system's underlying implementation can be based on cryptographic operators of choice. As mentioned above, in this paper we present a prototype \system that uses LHE and garbled circuits.

Second, SMC protocols can be prohibitively costly in practice unless they are carefully tuned to the application. \system supports optimized implementations for a small set of operators, which results in efficiency for all \system programs.

Third, a DP program can be typically divided into segments that (1) transform the private data, (2) perform noisy measurements, and (3) post-process the noisy measurements without touching the private data. A naive implementation may implement all the steps using SMC protocols even though post-processing can be performed in the clear. Given a DP program written in a general purpose programming language (like Python), automatically figuring out what can be done in the clear can be subtle. In \system programs, however, transformation and measurement are clearly delineated, as the data can be accessed only through a pre-specified set of operators. Thus, SMC protocols are only used for transformations and measurements, which improves 
performance. 

%Thus, in order to implement a given DP program using SMC operators, the data analyst has to first group the operations on the basis of whether it falls inside or outside the trust boundary. Otherwise, one has to implement the entire DP program inside a single protocol which will degrade performance. 
For  example, the AHP algorithm for histogram release \cite{AHP} works as follows: first, a noisy histogram, $\hat{H}$, is released using budget $\epsilon_1$. This is followed by post-processing steps of thresholding, sorting and clustering resulting in $\bar{H}$. Then a final histogram, $\tilde{H}$, is computed with privacy budget $\epsilon-\epsilon_1$. An implementation of the entire algorithm in a single SMC protocol using the EMP toolkit \cite{EMP} takes 810s for a dataset of size $\approx 30K$ and histogram size $100$.  In contrast, \system uses SMC protocols only for the first and third steps. \system automatically detects that the second post-processing step can be  performed in the clear.  A \system program for this runs in 238s ($3.4\times$ less time than that of the EMP implementation) for the same dataset and histogram sizes.

Last, the security (privacy) proofs for just stand-alone cryptographic and DP mechanisms can be notoriously tricky \cite{BellareCryptoError,DPSVTProof}. Combining the two thus exacerbates the technical complexity, making the design vulnerable to faulty proofs \cite{He:2017:CDP}. For example, given any arbitrary DP program written under the \cdp model, the distinction between intermediate results that can be released and the ones which have to be kept private is often ambiguous. An instance of this is observed in the Noisy-Max algorithm, where the array of intermediate noisy counts is private.  However, these intermediate noisy counts correspond to valid query responses. Thus, an incautious analyst, in a bid to improve performance, might reuse a previously released noisy count query output for a subsequent execution of the Noisy-Max algorithm leading to privacy leakage. In contrast, \system is designed to reveal nothing other than the outputs of the DP programs to the untrusted servers; every \system program comes with an automatic proof of security (privacy). Referring back to the aforementioned example, in \system, the Noisy-Max algorithm is implemented as a secure measurement operator thereby preventing any accidental privacy leakage. The advantages of a programming framework is further validated by the popularity of systems like PINQ \cite{PINQ}, Featherweight PINQ \cite{FWPINQ}, Ektelo \cite{ektelo} - frameworks for the \textsf{CDP} setting.

\stitle{Data Owners are Offline}:
 Recall, \system's goal is to mimic the \cdp model with untrusted  servers. Hence, it is designed so that the data owners are offline after submitting their encrypted records to the \textsf{AS}. %With online data owners, some computation can be offloaded to the data owners thereby improving efficiency. %This is beneficial as the \textsf{AS} does not need to maintain communication channels with the data owners over a long period of time. 
 %If the data owners were online, the efficiency of some programs could be improved as some of the computation %from the \textsf{AS} and \textsf{CSP} 
 %can be offloaded to the data owners.

\stitle{Low burden on CSP}: \system views the \textsf{AS} as an extension of the analyst; the \AS has a vested interest in obtaining the result of the programs. Thus, we require the \textsf{AS} to perform the majority of the work for any program; interactions with the \textsf{CSP} should be minimal and related to data decryption. Keeping this in  mind, the \textsf{AS} performs most of the data transformations by itself (Table \ref{perf}). Specifically, for every Crypt$\epsilon$ program, the \textsf{AS} processes the whole database and transforms it into concise representations (an encrypted scalar or a short vector) which is then decrypted by the \textsf{CSP}. An example real world setting can be when Google and Symantec assumes the role of the \AS and the \CSP respectively.

\stitle{Separation of logical programming framework and underlying physical implementation}: The programming framework is independent of the underlying implementation. This allows certain flexibility in the choice for implementation. 
 For example, we use one-hot-encoding as the input data format (Section \ref{sec:overview}). However, any other encoding scheme like range based encoding can be used instead. %Quite evidently, different encoding scheme will have different storage requirements. For instance, a k-attribute one-hot-encoding will be of size $d^k$ where $d$ is the domain size of each attribute. We choose one-hot-coding for our prototype \system implementation because it is a general representation that can be easily translated to other encoding schemes.  
 Another example is that in this paper, we use $\epsilon$-DP (pure DP) for our privacy analysis. However, other DP notions like $(\epsilon,\delta)$-DP, R\`enyi DP \cite{RDP} can also be used instead.  %an operator for adding Gaussian noise would suffice to enable RDP. 
 Similarly, it is straightforward to replace \textsf{LHE} with the optimized \textsf{HE} scheme in \cite{Blatt2019OptimizedHE} or garbled circuits with the ABY framework \cite{Demmler2015ABYA}. 
 
Yet another alternative implementation for Crypt$\epsilon$ could be where the private database is equally shared between the two servers and they engage in a secret share-based SMC protocol for executing the DP programs. This would require both the servers to do almost equal amount of work for each program. Such an implementation would be justified only if both the servers are equally invested in learning the DP statistics and is ill-suited for our context. A real world analogy for this can be if Google and Baidu decide to compute some statistics on their combined user bases. %We speculate that the above scenario is less likely than the other setting. Additionally, oblivious transfer (OT) is the  bottleneck for most SMC implementations \cite{OT:bottleneck:1,OT:bottleneck:2,OT:bottleneck:3} and each AND gate involves OT in a secret share based protocol. Thus a secret share based approach would be communication heavy degrading performance. %\am{One option is to not say anything about secret sharing in previous point, but say that we can change the implementation to a 2-party secret sharing ...} %Of course, as mentioned above, a two server secret share based implementation is also possible. 

\vspace{-1mm}\section{Background}\label{sec:background}
%In this section we give a brief introduction to definitions and primitives relevant to \system. \am{Can drop if you need space.}
\subsection{Differential Privacy}
\begin{definition} \label{def:dp}
An algorithm $\mathcal{A}$
satisfies $\epsilon$-differential privacy ($\epsilon$-DP), where $\epsilon > 0$ is a privacy parameter, iff
 for any two neighboring datasets $D$ and $D'$ such that $D = D' - {t}$ or $D' = D - {t}$, we have\vspace{-1mm}
\begin{gather}
\vspace{-2mm}
\small{\forall S \subset Range(\mathcal{A}), Pr \big[\mathcal{A}(D) \in S\big] \leq e^{\epsilon}Pr\big[\mathcal{A}(D') \in S\big]} \vspace{-2mm}
\end{gather}\vspace{-4mm}
\end{definition}
%\textit{where $Range(\mathcal{A})$ denotes the set of all possible outputs of$\mathcal{A}$}.
The above definition is sometimes called \textit{unbounded DP}. A variant is \textit{bounded-DP} where neighboring datasets $D$ and $D'$ have the same number of rows and differ in one row. Any $\epsilon$-DP algorithm also satisfies $2\epsilon$-bounded DP \cite{bookDP}.\vspace{-1mm}
\begin{theorem}(Sequential Composition) \label{theorem:seq}
If $\mathcal{A}_1$ and
$\mathcal{A}_2$ are $\epsilon_1$-DP and $\epsilon_2$-DP algorithms with independent randomness, then releasing $\mathcal{A}_1(D)$ and
$\mathcal{A}_2(D)$ on database $D$ satisfies $\epsilon_1+\epsilon_2$-DP.\vspace{-1mm}\end{theorem} 
%There exist advanced composition theorems that give tighter privacy lossbounds, but weuse Theorem 1 for our paper.
%Another important result is that any post-processing computation performed on the noisy output of a differentially private algorithm does not cause any loss in privacy.
\begin{theorem}(Post-Processing)\label{post}
Let $\mathcal{A}: D \mapsto R$ be a randomized
algorithm that is $\epsilon$-DP. Let $f : R \mapsto R'$ be an
arbitrary randomized mapping. Then $f \circ \mathcal{A} : D \mapsto R'$ is $\epsilon$-
DP. \vspace{-1mm}\end{theorem}
%For other basic properties of differential privacy we refer the readers to the brilliant book by Dwork et al. \cite{Dork}.
\subsection{Cryptographic Primitives}
\stitle{Linearly Homomorphic Encryption (\textsf{LHE}):}
If $(\mathcal{M}, +)$ is a finite group, an \textsf{LHE} scheme
for messages in $\mathcal{M}$ is: \squishlist
\item \textbf{Key Generation }($Gen$): This  algorithm takes the security parameter $\kappa$ as input and outputs
a pair of secret and public keys, $\langle s_k, p_k\rangle \leftarrow Gen(\kappa)$.
\item \textbf{Encryption} ($Enc$): This is a randomized algorithm that encrypts a message, $m \in \mathcal{M}$, using the public key, $p_k$, to generate the ciphertext, $\mathbf{c} \leftarrow Enc_{pk}(m)$.
\item \textbf{Decryption} ($Dec$): This uses the secret key, $s_k$, to
recover the plaintext, $m$, from the ciphertext, $\mathbf{c}$, deterministically.
\squishend
In addition, \textsf{LHE} supports the operator $\oplus$ that allows the summation of ciphers as follows:
\\ \textbf{Operator} $\oplus$: Let $c_1 \leftarrow Enc_{pk}(m1), \ldots, c_a \leftarrow Enc_{pk}(m_a)$ and $a \in \mathcal{Z}_{>0}$. Then we have  $Dec_{sk}(c_1\oplus c_2 ...\oplus c_a)=    m_1 + \ldots   + m_a$.  \\
One can multiply a cipher $c\leftarrow  Enc_{sk}(m)$ by a plaintext positive integer $a$ by $a$ repetitions of $\oplus$. We denote this operation by $cMult(a,c)$ such that $Dec_{sk}\big(cMult(a,c)\big)=a\cdot m$. 

\stitle{Labeled Homomorphic Encryption(\textsf{labHE})}:
Any \textsf{LHE} scheme can be extended to a \textsf{labHE} scheme \cite{Barbosa2017LabeledHE} with the help of a pseudo-random function.   In addition to the operations supported by an \textsf{LHE}  scheme, \textsf{labHE} supports multiplication of two \textsf{labHE} ciphers (details in \ifpaper 
 full paper \cite{anom}% using paper 
\else 
 Appendix \ref{app:background}
\fi).  \eat{Let $(Gen,Enc,Dec)$ be an \textsf{LHE} scheme with security parameter $\kappa$ and message space $\mathcal{M}$. Assume that a multiplication operation exists in $\mathcal{M}$, i.e., is a finite ring. Let $\mathcal{F}:\{0,1\}^s \times \mathcal{L}\rightarrow \mathcal{M}$ be a pseudo-random function with seed space $\{0,1\}^s$( s= poly($\kappa $)) and the label space $\mathcal{L}$. A \textsf{labHE} scheme is defined as
\squishlist
 \item $\textbf{labGen}(\kappa):$ Runs $Gen(\kappa)$ and outputs $(sk,pk)$.
\item $\textbf{localGen}(pk):$ For each user $i$ and with the public key as input, it samples a random seed $\sigma_i \in \{0,1\}^s$ and computes $pk_i = Enc_{pk}(\underline{\sigma_i})$ where $\underline{\sigma_i}$ is an  encoding of $\sigma_i$ as an  element of $\mathcal{M}$. It outputs $(\sigma_i,pk_i)$.
\item $\textbf{labEnc}_{pk}(\sigma_i, m , \tau):$ On input a message $m \in \mathcal{M} $ with label $\tau \in \mathcal{L}$  from user $i$, it computes $b=\mathcal{F}(\sigma_i, \tau)$ (mask) and outputs the labeled ciphertext $\mathbf{c}=(a,d) \in \mathcal{M} \times \mathcal{C}$ with $ a= m- b$  (hidden message) in $\mathcal{M}$ and $d=Enc_{pk}(b)$. For brevity we just use notation $\textbf{labEnc}_{pk}(m)$ to denote the above functionality, in the rest of paper. 
\item $\textbf{labDec}_{sk}(\mathbf{c}):$ This functions inputs a cipher $\mathbf{c}=(a,d) \in \mathcal{M} \times \mathcal{C}$  and decrypts it as $m=a-Dec_{sk}(d)$.\squishend
%\textsf{LHE} and \textsf{labHE} provides semantic security guarantee \cite{Katz}. 
}

\stitle{Garbled Circuit}:
Garbled circuit \cite{Yao,Yao2}  is a generic method for secure computation. Two data owners with respective private inputs $x_1$ and $x_2$ run the protocol such that, no data owner learns more  
than $f(x_1,x_2)$ for a function $f$.  One of the data owners, called
generator, builds a "garbled" version of a circuit for $f$ and sends it to the other data owner, called evaluator, alongside the garbled input values for $x_1$.  The evaluator, then, obtains the garbled input for $x_2$ from the generator via oblivious transfer and computes  $f(x_1, x_2)$.
%A more detailed discussion is presented in Appendix A.

%\vspace{-2em}
\section{\system System Description}\label{sec:sysdesc}
In this section, we describe \system's workflow (Section~\ref{sec:wf}), modules (Section~\ref{sec:modules}), and trust assumptions (Section~\ref{sec:trust}). 
\vspace{-1em}
\subsection{\system Workflow}\label{sec:wf}
\system operates in three phases: \\
(1) \textbf{Setup Phase}: At the outset, data owners initialize the \textsf{CSP} with a privacy budget, $\epsilon^B$, which is stored in its \textit{Privacy Engine} module. Next, the  \textsf{CSP}'s \textit{Key Manager} module generates key pair $(sk,pk)$ for labHE, publishes $pk$ and stores $sk$. 
\\(2) \textbf{Data Collection Phase}: In the next phase, each data owner encodes and encrypts his/her record using the \textit{Data Encoder} and \textit{Data Encryption} modules and sends the encrypted data records to the \textsf{AS}. The data owners are relieved of all other duties and can go completely offline. The \textit{Aggregator} module of the \textsf{AS}, then, aggregates these encrypted records into a single encrypted database, $\boldsymbol{\tilde{\mathcal{D}}}$.
\\(3) \textbf{Program Execution Phase}: In this phase, the \textsf{AS} executes a \system program provided by the data analyst. \system programs (details in Sections~\ref{sec:operators} and \ref{sec:implementation}) access the sensitive data via a restricted set of transformation operators, that filter, count or group the data, and measurement operators, which are DP operations to release noisy answers. Measurement operators need interactions with the \textsf{CSP} as they require (1) decryption of the answer, and (2) a check that the privacy budget is not exceeded. These  functionalities are achieved by \CSP's \textit{Data Decryption} and \textit{Privacy Engine} modules.

 The \emph{Setup} and \emph{Data Collection} phases occur just once at the very beginning, every subsequent program  is handled via the corresponding  \emph{Program Execution} phase. %We next detail the roles of the different modules in the data owners, the Analytics Server and the Cryptographic Service Provider.  
\eat{
\begin{table}[t]
\centering
\caption {Notations}
\scalebox{0.8}{
 \begin{tabular}{l|l}  \toprule
 \multicolumn{1}{c}{\textbf{Symbol} } &  \multicolumn{1}{c}{\textbf{Explanations}}\\\midrule
\textbf{Boldface}& \text{- represents encrypted data}\\
$\tilde{}$ & \text{- represents one-hot-coding}  \\  $\hat{}$ & - represents a differentially private output  \\ $A$ &- an attribute  \\ $s_A$ &- size of domain of attribute $A$
\\$dom(A)=\{v_1,\ldots,v_{s_A}\}$ & - domain of attribute $A$\\ $ct_{A,i}$  &- \# \text{records with value $v_i$ for attribute} A\\ $m$   &- \text{\# number of data onwers}\\ $\boldsymbol{\tilde{\mathcal{D}}}$  &- \text{encrypted database with records in}\\&\text{  per-attribute one-hot-coding } \\ %$\mathcal{A}=\{\mathcal{A}_1,...\mathcal{A}_l\}$   &- \text{set of attributes in the schema of $\boldsymbol{\tilde{\mathcal{D}}}$}\\
$x \times y \text{ table } \mathbf{T}$   &- \text{an encrypted table  with $x$ records in}\\&\text{ one-hot-coding and $y$ columns one for}\\&\text{ each attribute; serves as one of the }\\&\text{ inputs to a transformation operator}\\ $\mathbf{B}$&- \text{a vector of length $m$ such that each entry}\\&\text{ $\textbf{B}[i]$ represents whether record $r_i, i \in [m]$}\\& \text{is relevant to the program at hand} \\ $V$ & -\text{ represents a vector}\\$c$ &- \text{represents a scalar}\\$\mathcal{P}$ & - \text{represents a set}\\
 \bottomrule
 \end{tabular}}
 \label{Notations}
\end{table}}
\vspace{-.5em}
\subsection{\system Modules}\label{sec:modules}
\vspace{-.5em}
\stitle{Cryptographic Service Provider (\textsf{CSP})}\\
(1)\textbf{ Key Manager}:  The \textit{Key Manager} module initializes the \textsf{labHE} scheme for \system by generating its key pair, $\langle sk,pk\rangle$. It stores the secret key, $sk$, with itself and releases the public key, $pk$. The \textsf{CSP} has exclusive access to the secret key, $sk$, and is the only entity capable of decryption in \system.\\
(2)\textbf{ Privacy Engine}: Crypt$\epsilon$ starts off with a total privacy budget of $\epsilon^B$ chosen by the data owners. The choice of value for $\epsilon^B$ should be guided by social prerogatives \cite{abowd19:social,e1,e2}
and is currently outside the scope of Crypt$\epsilon$. For executing any program, the \textsf{AS} has to interact with the \textsf{CSP} at least once (for decrypting the noisy answer), thereby allowing the \textsf{CSP} to monitor the \textsf{AS}'s actions in terms of privacy budget expenditure. The \textit{Privacy Engine} module gets the program, $P$, and its allocated privacy budget, $\epsilon$, from the data analyst, and maintains a public ledger that records the privacy budget spent in executing each such program. Once the privacy cost incurred reaches
$\epsilon^B$, the \textsf{CSP} refuses to decrypt any further answers. This ensures that the total privacy budget is never exceeded.  The ledger is completely public allowing any data owner to verify it.\\ %as and when desired.
(3)\textbf{ Data Decryption}: The \textsf{CSP} being the only entity capable of decryption,  any measurement of the data (even noisy) has to involve the \textsf{CSP}. The \textit{Data Decryption} module is tasked with handling all such interactions with the \textsf{AS}. 

\stitle{Data Owners (\textsf{DO})}\\
(1)\textbf{ Data Encoder}: Each data owner, $\textsf{DO}_i, i \in [m]$, has a private data record, $D_i$, of the form $\langle A_1,...A_l\rangle$ where ${A}_j$ is an attribute. At the very outset, every data owner,  $\textsf{DO}_i$, represents his/her private record, $D_i$, in its respective per attribute one-hot-encoding format. The one-hot-encoding is a way of representation for categorical attributes and is illustrated by the following example. 
If the database schema is given by  $\langle Age,Gender\rangle$, then the corresponding one-hot-encoding representation for a data owner, $\textsf{DO}_i, i \in [m]$, with the record $\langle 30, Male\rangle$, is given by \scalebox{0.8}{${\tilde{D_i}=\langle[\underbrace{0,\ldots,0}_{29},1,\underbrace{0,\ldots,0}_{70}],[1,0]\rangle}$}. \\
(2)\textbf{ Data Encryption}: The \textit{Data Encryption} module stores the public key $pk$ of \textsf{labHE}  which is announced by the \CSP. Each data owner, $\textsf{DO}_i, i \in [m]$,  performs an element-wise encryption of his/her per attribute one-hot-encodings using $pk$ and sends the encrypted record, $\tilde{\mathbf{D_i}}$,  to the \textsf{AS} via a secure channel. \eat{In our aforementioned example, we get
\begin{eqnarray}
\mathbf{\tilde{D}}&=&\langle[\underbrace{labEnc_{pk}(0),\ldots}_{29},labEnc_{pk}(1),\nonumber \\
&& \underbrace{\ldots,labEnc_{pk}(0)}_{70}], [labEnc_{pk}(1),labEnc_{pk}(0)]\rangle. \nonumber \end{eqnarray}}
 This is the only interaction that a data owner ever participates in and goes offline after this.

\stitle{Analytics Server (\textsf{AS)}}\\
(1)\textbf{  Aggregator}: The \textit{Aggregator} collects the encrypted reco- rds, $\mathbf{\tilde{D}}_i$, from each of the data owners, $\textsf{DO}_i$, and collates them into a single encrypted database, $\boldsymbol{\tilde{\mathcal{D}}}$. %It also performs some pre-computations and creates certain data structures like DP index, DP range tree (details in Section \ref{sec:optimization}) that help in improving the performance of \system.%Note that in contrast, the server in the \textsf{CDP} model, being trusted, stores the data in the clear whereas in the \textsf{LDP} model the untrusted server stores appropriately randomized (noisy) data.
\\(2)\textbf{ Program Executor}: This module inputs a logical \system program, $P$, and privacy parameter, $\epsilon$, from the data analyst, translates $P$ to the implementation specific secure protocol and computes the noisy output with the \textsf{CSP}'s help.
\vspace{-3mm}
\subsection{Trust Model}\label{sec:trust}
%From the data owners perspective the trust assumption of \system is similar to that of \textsf{LDP}; the data owners need not place their trust in any external entity.
There are three differences in \system from the \textsf{LDP} setting:\\
\noindent(1) \textbf{Semi-honest Model}: We assume that the \textsf{AS} and the \textsf{CSP} are   \textit{semi-honest}, i.e., they follow the protocol honestly, but their contents and computations can be observed by an adversary. Additionally, each data owner has a private channel with the \textsf{AS}. For real world scenarios, the semi-honest behaviour can be imposed via legal bindings. Specifically, both the \AS and the \CSP can swear to their semi-honest behavior in legal affidavits; there would be loss of face in public and legal implications in case of breach of conduct.
\\(2) \textbf{Non-Collusion}: We assume that the \textsf{AS} and the \CSP are \emph{non-colluding}, i.e., they avoid revealing information \cite{non-collusion} to each other beyond what is allowed by the protocol definition. This restriction can be imposed via strict legal bindings as well. Additionally, in our setting the \CSP is a third-party entity with no vested interested in learning the program outputs. Hence, the \CSP has little incentive to collude with the \AS. Physical enforcement of the non-collusion condition can be done by implementing the \CSP inside a trusted execution environment (TEE) or via techniques which involve using a trusted mediator who monitors the communications between the servers  \cite{non_collusion}. %That is, they always follow the instructions of the protocol faithfully but strive to learn extra information about the private records from the messages received during the execution of the protocol. 
\\(3) \textbf{Computational Boundedness}: The adversary is \textit{computationally bounded}. Hence, the DP guarantee obtained is that of computational differential privacy or SIM-CDP \cite{CDP}. There is a separation between the algorithmic power of computational DP and information-theoretic DP in the multi-party setting \cite{CDP}. Hence, this assumption is inevitable in \system.

\eat{\textcolor{blue}{Table~\ref{DPCompare} summarizes the distinctions between \ldp , \cdp and \system.}
\begin{table}[t]
\centering
\caption {\textcolor{blue}{Comparative analysis of different DP models}}
\scalebox{0.7}{ \begin{tabular}{|c| c c c|}  \toprule
\multicolumn{1}{|c}{\textbf{Features}} & \textbf{LDP}  & \textbf{CDP}  & \textbf{Crypt$\epsilon$}  \\ [0.5ex]
 \hline \hline\# Centralized Servers & 1& 1 & 2\\\hline
Trust Assumption & & & Untrusted \\   for Centralized & Untrusted & Trusted & Semi Honest \\ Server &  &   &  Non-Colluding  \\ \hline
Data Storage & \multirow{2}{*}{N/A} & \multirow{2}{*}{Clear} & \multirow{2}{*}{Encrypted} \\in Server & &  &  \\\hline
\multirow{2}{*}{Adversary} & Information & Information & Computationally \\& Theoretic & Theoretic & Bounded\\\hline
 Error on Statistical Counting Query& $O\Big(\frac{\sqrt(n)}{\epsilon}\Big)$& $O\Big(\frac{1}{\epsilon}\Big)$ & $O\Big(\frac{1}{\epsilon}\Big)$\\
  [1ex]
 \bottomrule
 \end{tabular}}\label{DPCompare}
\end{table}}

\eat{\subsection{\system Design Principles}\label{sec:discuss-arch}
The design of \system is guided by the following principles. 

\stitle{Expressibility:} \system is designed to ensure that state-of-the-art DP programs can be executed on sensitive data by the \textsf{AS}. This necessitates adding noise to functions computed on the entire database (like in \cdp), and not just to individual records (like in \ldp). In principle, any function can be computed on encrypted data via secure computation, and thus \system can support any DP algorithm. However, we currently limit the expressibility of the programs supported in \system to those that operate on the sensitive data with efficiently implementable operators, and whose privacy can be easily tracked by the \textsf{CSP}. Nevertheless, as shown later in the paper, \system can already support a variety of state-of-the-art DP programs that provide orders of magnitude higher error than their \ldp counterparts. %The separation between \textsf{LDP} and \system is discussed further in Appendix \ref{app:sepldp}.

\stitle{Minimal trust assumptions:} The only assumptions we make are that adversaries are computationally bounded and that the \textsf{AS} and  \textsf{CSP} do not collude. The former allows us to use cryptographic tools, and is in tune with a growing line of work that seeks to address \cdp's trust assumptions via cryptographic operators \cite{Prochlo,mixnets,amplification,Shi,Shi2,kamara,Rastogi}. The latter assumption of non-colluding servers is a popular model for privacy preserving computations \cite{Boneh1,Boneh2,Ridge2,Matrix2,secureML,LReg,Ver}.

\stitle{Data owners are offline:} \system's goal is to mimic the \cdp model with untrusted centralized servers. Hence, it is designed so that the data owners are offline once they submit their encrypted records to the \textsf{AS}. This is beneficial as the \textsf{AS} does not need to maintain communication channels with the data owners over a long period of time. If the data owners were online, some of our programs can be made more efficient as some of the computation from the \textsf{AS} and \textsf{CSP} can be offloaded to the data owners.

\stitle{Low burden on \textsf{CSP}:} \system views the \textsf{AS} as an extension of the analyst, and that it has a vested interest in obtaining the result of the computation.   Thus in \system, we require the \textsf{AS} to  shoulder the major chunk of the workload for any \system program execution; interactions with the \textsf{CSP} should be minimal and only related to data decryption. Keeping this in  mind, we design the \textsf{AS} to perform most of the data transformations by itself (Table \ref{perf}). Specifically for every Crypt$\epsilon$ program, the \textsf{AS} processes the whole database and transforms it into concise representations (like an encrypted scalar or a short vector) which is then decrypted with the help of the \textsf{CSP}. An example setting in the real world can be when Google assumes the role of the \AS and Symantec can provide the services of the \CSP. It is interesting to note that we could have had an alternative implementation for Crypt$\epsilon$ where the private database is equally shared between the two servers and they engage in a secret share based secure computation protocol for computing the differentially private answers. However, this would require both the servers to do almost equal amount of work for every program. Such an arrangement would be justified only if both the servers are equally invested in learning the differentially private statistics and is ill-suited for Crypt$\epsilon$. A real world analogy for this can be when Google and Baidu decide to compute some statistics on their combined user bases. We speculate that the above scenario is much less likely than the other setting. Additionally, oblivious transfer (OT) is the major bottleneck for most SMPC implementations \cite{OT:bottleneck:1,OT:bottleneck:2,OT:bottleneck:3} and in a secret share based computation, each AND (equivalently multiplication) gate involves an OT step. Thus a secret share based implementation would be much more communication intensive resulting in a performance hit. 

\stitle{Separation of the logical programming framework and the underlying physical implementation:} The programming framework (i.e., the set of \system operators) is independent from the underlying cryptographic operator specific implementation. This separation between the logical and physical layers allows certain flexibility in the choice for lower level physical implementation as follows --
\\(1) \stitle{Choice of input data representation: } For the prototype \system presented in this paper, as discussed above the input data is represented in per attribute one-hot-encoding. However, we can easily switch to any other encoding scheme like multi-attribute one-hot-encoding, range based encoding etc. Quite evidently, different encoding scheme will have different storage requirements. For example, a k-attribute one-hot-encoding will be of size $d^k$ where $d$ is the domain size of each attribute. We choose one-hot-encoding for our prototype \system implementation because it is a general representation that can be easily translated to other encoding schemes.  
\\(2) \stitle{Choice of cryptographic operators: } As discussed above, due to our design choice of having low burden on the \CSP, we implement \system using LHE and garbled circuits. However translation to an implementation based on other cryptographic operators is straighforward. For example, the optimized HE scheme in \cite{Blatt2019OptimizedHE} can be used in place of LHE. Similarly, garbled circuits can be replaced by the mixed protocol ABY framework \cite{Demmler2015ABYA}. Of course, as mentioned above, a two server secret share based implementation is also possible. 
\\ (3) \stitle{Choice of DP operators: }Although in this paper, we use $\epsilon$-DP (pure DP) for our privacy analysis, \system can be easily extended to accommodate other DP notions like $(\epsilon,\delta)$-DP, R\`enyi differential privacy (RDP) \cite{RDP} etc. For example, designing a operator for adding Gaussian noise would suffice to allow RDP analysis in \system.
}
%!TEX root = main.tex
%\vspace{-2mm}
\section{\system Operators}\label{sec:operators}
Let us consider an encrypted instance of a database, $\encD$, with   schema $\langle A_1,\ldots,A_l \rangle$. %\am{Do we need to use the notation $\tilde{A}_i$ to denote a one-hot-encoded and encrypted version of the original attribute A?} %, \system permits data analysts to author \textit{logical} programs on $\encD$ with DP guarantee. % The logical programs mainly consist of data transformation operators inspired by relational algebra and differentially private measurement operators. These programs can  have constructs like looping and conditionals, and can arbitrarily post-process outputs of measurement operators. \system compiles these logical programs into \system protocols that can work on the encrypted data on the \textsf{AS} and \textsf{CSP}. Though these logical \system programs are designed to run on encrypted data, when operating them on plaintext data, they give differential privacy under \textsf{CDP}.
%\am{Should we say that logical \system program when operating on the raw data gives you DP under CDP? }
In this section, we define the \system operators (summarized in Table~\ref{tab:operators}) and illustrate how to write logical \system programs for DP algorithms on $\encD$. The design of \system operators are inspired by previous work \cite{ektelo, PINQ}. %These programs allow constructs like looping and conditionals, and can arbitrarily post-process outputs of measurement operators. 
% \usepackage{multirow}
\begin{table*}
\caption {\system Operators}\label{tab:operators}%\vspace{-3mm}
\scalebox{0.8}{\begin{tabular}{|l|l|l|l|l|l|}
\hline
\bf{Types}                           & \bf{Name}         & \bf{Notation} & \bf{Input} & \bf{Output} & \bf{Functionality} \\ \hline \hline
\multirow{7}{*}{Transformation} & \multirow{2}{*}{\textsf{CrossProduct}} &  \multirow{2}{*}{$\crossproduct_{A_i,A_j\rightarrow A'}(\cdot)$} & 

\multirow{2}{*}{$\encT$}   &  \multirow{2}{*}{$\encT'$}   & Generates a new attribute  $A'$ (in one-hot-coding) to represent\\ & & & & & the data for both the attributes $A_i$ and $A_j$   \\ \cline{2-6}
                                & \textsf{Project}     & $\project_{A^*}(\cdot)$  &  $\encT$       &  $\encT'$       &  Discards all attributes but $A^*$ \\ \cline{2-6}
                                & \textsf{Filter}       & $\filter_{\phi}(\cdot)$   &  $\encT$      &   $\encB'$     &  Zeros out records not satisfying $\phi$    in $\encB$          \\ \cline{2-6}
                                & \textsf{Count}             & $\countagg(\cdot)$         & $\encT$  &  $\encC$    &  Counts the number of 1s in $\encB$               \\ \cline{2-6}
                                & \textsf{GroupByCount}             & $\groupbystar_{A}(\cdot)$ &  $\encT$  & $\encV$    & Returns encrypted histogram of $A$                \\ \cline{2-6}
                                & \textsf{GroupByCountEncoded}              & $\groupby_{A}(\cdot)$    &$\encT$       & $\tilde{\encV}$        &  Returns encrypted histogram of $A$ in one-hot-encoding    \\ \cline{2-6}
                                & \textsf{CountDistinct}             & $\countdistinct(\cdot)$     &    $\encV$   & $\encC$   &  Counts the number of non-zero values in $\encV$       \\ \hline
\multirow{2}{*}{Measurement}    & \textsf{Laplace}     & $\lap_{\epsilon,\Delta}(\cdot)$    &  $\encV\textbackslash \encC$      &   $\hat{V}$    &  Adds Laplace noise to $\encV$    \\ \cline{2-6}
                                & \textsf{NoisyMax}     & $\noisymax_{\epsilon,\Delta}^k(\cdot)$         & $\encV$       &  $\hat{\mathcal{P}}$      &  %Adds Laplace noise to $\encV$ and 
                                Returns indices of the top $k$ noisy values              \\ \hline
\end{tabular}
}%\vspace{-1mm}
\end{table*}
\eat{
\begin{table*}[h!]
\small
\caption {\system Primitives}
 \begin{tabular}{@{}l c@{}c@{}l@{}}  \toprule
\multicolumn{1}{c}{\textbf{Primitives}} & \textbf{Input}  & \textbf{Output}  & \multicolumn{1}{c}{\textbf{Functionality}}  \\ [0.5ex] 
 \midrule \midrule \textsf{CrossProduct} &$\tilde{\mathbf{T}}, A_i, A_j$ & $\tilde{\mathbf{T'}}$ & Generates  one-hot-coding for the\\&& & attribute $A_i\times A_j$ 
 \\\textsf{Project}&$\tilde{\mathbf{T}}, A^*$&$\tilde{\mathbf{T'}}$ &Discards all attributes but $A^*$
 \\\textsf{Filter}&$\tilde{\mathbf{T}},\phi$&$\tilde{\mathbf{B}}$&Zeros out records not satisfying $\phi$
 \\\textsf{Count}&$\tilde{\mathbf{T}}$&$\mathbf{C}$&Counts the number of records in $\textbf{B}$ 
 \\$\textsf{GroupBy}^*$ &$\tilde{\mathbf{T}},A$&$\mathbf{V}$& Returns encrypted histogram of $A$
 \\\textsf{GroupBy} &$\tilde{\mathbf{T}},A$ &$\tilde{\mathbf{V}}$ &Returns encrypted histogram of $A$\\&&& in one-hot-coding
 \\\textsf{CountDistinct} &$\mathbf{V}$&$\mathbf{C}$ &Counts the number of non-zero values in $\textbf{V}$
 \\\textsf{Laplace}&$\textbf{V},\epsilon$ &$\hat{\textbf{V}}$ & Both \textsf{CSP} and {AS} adds Laplace noise to $\textbf{V}$ 
 \\\textsf{NoisyMax} & $\textbf{V},\epsilon, k$&$P$ &Outputs top k values from noisy vector $\textbf{V}$\\
  [1ex] 
 \bottomrule
 \end{tabular}
\end{table}
}

%\vspace{-3mm}
\subsection{Transformation operators}\label{sec:transformation_operators}
Transformation operators input encrypted data and output a transformed encrypted data.  These operators thus work completely on the encrypted data without expending any privacy budget. Three types of data are considered in this context: (1) an encrypted table, $\encT$, of $x$ rows and $y$ columns/attributes where each attribute value is represented by its encrypted one-hot-encoding; (2) an encrypted vector, $\encV$; and (3) an encrypted scalar, $\encC$. In addition, every encrypted table, $\encT$, of $x$ rows has an encrypted bit vector, $\encB$, of size $x$ to indicate whether the rows are relevant to the program at hand. The $i$-th row in $\encT$ will be used for answering the current program only if the $i$-th bit value of $\encB$ is $1$. The input to the first transformation operator in \system program is $\encD$ with all bits of $\encB$ set to $1$. For brevity, we use just $\encT$ to represent both the encrypted table, $\encT$, and $\encB$. The transformation operators are:

\stitle{{\normalfont(1)} \textsf{CrossProduct}} $\crossproduct_{(A_i,A_j)\rightarrow A'}(\encT)$: This operator transforms the two encrypted one-hot-encodings for attributes $A_i$ and $A_j$ in $\encT$ into a single encrypted one-hot-encoding of a new attribute, $A'$. The domain of the new attribute, $A'$, is the cross product of the domains for $A_i$ and $A_j$. The resulting table, $\encT'$, has one column less than $\encT$. Thus, the construction of the one-hot-encoding of the entire $y$-dimensional domain can be computed by repeated application of this operator. 	
%1) \textbf{\textsf{CrossProduct}}:$\crossproduct_{A_1,A_2}(\tilde{\mathbf{T}})$ - Given encrypted one-hot-codings for two different attributes $A_1$ and $A_2$ of domain sizes $s_{A_1}$ and $s_{A_2}$ respectively, the goal of this transformation is to compute the encrypted one-hot-coding for the entire two-dimension domain of the new $\lq$attribute' $A_1\times A_2$ of size $s_{A_1}\cdot s_{A_2}$. Thus this transformation takes as input a $x \times y $ table, $\tilde{\mathbf{T}}$ defined over attribute set $A=\{A_1,A_2,...,A_y\}$ where each cell $\tilde{\mathbf{T}}[i,j] , i \in [x], j \in [y], 2 \leq y \leq k$ corresponds to the encrypted one-hot-coding for attribute $A_j$ for the data owner $\textsf{DO}_i$ and outputs a $x \times (y-1)$ table with attribute set $\{A_1\times A_2,A_3,\ldots,A_{y}\}$.  Note that the construction of the one-hot-coding of the full $y$-dimension domain can be computed by repeated application of this transform. 

\stitle{{\normalfont(2)} \textsf{Project}} $\project_{\bar{A}}(\encT)$: This operator projects $\encT$ on  a subset of attributes, $\bar{A}$, of the input table. All the attributes that are not in $\bar{A}$ are discarded from the output table $\encT'$.
	%2) \textbf{\textsf{Project}} : $\project_{A^*}(\tilde{\mathbf{T}})$- In addition to the $ x \times y$ table, $\tilde{\mathbf{T}}$ over attribute set $\{A_1, A_2, ..., A_y\}$, the \textsf{Project} transformation takes a set of attributes $A*=\{A^*_1,...A^*_p\}, p < y$ as inputs. The result of the transformation is defined as the $x \times p$ data source table where each record is just restricted to the attribute set $A^*$, i.e., it discards all other attributes. 
	%Infact it is analogous to the operation of marginalization which is described as follows.
	%Assuming  $A$ and $B$ to be two attributes with finite domains, let $x$ be a vector of counts representing a histogram over the cross product of the domain (with $|A|*|B|$ entries).
	%Marginalization over the attribute $B$ results in a vector of counts on the attribute $A$ alone by adding up counts corresponding to the same value of $A$.  

\stitle{{\normalfont(3)} \textsf{Filter}} $\filter_{\phi}(\encT)$: This operator specifies a filtering condition that is represented by a Boolean predicate, $\phi$, and defined over a subset of attributes, $\bar{A}$, of the input table, $\encT$. The predicate can be expressed as a conjunction of range conditions over $\bar{A}$, i.e., for a row $r \in  \encT$, $\phi(r) = \bigwedge_{A_i \in \bar{A}} ~~(r.{A_i} \in V_{A_i})$,  where $r.A_i$ is value of attribute $A_i$ in row $r$ and $V_A$ is a subset of values (can be a singleton) that $A_i$ can take.  For example, $Age\in [30,40]\wedge Gender=M$ can be a filtering condition. The \textsf{Filter} operator affects only the associated encrypted bit vector of $\encT$ and keeps the actual table untouched. If any row, $r \in \encT$, does not satisfy the filtering condition, $\phi$, the corresponding bit in $\encB$ will be set to $labEnc_{pk}(0)$; otherwise, the corresponding bit value in $\encB$ is kept unchanged.  Thus the \textsf{Filter} transformation suppresses all the records that are extraneous to answering the program at hand (i.e., does not satisfy $\phi$) by explicitly zeroing the corresponding indicator bits and outputs the table, $\encT'$, with the updated indicator vector.

\stitle{{\normalfont(4)} \textsf{Count}} $\countagg(\encT)$: This operator simply counts the number of rows in $\encT$ that are pertinent to the program at hand, i.e. the number of $1$s in its associated bit vector $\encB$.  This operator outputs an encrypted scalar, $\encC$. 
%Typically, this transformation is followed by a measurement operator and is immediately preceded by a \textsf{Filter} operator. \\
% 4) \textbf{\textsf{Count}}$:\countagg(\mathbf{T})$  - The \textsf{Count} transformation outputs the encrypted value of the non-noisy true count for the program at hand. For answering linear counting queries, typically \textsf{Count}  is the last transformation to be applied and is immediately preceded by a \textsf{Filter} transformation. Recall that the \textsf{Filter} transformation sets bit $i \in [m]$ to be 1 (encrypted) if the $i^{th}$ record satisfies the filter condition and 0 otherwise and outputs this encrypted $m\times 1$ vector. Hence the \textsf{Count} operator simply adds up all the entries of this bit vector $\mathbf{B}$ and  outputs the sum which is a single encrypted value. 

\stitle{{\normalfont(5)} \textsf{GroupByCount}} $\groupbystar_{A}(\mathbf{\tilde{T}})$: The  \textsf{GroupByCount} operator partitions the input table, $\mathbf{\tilde{T}}$, into groups of rows having the same value for an attribute, $A$. The output of this transformation is an encrypted  vector, $\mathbf{V}$, that counts the number of unfiltered rows for each value of $A$. This operator serves as a preceding transformation for other Crypt$\epsilon$ operators specifically, \textsf{NoisyMax}, \textsf{CountDistinct} and \textsf{Laplace}.

\stitle{{\normalfont(6)} \textsf{GroupByCountEncoded}} $\groupby_{A}(\mathbf{\tilde{T}})$: %This operator is similar to the aforementioned \textsf{GroupByCount*} operator. The only difference between the two is that, \textsf{GroupByCount} outputs the encrypted histogram of attribute $A$ with each count represented in one-hot-coding, $\tilde{V}$.  This transformation allows us to answer queries based on the count of a particular value for attribute $A$.
This operator is similar to \textsf{GroupByCount}. The only difference between the two is that \textsf{GroupByCountEncoded} outputs a new table that has two columns -- the first column corresponds to $A$ and the second column corresponds to the number of rows for every value of $A$ (in one-hot-encoding). This operator is useful for expressing computations of the form ``count the number of age values having at least 200 records" (see P7 in Table~\ref{tab:programexamples}).

\stitle{{\normalfont(7)} \textsf{CountDistinct } }$\countdistinct(\mathbf{V})$: This operator is always preceded by \textsf{GroupByCount}. Hence the input vector, $\mathbf{V}$, is an encrypted histogram for attribute, $A$, and this operator returns the number of distinct values of $A$ that appear in  $\boldsymbol{\tilde{\mathcal{D}}}$ by counting the non-zero entries of $V$. %\am{Can we change $count*$ to $countD$ just to be clear?}
%The input to Crypt$\epsilon$ is an encrypted instance of a database $\boldsymbol{\tilde{\mathcal{D}}}$ with a single relational schema $\langle \mathcal{A}_1,\mathcal{A}_2, . . . ,\mathcal{A}_l\rangle$. Each attribute $\mathcal{A}_i$ is assumed to be discrete (or suitably discretized) and represented in one-hot-coding form.

\eat{
Transformation operators take as input an encrypted source variable (a table of size $x \times y, x,y \in \mathcal{Z}_{\geq 0}$) and output a transformed data source (again  a table $x' \times y', x',y' \in \mathcal{Z}_{\geq 0}$) that is still encrypted. Typically $x$ and $x'$ are equal to $m$, the total number of data owners, i.e., every tuple in the data source tables corresponds to the record of a single data owner. In case $x=1$ or $x'=1$ the data source is an encrypted vector and we represent it as $\mathbf{V}$.
The transformation operators are mostly carried out by the \textsf{AS} on its own; this is enabled by our use of labeled homomorphic encryption scheme which allows us to perform certain operations, specifically multiplication and addition, directly over the encrypted data. %Only two transformations  (\textsf{GroupBy} and \textsf{CountDistinct}) need to be computed via a secure computation protocol between the \textsf{AS} and the \textsf{CSP}. 
Since these operators work entirely on encrypted data, they do not expend the privacy budget. However these operators can affect the privacy analysis through their stability. Every transformation in Crypt$\epsilon$ has a well-established stability.
For each record of the database $\boldsymbol{\tilde{\mathcal{D}}}$ (i.e., data corresponding to a single data owner) we maintain an encrypted bit which indicates whether the record is relevant to the program at hand. Let $\mathbf{B}$ represent this bit vector where $\mathbf{B}[i]$ corresponds to this indicator bit for the $i^{th}$ record.  If $\mathbf{B}[i] =Enc_{pk}(1)$, then the $i^{th}$ record is to be considered for answering the current program and vice versa. Only one of the transformation, \textsf{Filter} alters the bit vector $\mathbf{B}$. Before every program execution, $\mathbf{B}$ is initialized to a 1-vector. 
\begin{enumerate}

	\item \textsf{CrossProduct} ($\tilde{\mathbf{T}}, A_i, A_j$) - Given encrypted one-hot-encodings for two different attributes $A_1$ and $A_2$ of domain sizes $s_{A_1}$ and $s_{A_2}$ respectively, the goal of this transformation is to compute the encrypted one-hot-encoding for the entire two-dimension domain of the new $\lq$attribute' $A_1\times A_2$ of size $s_{A_1}\cdot s_{A_2}$. Thus this transformation takes as input a $x \times y $ table, $\tilde{\mathbf{T}}$ defined over attribute set $A=\{A_1,A_2,...,A_y\}$ where each cell $\tilde{\mathbf{T}}[i,j] , i \in [x], j \in [y], 2 \leq y \leq k$ corresponds to the encrypted one-hot-encoding for attribute $A_j$ for the data owner $\textsf{DO}_i$ and outputs a $x \times (y-1)$ table with attribute set $\{A_1\times A_2,A_3,\ldots,A_{y}\}$.  Note that the construction of the one-hot-encoding of the full $y$-dimension domain can be computed by repeated application of this transform.

	\item \textsf{Project}($\tilde{\mathbf{T}},A^*$)- In addition to the $ x \times y$ table, $\tilde{\mathbf{T}}$ over attribute set $\{A_1, A_2, ..., A_y\}$, the \textsf{Project} transformation takes a set of attributes $A*=\{A^*_1,...A^*_p\}, p < y$ as inputs. The result of the transformation is defined as the $x \times p$ data source table where each record is just restricted to the attribute set $A^*$, i.e., it discards all other attributes. \\
	%Infact it is analogous to the operation of marginalization which is described as follows.
	%Assuming  $A$ and $B$ to be two attributes with finite domains, let $x$ be a vector of counts representing a histogram over the cross product of the domain (with $|A|*|B|$ entries).
	%Marginalization over the attribute $B$ results in a vector of counts on the attribute $A$ alone by adding up counts corresponding to the same value of $A$.  
  \item \textsf{Filter}($\tilde{\mathbf{T}},\phi$) - Let $\tilde{\mathbf{T}}$ be an encrypted table of one-hot-encodings over attribute set $A=\{A_1,...,A_k\}$, $\phi$ be a  predicate defined over a subset of attributes $A^*\subseteq A$ and $\mathbf{B}$ be the current state of the indicator vector which is stored by the \textsf{AS}. The predicate $\phi$ has to be expressed as a conjunction of range conditions over $A^*$, i.e.,\begin{gather}\phi = \bigwedge_{A \in A^*}(A \in \{v_{1},\ldots,v_{t}\} ) \label{phi} \end{gather} If for some attribute $A \in A^*$, the condition is a equality condition as $A==v$ instead of a range condition, then simply put $v_1=v_t$. For e.g., a condition of this form is find the number of records that satisfy $Age$ in range [30,40] and $Gender$ is male. For each record $r_i, i \in [m]$, the Filter transformation zeros the corresponding indicator bit $\mathbf{B}[i] $ if $\phi(r_i)=False$. $\mathbf{B}[i] $ is kept unchanged otherwise. Thus the \textsf{Filter} transformation suppresses all the records that are extraneous to answering the program at hand (i.e., does not satisfy $\phi$) by explicitly zeroing the corresponding indicator bits and outputs the updated indicator vector. %It takes as input a $x \times 1$ table $\tilde{\mathbf{T}}$, whose every row is an encrypted one-hot-encoding (of the form $\mathbf{\tilde{R}}$) for the attribute of concern $A$, and a vector $\mathbf{C}$ which has encryptions of appropriate non-zero weights for indices that satisfy $\phi$. %Note that $A$ need not be an attribute of the original attribute set $\mathcal{A}$ but can be a new multi-dimension 'attribute' constructed over $\mathcal{A}^* \subseteq \mathcal{A}$, i.e., $A= \prod_{A*_i \in \mathcal{A}^*  }A^*_i$. 
    \item{\textsf{Count}($\mathbf{T}$) } - The \textsf{Count} transformation outputs the encrypted value of the non-noisy true count for the program at hand. For answering linear counting queries, typically \textsf{Count}  is the last transformation to be applied and is immediately preceded by a \textsf{Filter} transformation. Recall that the \textsf{Filter} transformation sets bit $i \in [m]$ to be 1 (encrypted) if the $i^{th}$ record satisfies the filter condition and 0 otherwise and outputs this encrypted $m\times 1$ vector. Hence the \textsf{Count} operator simply adds up all the entries of this bit vector $\mathbf{B}$ and  outputs the sum which is a single encrypted value. 
    \item{\textsf{GroupBy*}($\mathbf{\tilde{T}},A$)}- The purpose of the \textsf{GroupBy*} transformation is to essentially bucket the input $x\times y$ table $\mathbf{\tilde{T}}$ into groups of records having the same value for an attribute of choice $A$. The output of this transformation is an $1\times s_A$ encrypted  vector $\mathbf{V}$  where each vector element $\mathbf{V}[i], i \in [s_A]$ represents the encrypted count of the number of records in $\boldsymbol{\tilde{\mathcal{D}}}$ having value $v_{i}$ for attribute $A$. Thus \textsf{GroupBy*} essentially returns an encrypted histogram for $A$.
    This operator serves as a preceding transformation for other Crypt$\epsilon$ operators like \textsf{NoisyMax}, \textsf{CountDistinct} et al.
     \item{\textsf{GroupBy}($\mathbf{\tilde{T}},A$)-} The \textsf{GroupBy} transformation is similar to the aforementioned \textsf{GroupBy*} transformation. The only difference between the two is that, the former outputs the encrypted one-hot-encoding of the respective counts. That is, the output of \textsf{GroupBy}($A$)  is an $s_A$ lengthed encrypted vector $\tilde{\mathbf{V}}$ such that each element, $\tilde{\mathbf{V}}[i], i \in [s_A]$ represents the encrypted one-hot-encoding of the number of records in $\boldsymbol{\tilde{\mathcal{D}}}$ having value $v_{i}$ for attribuet $A$. This transformation allows us to answer queries based on the count of a particular value for attribute $A$.
     %Note that since for \textsf{GroupBy} we need to create the one-hot-coding of the counts, this requires an interaction with the \textsf{CSP}.
     \item {\textsf{CountDistinct}($\mathbf{V}$)-} As mentioned before, the \textsf{CountDistinct} operator takes as input an encrypted vector $\mathbf{V}$ which is the output of a \textsf{GroupBy*}($A$) operator for some attribute $A$. Thus the \textsf{CountDistinct} operator  returns the number of distinct values of $A$ that appear in the records of $\boldsymbol{\tilde{\mathcal{D}}}$ by counting the non-zero entries of $V$.  
\end{enumerate}
%Note that the first four transformations namely \textsf{CrossProduct, Project, Filter} and Count are performed by the \textsf{AS} alone. Only for transformation \textsf{GroupBy} the \textsf{AS} engages in a secure computation protocol with the \textsf{CSP}.

}%%%ENDOF EAT
%\vspace{-0.3cm}
\subsection{Measurement operators} \label{sec:measurement_operators}
The measurement operators take encrypted vector of counts, $\encV$ (or a single count, $\encC$), as input and return noisy measurements on it in the clear. These two operators correspond to two classic DP mechanisms -- Laplace mechanism and Noisy-Max ~\cite{Dork}. Both mechanisms add Laplace noise, $\eta$, scaled according to the transformations applied to $\encD$. 

Let the sequence of transformations applied on $\encD$ to get $V$ be $\bar{\mathcal{T}}(D) = \mathcal{T}_l(\cdots \mathcal{T}_2((\mathcal{T}_1(D))))$. The \emph{sensitivity} of a sequence of transformations is defined as  the maximum change to the output of this sequence of transformations \cite{PINQ} when changing a row in the input database, i.e., $\Delta_{\bar{\mathcal{T}}} = \max_{D,D'} \|\bar{\mathcal{T}}(D)-\bar{\mathcal{T}}(D')\|_1$ where $D$ and $D'$ differ in a single row. The sensitivity of $\bar{\mathcal{T}}$ can be upper bounded by the product of the stability \cite{PINQ} of these transformation operators, i.e., $\Delta_{\bar{\mathcal{T}}=(\mathcal{T}_l,\ldots,\mathcal{T}_1)} = \prod_{i=1}^l \Delta \mathcal{T}_i$.  The transformations in Table~\ref{tab:operators} have a stability of 1, except for \textsf{GroupByCount} and \textsf{GroupByCountEncoded} which are 2-stable. Given $\epsilon$ and $\Delta_{\bar{\mathcal{T}}}$, we define the measurement operators:

\stitle{{\normalfont(1)} \textsf{Laplace}} $\lap_{\epsilon,\Delta}(\mathbf{V}/\encC)$:  This operator implements the classic Laplace mechanism \cite{Dork}. Given an encrypted vector, $\encV$, or an encrypted scalar, $\encC$, a privacy parameter $\epsilon$ and sensitivity $\Delta$ of the preceding transformations, the operator adds noise drawn from $Lap(\frac{2\Delta}{\epsilon})$ to $\encV$ or $\encC$ and outputs the noisy answer.

\stitle{{\normalfont(2)} \textsf{NoisyMax}} $\noisymax^k_{\epsilon,\Delta}(\mathbf{V})$:  Noisy-Max is a  differentially private selection mechanism \cite{Dork,APEx} to determine the top $k$ highest valued queries. This operator takes in an encrypted vector $\encV$ and adds independent Laplace noise from $Lap(\frac{2 k\Delta}{\epsilon})$ to each count. The indices for the top $k$ noisy values, $\hat{\mathcal{P}}$, are reported as the desired answer. \\
%The measurement operators satisfy $\epsilon/2$-DP  (Definition \ref{def:dp}) \cite{Dork,APEx} and $\epsilon$-bounded DP (see Section \ref{sec:background}) \cite{bookDP}.
\eat{
Thus they expend the privacy budget and require secure computation between the \textsf{AS} and the \textsf{CSP}. 
\begin{comment} All measurement operators must involve joint computation with the \textsf{CSP}. Note that the requisite noise to be added to ensure differentially privacy has to be jointly added by both the \textsf{AS} and the \textsf{CSP}. It is so because, had only either one of the servers added the noise, then that server would be able to retrieve the true non-noisy answer by simply de-noising the published differentially private answer. This means that the sensitivity of the program being executed should be known to both the servers. This poses no hindrance in our setting  since the program is public, the  sensitivity computation can be performed very easily by observing the sequence of the preceding transformations.
\end{comment}
\begin{enumerate}
	\item \textsf{Laplace}($\mathbf{V},\epsilon$) - In the Laplace mechanism, in order
to publish $f(D)$ where $f : D \mapsto R$, $\epsilon$-differentially private mechanism $\mathcal{M()}$ 
publishes $f(D) + Lap\Big(\frac{\Delta f}{\epsilon}\Big)$  
where $\Delta f = \max_{D,D'}||f(D)-f(D')||_1$ is known as the sensitivity of the query. The p.d.f of $Lap(b)$ is given by\begin{gather}\mathbf{f}(x)={\frac  {1}{2b}}e^{ \left(-{\frac  {|x-\mu |}{b}}\right)}\end{gather} The sensitivity of the function $f$ basically captures the magnitude by which a single individual's data can change the function $f$ in the worst case. Therefore, intuitively, it captures the uncertainty in the response that we must introduce in order to hide the participation of a single individual. For counting queries, the sensitivity is 1. The \textsf{Laplace} operator enables the \textsf{AS} and the \textsf{CSP} to add two separate instances of random Laplace noise to the true result of a counting query for generating a differentially private output. It takes an input an encrypted vector $\mathbf{V}$ (could be a scalar too) and adds two instances of noise drawn from $[Lap(\frac{1}{\epsilon})]^{|V|}$ to it.

	\item \textsf{NoisyMax}($\mathbf{V},\epsilon, k$)-Noisy-Max is a type of differentially-private selection mechanism where the goal is to determine the counting query with the highest value out of $n$ different counts.  
	The algorithm works as follows. First, generate each of the counts and then add independent Laplace noise from the distribution $Lap(\frac{1}{\epsilon})$ to each of them. The index of the largest noisy count is then reported as the noisy max.
	This has two fold advantage over the naive implementation of finding the maximum count.
First, noisy-max applies "information minimization" as rather than releasing all the noisy counts
and allowing the analyst to find the max and its index, only the
index corresponding to the maximum is made public.
Secondly, the noise added is much smaller than that in the case of the naive implementation (it has sensitivity $\Delta f=m$). Thus the \textsf{NoisyMax} operator takes as input of encrypted vector where each vector element is a count. It then adds noise drawn from $Lap(\frac{1}{\epsilon})$ to each vector element and computes the indices of the top k elements.
\end{enumerate}
}

\vspace{-1.1cm}
\subsection{Program Examples}
A \system program is a sequence of transformation operators followed by a measurement operator and arbitrary post-processing. Consider a database schema $\langle Age$, $Gender$, $NativeCountry$, $Race\rangle$. We show 7 \system program examples in Table~\ref{tab:programexamples} over this database.  %Note that the sensitivities are explicitly shown in Table~\ref{tab:programexamples}, but the data analysts do not need to specify them in the \system programs; the \textsf{AS} can analyze the sequence of transformations and derive the sensitivity automatically.

We will use P1 in Table~\ref{tab:programexamples} to illustrate how a \system program can be written and  analyzed. Program P1 aims to compute the cumulative distribution function (c.d.f.) of attribute $Age$ with domain $[1,100]$. The first step is to compute 100 range queries, where the $i$-th query computes the the number of records in $\encD$ having $Age \in [1,i]$ with privacy parameter $\epsilon_i$.  The sequence of transformation operators for each range query is $\countagg(\filter_{Age\in [1,i]}(\project_{Age}(\encD)))$. All these three operators are 1-stable and hence, the sensitivity of the resulting range query is upper bounded by the product of these stability values, $1$ \cite{PINQ}. Thus, the subsequent measurement operator $\textsf{Laplace}$ for the $i$-th range query takes in privacy budget $\epsilon_i$ and sensitivity $\Delta=1$, and outputs a noisy plaintext count, $\hat{c}_i$.  At this stage the program is  $\sum_{i=1}^{100} \epsilon_i/2$-DP by Theorem~\ref{theorem:seq}\cite{Dork} (recall we add noise from $Lap(2\cdot\Delta/\epsilon_i)$ in Section \ref{sec:measurement_operators}). After looping over the 100 ranges, P1 obtains a  noisy plaintext output $\hat{V}=[\hat{c}_1,...,\hat{c}_{100}]$ and applies a post-processing step, denoted by $post_{c.d.f}(\hat{V})$. This operator inputs a noisy histogram, $\hat{V}$, for attribute $A$ and computes its c.d.f $\hat{V}'=[\hat{c}'_1,...,\hat{c}'_{100}]$ via isotonic regression \cite{cdf}
$\min_{\hat{V}'} \|\hat{V}'-\hat{V}\|_2 ~~~~~~s.t.~~~ 0\leq\hat{c}'_1\leq \cdots \leq \hat{c}'_{100} \leq |\encD|.$
Hence, by Theorem~\ref{post}, P1 is $\epsilon/2$-DP, where $\epsilon=\sum_{i=1}^{100} \epsilon_i$. However, since \system also reveals the total dataset size, the total privacy guarantee is $\epsilon$-bounded DP (see Section \ref{sec:security} for details).
%\arc{Now the datanalyst reports the privacy budget to be used for progam i. the CSP decrypts the only if $\sum < \epsilon^B$. This ensures that  AS reports. At any point Under he semi-honest model assumption . Here we The sum of  }
\eat{We also show how privacy parameter allocation and sensitivity analysis is done in \system via one of the programs, P1 as follows. \\
\textbf{Query} - Compute the c.d.f over the attribute $Age$\\
\textbf{\system Program }- \\
$\forall i\in [1,100],\hat{c}_i \leftarrow \lap_{\epsilon_i,1}(\countagg(\filter_{Age \in (0, c_i]}(\project_{Age}(\encD))))$;\\ $post_{c.d.f}([\hat{c}_1,\ldots,\hat{c}_{100}])$\\
Now we provide a step-by-step breakdown of the program
\squishlistnum \item The first step is to compute 100 range queries, where the $i$-th query computes the the number of records in with age in $[1,i]$. Thus, if the total privacy budget is $\epsilon$, each individual range query gets a privacy budget of $\frac{\epsilon}{100}$. 
\item  The sequence of transformation operators for each such query is $count(\sigma(\pi(\encD))$. Thus the sensitivity of each range query is upper bounded by the product of the stabilities of these three operators which is $1$. \item Thus the input to the measurement operator $Lap$ is privacy parameter $\frac{\epsilon}{100}$ and sensitivity 1. \item Finally, the program performs post processing over the noisy plaintext output $\hat{V}=[\hat{c}_1,...,\hat{c}_{100}]$. We denote this operator by $post_{c.d.f}(\hat{V})$. This operator input a noisy histogram $\hat{V}$
for an attribute $A$ and computes its c.d.f $\hat{V}'$ via isotonic regression. By theorem 2, this step does not amount to any privacy loss.\item Thus by Theorem 1, the above program is $\epsilon$-DP. \squishendnum }
\begin{table*}[t]
\caption{Examples of \system Program}\label{tab:programexamples} \vspace{-3mm}
\scalebox{0.67}{
\begin{tabular}{|l|l|}
\hline
 {\bf \system Program} & {\bf Description} \\ \hline \hline
%P1:  $\hat{c} \leftarrow \lap_{\epsilon,1}(\countagg(\filter_{Age\in [50,60]}(\project_{Age}(\encD))))$ &  Counts the number of records satisfying $Age \in [50,60]$.      \\ \hline
P1:  $\forall i\in [1,100],\hat{c}_i \leftarrow \lap_{\epsilon_i,1}(\countagg(\filter_{Age \in (0,i]}(\project_{Age}(\encD))))$; $post_{c.d.f}([\hat{c}_1,\ldots,\hat{c}_{100}])$ &  Outputs the c.d.f of $Age$ with domain $[1,100]$.  \\ \hline
P2: $\hat{P} \leftarrow \noisymax^5_{\epsilon,1}(\groupbystar_{Age}(\encD))$               &       Outputs the 5 most frequent age values. \\ \hline
P3: $\hat{V} \leftarrow \lap_{\epsilon,2}(\groupbystar_{Race\times Gender}(\project_{Race \times Gender}(\crossproduct_{Race,Gender\rightarrow{Race \times Gender}}(\encD))))$     &       Outputs the marginal over the attributes $Race$ and $Gender$. \\ \hline
P4: $\hat{V} \leftarrow \lap_{\epsilon,2}(\groupbystar_{Age\times Gender}(\filter_{NativeCountry=Mexico}(\project_{Age\times Gender, NativeCountry}(\crossproduct_{Age,Gender\rightarrow{Age \times Gender}}(\encD)))))$ & \multirow{1}{*}{Outputs the marginal over $Age$ and $Gender$ for Mexican employees.} %\\$\hspace{3.5cm}\project_{Age\times Gender, NativeCountry}(\crossproduct_{Age,Gender\rightarrow{Age \times Gender}}(\encD)))))$ &                         
\\ \hline
P5: $\hat{c} \leftarrow \lap_{\epsilon,1}(\countagg(\filter_{Age=30 \wedge Gender=Male \wedge NativeCountry=Mexico}(\project_{Age,Gender,NativeCountry}(\encD))))$&\multirow{1}{*}{Counts the number of male employees of Mexico having age 30.}
%\\$\hspace{7.3cm}\project_{Age,Gender,NativeCountry}(\encD))))$                                       &       
\\ \hline
P6:   $\hat{c} \leftarrow \lap_{\epsilon,2}(\countdistinct(\groupbystar_{Age}(\filter_{Gender=Male}(\project_{Age , Gender}(\encD)))))$ & Counts the number of distinct age values for the male employees.       \\ \hline
P7:  $\hat{c} \leftarrow \lap_{\epsilon,2}(\countagg(\filter_{Count\in[200,m]}(\groupby_{Age}(\project_{Age}(\encD)))))$
                                     & Counts   the number of  age values having at least 200 records.   \\ \hline

\end{tabular}}\vspace{-2mm}
\end{table*}

\section{Implementation}\label{sec:implementation}
%To demonstrate the use of Crypt$\epsilon$ operators let us look at the following example. 
%In Section~\ref{sec:implementation} we describe how \system compiles its operators down to protocols that work on encrypted data.
In this section, we describe the implementation of \system. 
First, we discuss our proposed technique for extending the multiplication operation of \textsf{labHE} to support $n > 2$ multiplicands which will be used for the \textsf{CrossProduct} operator. Then, we describe the implementations of \system operators.  %Based on these implementations, \system can compile a logical program written by the data analyst into a protocol run by the \textsf{AS} and \textsf{CSP} on the encrypted data. 
%Note that all the post-processing on the output of differentially private measures are executed in clear. Finally, we present a classification of the \system programs.
\eat{
\begin{algorithm}[b]
\caption{$genLabMult$ - generate label for $labMult$}\label{algo:genlabmult}
\small
\begin{algorithmic}[1]
\STATEx
\textbf{Input}: $\mathbf{c_1}=(a_1,d_1)=labEnc_{pk}(m_1)$ and $\mathbf{c_2}=labEnc_{pk}(m_2)$ 
\STATEx where $a_1= m_1-b_1, d_1=Enc_{pk}(b_1)$, $a_2= m_2-b_2, d_2=Enc_{pk}(b_2)$
\STATEx \textbf{Output}: $\mathbf{e}=labEnc_{pk}(m_1\cdot m_2)$
\STATEx \textbf{\textsf{AS}:} 
\STATE Computes $\textbf{e}'=labMult(\mathbf{c_1,c_2}) \oplus Enc_{pk}(r)$ where $r$ is a random mask 
\STATEx  //$e'$ corresponds to $m_1\cdot m_2-b_1\cdot b_2+r$
\STATE Sends $\mathbf{e'},d_1,d_2$ to \textsf{CSP}
\STATEx \textbf{\textsf{CSP}:}
\STATE Computes $e''= Dec_{sk}(\mathbf{e'}) + Dec_{sk}(d_1)\cdot Dec_{sk}(d_2)$
\STATEx //$e''$ corresponds to $m_1\cdot m_2 + r$ 
%\STATE Decrypts $\mathbf{e'}$, to get $Dec_{sk}(\mathbf{e}')=m_1\cdot m_2 -b_2\cdot b_1 + r$
%\STATE Computes $b_1 \cdot b_2$ from $d_1$ and $d_2$.
%\STATE Removes $b_1\cdot b_2$ from $e'$ to compute $e''=m_1\cdot m_2+r$
\STATE Picks a seed $\sigma'$ and label $\tau'$ and computes $b'=\mathcal{F}(\sigma',\tau')$ 
%\STATE Computes $\bar{a}=e''-b'=m_1\cdot m_2 +r -b'$,  and $d'=Enc_{pk}(b')$
\STATE Sends $\bar{e}=(\bar{a},d')$ to \textsf{AS}, where $\bar{a} = e''-b'$ and $d' = Enc_{pk}(b')$
\STATEx //$\bar{a}$ corresponds to $m_1\cdot m_2 + r-b'$.
\STATEx \textbf{\textsf{AS}:}
\STATE Computes true cipher $\mathbf{e}=(a',d')$ where $a'=\bar{a}-r$ %m_1\cdot m_2 - b'$
 \end{algorithmic}
\end{algorithm}
}

\vspace{-.5em} 
\subsection{\textbf{General} \textsf{labHE} $n$-way Multiplication }\label{genlab}
The \textsf{labHE} scheme is an extension of a \textsf{LHE} scheme where every ciphertext is now associated with a ``label'' \cite{Barbosa2017LabeledHE}. This extension enables \textsf{labHE} to support multiplication of two \textsf{labHE} ciphertexts via the $labMult()$ operator (without involving the \CSP).  
\eat{\squishlist
\item $\textbf{labMult}(\mathbf{c}_1,\mathbf{c}_2)$ - On input two \textsf{labHE} ciphertexts $\mathbf{c}_1=(a_1,d_1)$ and $\mathbf{c}_2=(a_2,d_2)$, it computes a "multiplication" ciphertext  $\mathbf{e}=labMult(\mathbf{c_1,}$ $\mathbf{c_2})=Enc_{pk}(a_1,a_2)\oplus cMult(d_1,a_2) \oplus cMult(d_2,a_1)$. Observe that $Dec_{sk}(\mathbf{e})=m_1\cdot m_2 -b_1 \cdot b_2$.
\item $\textbf{labMultDec}_{sk}(d_1,d_2,\mathbf{e})$ - On input two encrypted masks $d_1,d_2$ of two \textsf{labHE} ciphertexts $\mathbf{c_1},\mathbf{c_2}$, this algorithm decryts the output $\mathbf{e}$ of $labMult(\mathbf{c_1},\mathbf{c_2})$ as $m_3=Dec_{sk}(\mathbf{e})+Dec_{sk}(d_1)\cdot Dec_{sk}(d_2)$ which is equals to $m_1\cdot m_2$.   
\squishend}
However, it cannot support multiplication of more than two ciphertexts because the``multiplication'' ciphertext $\mathbf{e}=labMult(\mathbf{c_1},\mathbf{c_2})$, ($\mathbf{c_1}$ and $\mathbf{c_2}$ are \textsf{labHE} ciphertexts) does not have a corresponding label, i.e., it is not in the correct \textsf{labHE} ciphertext format. Hence, we propose an algorithm $genLabMult$ to generate a label for every intermediary product of two multiplicands to enable generic $n$-way multiplication (details are in the\ifpaper 
the full paper \cite{anom})
\else 
Algorithm~\ref{algo:genlabmult}) 
\fi. %An $n-$way multiplication can be parallelized and requires  $\lceil \log n\rceil$ rounds of communication with the \textsf{CSP}. %Furthermore, it can be shown that the order of multiplication can be parallelized as shown in Figure ~\ref{genlab-fig} (Appendix) and a $n-$way multiplication requires a total of $\lceil \log n\rceil$ rounds of communication with the \textsf{CSP}.
\eat{Note that the mask $r$ protects the value of $(m_1\cdot m_2)$ from the \textsf{CSP} (Step 3) and $b'$ hides $(m_1\cdot m_2)$ from the \textsf{AS} (Step 6). 
For example, suppose we want to multiply the respective ciphers of  $4$ messages $\{m_1,m_2,m_3,m_4\} \in \mathcal{M}^4$ and obtain $\mathbf{e}=labEnc_{pk}(m_1\cdot m_2\cdot m_3 \cdot m_4)$. For this, the \textsf{AS} first generates $\mathbf{e_{1,2}}=labEnc_{pk}(m_1\cdot m_2)$ and $\mathbf{e_{3,4}}=labEnc_{pk}(m_3\cdot m_4)$ using Algorithm~\ref{algo:genlabmult}. Both operations can be done in parallel in just one interaction round between the \textsf{AS} and the \textsf{CSP}. In the next round,  the \textsf{AS} can again use Algorithm~\ref{algo:genlabmult} with inputs $\mathbf{e_{1,2}}$ and $\mathbf{e_{3,4}}$ to obtain the final answer $\mathbf{e}$. %consider a case of mu Now with the true \textsf{labHE} cipher $\mathbf{c}=(a',d')$ for the product the \textsf{AS} can compute further multiplications on it. 
Thus for a generic $n-way$ multiplication the order of multiplication can be, in fact, parallelized as  shown in Figure ~\ref{genlab-fig} (Appendix\cite{anom}) to require a total of $\lceil \log n\rceil$ rounds of communication with the \textsf{CSP}.}

\vspace{-.5em} 
\subsection{Operator Implementation}\label{sec:operator_implementation}
%Now let us explain the implementation details of the aforementioned Crypt$\epsilon$ operators.  
We now summarize how \system operators are translated to protocols that the \textsf{AS} and \textsf{CSP} can run on encrypted data.  %Details are in \ifpaper the full paper \cite{anom}% using paper \else Appendix~\ref{app:implement_operators}\fi.

\stitle{\textsf{Project}} $\project_{\bar{A}}(\encT)$: The implementation of this operator simply drops off all but the attributes in $\bar{A}$ from the input table, $\encT$, 
and returns the truncated table, $\encT'$.

\stitle{\textsf{Filter}} $\filter_{\phi}(\encT)$: Let $\phi$ be a predicate  of the form $r.A_j$  $\in V_{A_j}$. Row $i$ satisfies the filter if one of the bits corresponding to positions in $V_{A_j}$ is 1. Thus, the bit corresponding to row $i$ is set as: $\encB[i] = labMult(\encB[i], \bigoplus_{l\in V_{A_j}}\tilde{\bf{v}}_j[l])$. The multi-attribute implementation is detailed in  \ifpaper 
the full paper \cite{anom}% using paper 
\else 
Appendix~\ref{app:implement_operators}
\fi.

\stitle{\textsf{CrossProduct}} $\crossproduct_{A_i,A_j\rightarrow A'}(\encT)$: The crossproduct between two attributes are computed using $genLabMult()$ described above.

\stitle{\textsf{Count}} $\countagg(\encT)$:  This operator simply  adds up the bits in $\encB$ corresponding to input table $\encT$, i.e., $\bigoplus_{i} \encB[i]$.

$\hspace{-3.5mm}$\stitle{\textsf{GroupByCount}} $\groupbystar_{A}(\tilde{\mathbf{T}})$: The implementations for \textsf{Project}, \textsf{Filter} and \textsf{Count} are reused here. First, \system projects the input table $\encT$ on attribute $A$, i.e. $\encT_1 = \project_A(\encT)$. Then, \system loops each possible value of $A$. For each value $v$, \system initializes a temporary $\encB_v=\encB$ and filters $\encT'$ on $A=v$ to get an updated $\encB'_v$. Finally, \system outputs the number of 1s in $\encB'_v$. 

$\hspace{-3.5mm}$\stitle{\textsf{GroupByCountEncoded}} $\groupby_A(\mathbf{\tilde{T}})$: For this operator, the \textsf{AS} first uses $\textsf{GroupByCount}$ to generate the encrypted histogram, $\encV$, for attribute $A$. Since each entry of  $\mathbf{V}$ is a count of rows, its value ranges from $\{0,...,|\encT|\}$. The \textsf{AS}, then, masks $\encV$ and sends it to the \textsf{CSP}. The purpose of this mask is to hide the true histogram from the \textsf{CSP}. Next, the \textsf{CSP} generates the encrypted one-hot-coding representation for this masked histogram $\boldsymbol{\tilde{\mathcal{V}}}$ and returns it to the \textsf{AS}. The \textsf{AS} can simply rotate $\boldsymbol{\tilde{\mathcal{V}}}[i], i \in [|V|]$ by its respective mask value $M[i]$ and get back the true encrypted histogram in one-hot-coding $\tilde{\encV}$. The details are presented in \ifpaper 
the full paper \cite{anom}% using paper 
\else 
in in Algorithm~\ref{groupby-imp} Appendix~\ref{app:implement_operators} 
\fi. %Notice that each entry of $\boldsymbol{\tilde{\mathcal{V}}}$ is a $m$-length vector. 
%Note that the \textsf{GroupByCount} operator could have an alternative implementation using a garbled circuit that takes as input the encrypted vector and outputs the corresponding one-hot-coding representation. However this would require the circuit to decrypt and re-encrypt $O(m)$ data inside it which would be very costly. 

\stitle{\textsf{CountDistinct}}  $\countdistinct(\encV)$: This operator is implemented by a garbled circuit (details are in \ifpaper 
the full paper \cite{anom}% using paper 
\else 
Appendix~\ref{app:implement_operators}
\fi).

\stitle{\textsf{Laplace }}$\lap_{\epsilon,\Delta}(\encV\textbackslash\encC)$:  The \textsf{Laplace} operator  has two phases (since both the \AS and the \CSP adds Laplace noise). In the first phase,  the \textsf{AS} adds an instance of encrypted Laplace noise, $\eta_1 \sim Lap(\frac{2\Delta}{\epsilon})$, to the encrypted input \eat{(step 1 in Algorithm ~\ref{lap})} to generate $\mathbf{\hat{c}}$. In the second phase, the \CSP first checks whether $\sum_{i=1}^t \epsilon_i +\epsilon \leq \epsilon^B$ where $\epsilon_i$ represents the privacy budget used for a previously executed program, $P_i$ (presuming a total of $t \in \mathbb{N}$ programs have been executed hitherto the details of which are logged into the \CSP's public ledger). Only in the event the above check is satisfied, the \CSP proceeds to decrypt $\mathbf{\hat{c}}$, and records $\epsilon$ and the current program details (description, sensitivity) in the public ledger. Next, the \CSP adds a second instance of the Laplace noise, $\eta_2 \sim Lap(\frac{2\Delta}{\epsilon})$, to generate the final noisy output, $\hat{c}$, in the clear. %(steps 3-4)
The \textsf{Laplace} operator with an encrypted scalar, $\encV$, as the input is implemented similarly.
 %Another observation is that this double noise addition does not affect the differential privacy guarantee. After the addition of the first instance of noise by the \textsf{AS}, the second can be seen as a post-processing step. Hence our results Crypt$\epsilon$ programs are still differentially private by Theorem 2.
%\textit{Note:} Following our discussion on the operator implementations in this section and Appendix C, we see that the major chunk of the work for almost all the transformation operators is carried out the \textsf{AS} by itself. This conforms to our second requirement in section.

\stitle{\textsf{NoisyMax}} $\noisymax^k_{\epsilon,\Delta}(\encV)$: This operator is implemented via a two-party computation between the \textsf{AS} and the \textsf{CSP} using garbled circuits (details are in \ifpaper 
the full paper \cite{anom}% using paper 
\else 
Appendix~\ref{app:implement_operators}
\fi).

\stitle{Note}: \system programs are grouped into three classes based on the number and type of interaction between the \textsf{AS} and the \textsf{CSP}. For example, P1, P2 and P3 (Table \ref{tab:programexamples}) have just one interaction with the \CSP for decrypting the noisy output. P4 and P5, on the other hand, require additional interactions for the $n$-way multiplication of ciphers in the \textsf{CrossProduct} operator. Finally, P6 and P7 require intermediate intercations due to operators \textsf{CountDistinct} and \textsf{GroupByCountEncoded} respectively. The details are  %A classification of \system programs based on the number and type of interaction between the \textsf{AS} and the \textsf{CSP} is 
presented in \ifpaper the full paper \cite{anom} \else Appendix \ref{app:sec:classification} \fi.
\eat{\begin{algorithm}
\small{
\caption{\textsf{Laplace }$\lap_{\epsilon,\Delta}(\mathbf{V})$}
\begin{algorithmic}[1]
\STATEx
\textbf{Input}: $\encV$
\STATEx \textbf{Output}: $\hat{V}$
\STATEx \textbf{\textsf{AS}:} \STATE Generates a noisy vector $\hat{\encV}$  as \begin{gather*}\hat{\mathbf{V}}[i] = \mathbf{V}[i]\oplus labEnc_{pk}(\eta[i]),\\ \eta \sim [Lap(\frac{1}{\epsilon})]^{|V|}, i \in [|V|] \end{gather*}
\STATE Sends $\hat{\mathbb{\mathcal{V}}}$  to \textsf{CSP}
\STATEx \textbf{\textsf{CSP}:}
\STATE Decrypts $\mathbf{\hat{\mathcal{V}}}$ to get $\hat{\mathcal{V}}[i]=labDec_{sk}(\mathbf{\hat{\mathcal{V}}}[i]), i \in [|V|]$
\STATE Generates a the final noisy vector $\hat{V}$ as follows 
\begin{gather*} \hat{V}[i]=\hat{\mathcal{V}}[i]+\eta'[i], i \in [|V|], \eta' \sim [Lap(\frac{1}{\epsilon})]^{|\hat{V}|} \end{gather*}
\STATE Returns $\hat{V}$ to \textsf{AS}
 \end{algorithmic} \label{lap}
}
\end{algorithm}} %The implementation for the \textsf{Laplace} operator is given by Algorithm ~\ref{lap}.
\eat{\vspace{-4mm}\subsection{Classification of \system Programs}
\system programs are grouped into three classes based on the number and type of interaction between the \textsf{AS} and the \textsf{CSP}.

\stitle{Class I: Single Decrypt Interaction Programs}\\
%Recall that the transformation operators output encrypted data.  Since  the \textsf{CSP} has exclusive access to the secret key, it is the only party in \system capable of decryption. Thus f
For releasing any result (noisy) in the clear, the \textsf{AS} needs to interact at least once with the \textsf{CSP} (via the two measurement operators) as the latter has exclusive access to the secret key. \system programs like P1, P2 and P3 (Table~\ref{tab:programexamples}) that require only a single interaction of this type fall in this class. %Typically these programs filter the database on a single attribute. 
%Examples of this type of programs are . %The post-processing step in P1 is done in the clear and hence requires no more interactions with the \textsf{CSP}.

\stitle{Class II: \textsf{LabHE} Multiplication Interaction Programs}\\
\system supports a $n$-way multiplication of ciphers for $n > 2$ as described in Section~\ref{genlab} which requires intermediate interactions with the \textsf{CSP}. Thus all \system programs that require multiplication of more than two ciphers need interaction with the \textsf{CSP}. Examples include %All programs that filter the database on more than three attributes, such as 
P4 and P5 (Table~\ref{tab:programexamples}).

\stitle{Class III: Other Interaction Programs}\\
 The \textsf{GroupByCountEncoded} operator requires an intermediate interaction with the \textsf{CSP}. %(for generating the encrypted one-hot-coding for the new attribute). 
 The \textsf{CountDistinct} operator also uses a garbled circuit (\ifpaper 
details in full paper \cite{anom}% using paper 
\else 
details in Appendix~\ref{app:implement_operators}
\fi) and hence requires interactions with the \textsf{CSP}. %This is in addition to the interaction required for decrypting the noisy answer (as explained in Class I above). 
Therefore, any program with the above two operators, like P6 and P7 (Table~\ref{tab:programexamples}), requires at least two rounds of interaction. 
}

\eat{
\begin{frame}{}
  \begin{center}
    \scalebox{0.75}{
    \begin{minipage}{1\linewidth}
\begin{algorithm}[H]
\caption{\textsf{GroupByCount }$\groupby_A(\mathbf{\tilde{T}})$}
\begin{algorithmic}[1]
\STATEx
\textbf{Input}: $\mathbf{\tilde{T}}$
\STATEx \textbf{Output}: $\tilde{\encV}$
\STATEx \textbf{\textsf{AS}:} \STATE Computes $\mathbf{V}=\groupbystar_{A}(\encT)$.
\STATE Masks the encrypted histogram $\mathbf{V}$ for attribute $A$ as follows \begin{gather*}\boldsymbol{\mathcal{V}}[i]= \mathbf{V}[i] \oplus labEnc_{pk}(M[i])\\M[i] \in_R [m], i \in [|V|]\end{gather*}
\STATE Sends $\boldsymbol{\mathcal{V}}$ to \textsf{CSP}.
\STATEx \textbf{\textsf{CSP}:}
\STATE Decrypts  $\boldsymbol{\mathcal{V}}$ as $\mathcal{V}[i]=labDec_{sk}(\boldsymbol{\mathcal{V}}), i \in [|V|]$.\STATE Converts each entry of $\mathcal{V}$ to its corresponding one-hot-coding and encrypts it, $\boldsymbol{\tilde{\mathcal{V}}}[i]=labEnc_{pk}(\tilde{\mathcal{V}[i]}), i \in [|V|]$
\STATE Sends $\boldsymbol{\tilde{\mathcal{V}}}$ to \textsf{AS}.
\STATEx \textbf{\textsf{AS}}:
\STATE  Rotates every entry by its corresponding mask value to obtain the desired  encrypted one-hot-coding $\boldsymbol{\tilde{V}}[i]$. \begin{gather*}\boldsymbol{\tilde{V}}[i]=RightRotate(\boldsymbol{\tilde{\mathcal{V}}},M[i]), i \in [|V|]\end{gather*} 
 \end{algorithmic} \label{groupby-imp}
\end{algorithm} 
 \end{minipage}}
  \end{center}
\end{frame}
}

%\vspace{-1.5em}
\section{\system Security Sketch} \label{sec:security}
In this section, we provide a sketch of the security proof in the semi-honest model using the well established
simulation argument~\cite{Oded}. %We assume that the reader is familiar with standard concepts, such when two distributions are computationally indistinguishable ($\equiv_c$).
\eat{Let pro-
gram P be executed on a dataset D with privacy parameter
ε and let PCDP(D,ε/2) denote the random variable corre- U
sponding to the output of running P in the CDP model un-
der unbounded ε/2 - DP (data size is unknown, Defn. 3 in
the full paper [6]). Now, let PCDP (D,ε) denote the random
variable corresponding to the output of running P in the
CDP model under bounded ε - DP such that PCDP (D,ε) ≡
(PCDP (D,ε/2), |D|). Clearly by Theorem 7 in the full paper U
[6], PCDP (D,ε) ensures overall bounded ε - DP.}
\system takes as input a DP program, $P$, and a privacy parameter, $\epsilon$, and translates $P$ into a protocol, $\Pi$, which in turn is executed by the \AS and the \CSP.    In addition to revealing the output of the program $P$,  $\Pi$ also reveals the number of records in the dataset, $\mathcal{D}$. Let $P^{CDP}(\mathcal{D},\epsilon/2)$ denote the random variable corresponding to the output of running $P$ in the \textsf{CDP} model under $\epsilon/2$-DP (Definition \ref{def:dp}). We make the following claims: 
\squishlist
\item The views and outputs of the \AS and \CSP are computationally indistinguishable from that of simulators with access to only
$P^{CDP}(\mathcal{D},\epsilon/2)$ and the total dataset size $|\mathcal{D}|$.
\item For every $P$ that satisfies $\epsilon/2$-DP (Definition \ref{def:dp}), revealing its output (distributed identical to $P^{CDP}(\mathcal{D},\epsilon/2)$) as well as $|\mathcal{D}|$ satisfies $\epsilon$-bounded DP, where neighboring databases have the same size but differ in one row.
\item Thus, the overall protocol satisfies computational differential privacy under the SIM-CDP model. 
\squishend
Now, let $P^{CDP}_B(\mathcal{D},\epsilon)$ denote the random
variable corresponding to the output of running $P$ in the
\textsf{CDP} model under $\epsilon$-bounded DP such that $P^{CDP}_B(\mathcal{D},\epsilon) \equiv
(P^{CDP}(\mathcal{D},\epsilon/2), |\mathcal{D}|)$. 
We state the main theorems here and refer the reader to the \ifpaper 
the full paper \cite{anom}% using paper 
\else 
Appendix \ref{app:background} \fi for formal proofs.
\vspace{-2mm}
\begin{theorem}\label{thm:security}
\rm
Let protocol $\Pi$ correspond to the execution of program $P$ in \system. The
views and outputs of the \textsf{AS} and the \textsf{CSP} are denoted as $
View_1^{\Pi}(P,\mathcal{D},\epsilon), Output_1^{\Pi}(P,\mathcal{D},\epsilon)$ and $
View_2^{\Pi}(P,\mathcal{D},\epsilon), Output_2^{\Pi}(P,\mathcal{D},\epsilon)$ respectively.
There exists Probabilistic Polynomial Time (PPT) simulators, $Sim_1$
and $Sim_2$, such that:
\squishlist
\item \scalebox{0.9}{$Sim_1 (P^{CDP}_B(\mathcal{D},\epsilon))$} is computationally indistinguishable ($\equiv_c$)
from \scalebox{0.9}{$(View_1^{\Pi}(P,\mathcal{D},\epsilon),Output^{\Pi}(P,\mathcal{D},\epsilon))$}, and
\item \scalebox{0.9}{$Sim_2 (P^{CDP}_B(\mathcal{D},\epsilon))$} is $\equiv_c$ to
  \scalebox{0.9}{$(View_2^{\Pi}(P,\mathcal{D},\epsilon),Output^{\Pi}(P,\mathcal{D},\epsilon))$}.
  \squishend \scalebox{0.9}{$Output^{\Pi}(P,\mathcal{D},\epsilon))$} is the combined
  output of the two parties\footnote{Note that the simulators are
    passed a random variable $P^{CDP}_B(\mathcal{D},\epsilon))$, i.e., the simulator is given the ability to sample
    from this distribution.}.
\end{theorem}
The main ingredient for the proof is the composition theorem~\cite{Oded}, which informally states: suppose a protocol, $\Pi_f^g$,
implements functionality $f$ and uses function $g$ as an oracle (uses only input-output behavior of $g$).  Assume that protocol $\Pi_g$ implements $g$ and calls to $g$ in $\Pi_f^g$ are replaced
by instances of $\Pi_g$ (referred to as the composite
protocol). If $\Pi_f$ and $\Pi_g$ are correct (satisfy the above simulator
definition), then the composite protocol is
correct. Thus, the proof can be done in a modular
fashion as long as the underlying operators are
used in a blackbox manner (only the input-output behavior are used
and none of the internal state are used).

Next, every \system program expressed as a sequence of  transformation operators
 followed by a measurement operator, satisfies $\epsilon/2$-DP (as in Definition~\ref{def:dp}). It is so because recall that the measurement operators add noise from $Lap(\frac{2\Delta}{\epsilon})$ (Section \ref{sec:measurement_operators}) where $\Delta$ denotes the sensitivity of $P$ (computed w.r.t to Definition \ref{def:dp}) \cite{Dork,APEx}.  However, \system reveals both the output of the program as well as the total size of dataset $\mathcal{D}$. While revealing the size exactly would violate Definition~\ref{def:dp}, it does satisfy \textit{bounded}-DP albeit with twice the privacy parameter, $\epsilon$ -- changing a row in $\mathcal{D}$ is equivalent to adding a row and then removing a row.

Finally, since every program $P$
executed on \system satisfies $\epsilon$-bounded DP,
it follows from Theorem~\ref{thm:security} that every execution of
\system satisfies computational DP.
 \noindent
\begin{corollary} \vspace{-1mm}
	Protocol $\Pi$ satisfies computational differential privacy under the \textsf{SIM-CDP} notion \cite{CDP}.
\end{corollary}

Note that Theorem \ref{thm:security} assumes that \textsf{AS} and the \textsf{CSP} do
not collude with the users (data owners). However, if the \textsf{AS}
colludes with a subset of the users, $U$, then $Sim_1$
($Sim_2$) has to be given the data corresponding to users in $U$ as
additional parameters. This presents no complications in the proof (see the proof in~\cite{LReg}). If a new user $u$ joins, their
data can be encrypted and simply added to the database. We discuss extensions to handle malicious adversaries in Section \ref{sec:malicious}.
\section{\system Optimizations}\label{sec:optimization}
In this section, we present the optimizations used by \system. %These optimizations do not alter the end-to-end privacy guarantees of the system. 
%\vspace{-0.2cm}
\subsection{DP Index Optimization}\label{sec:dp_optimization}
 This optimization is motivated by the fact that several programs, first, filter out a large number of rows in the dataset. For instance, P5 in Table~\ref{tab:programexamples} constructs a histogram over $Age$ and $Gender$ on the subset of rows for which $NativeCountry$ is Mexico. \system's filter implementation retains all the rows as the \textsf{AS} has no way of telling whether the filter condition is satisfied. As a result, the subsequent \textsf{GroupbyCount} is run on the full dataset. If there were an index on $NativeCountry$,  \system could run the \textsf{GroupbyCount} on only the subset of rows with $NativeCountry$=Mexico. But an exact index would violate DP. Hence, we propose a DP index to bound the information leakage while improving the performance.
 
At a high-level, the DP index on any ordinal attribute $A$ is constructed as follows:
(1) securely sort the input encrypted database, $\encD$, on $A$ and (2) learn a mapping, $\mathcal{F}$, from the domain of $A$ to  $[1,|\encD|]$ such that most of the rows with index less than $\mathcal{F}(v), v \in domain(A)$, have a value less than $v$.  The secure sorting is done via the following garbled circuit that (1) inputs $\encD$ (just the records without any identifying features)  and indexing attribute $A$ from the \AS (2) inputs the secret key $sk$ from the \CSP
(3) decrypts  and sort $\mathcal{D}$ on $A$ 
(4) re-encrypt the sorted database using $pk$ and outputs $\encD_s = labEnc_{pk}(sort(\mathcal{D}))$. 
The mapping, $\mathcal{F}$, must be learned under DP, and we present a method for that below. Let $P=(P_1,\ldots,P_k)$ be an equi-width partition on the sorted domain of $A$ such that each partition (bin) contains $\frac{s_A}{k}$ %(if $s_A$ is not a multiple of $k$, then $P_k$ has a smaller size) 
consecutive domain values where $s_A$ is the domain size of $A$. The index is constructed using a \system program that firstly computes the noisy prefix counts, $\hat{V}[i]=\sum_{v \in \cup_{l=1}^i P_l} ct_{A,v}+\eta_i$ for $i \in [k], \mbox{ where } \eta_i\sim Lap(2k/\epsilon_A)$ and $ct_{A,v}$ denotes the number of rows with value $v$ for $A$. Next, the program uses isotonic regression \cite{cdf} on $\hat{V}$ to generate a noisy cumulative histogram $\tilde{\mathcal{C}}$ with non-decreasing counts. Thus, each prefix count in $\tilde{\mathcal{C}}$ gives an \emph{approximate index} for the sorted database where the values of attribute $A$ change from being in $P_i$ to a value in $P_{i+1}$. When a \system program starts with a filter $\phi=A \in [v_s, v_e]$, we compute two indices for the sorted database, $i_s$ and $i_e$, as follows. Let $v_s$ and $v_e$ fall in partitions $P_i$ and $P_j$ respectively. If $P_i$ is the first partition, then we set $i_s=0$; otherwise set $i_s$ to be $1$ more than the $i-1$-{th} noisy prefix count from $\tilde{\mathcal{C}}$. Similarly, if $P_j$ is the last partition, then we set $i_e=|\encD|$; otherwise, we set $i_e$ to be the $j+1$-th noisy prefix count from $\tilde{\mathcal{C}}$. This gives us the DP mapping $\mathcal{F}$. %: (a) the noisy starting index, for the bin $P_s$ such that $v_s \in P_s$, $i_{s}=\hat{\mathcal{C}}[s-1]+1, v_s \in [\frac{s_A}{k}\cdot(s-1)+1,\frac{s_A}{k}\cdot s], s \in [k]$,   and  (b) the noisy end index for the bin $P_e$ such that $v_e \in P_e$, $i_{e}=\hat{\mathcal{C}}[e], v_e \in [\frac{s_A}{k}\cdot(e-1)+1,\frac{s_A}{k}\cdot e],e \in [k]$. 
We then run the program on the subset of rows in $[i_{s},i_{e}]$.  
 For example, in Figure \ref{fig:index}, the indexing  attribute with domain $\{v_1,\cdots,v_{10}\}$ has been partitioned into $k=5$ bins and  if $\phi \in [v_3,v_6]$, $i_{s}=\tilde{\mathcal{C}}[1]+1=6$ and $i_{e}=\tilde{\mathcal{C}}[3]=13$. \vspace{-6mm}
\begin{lemma}\label{lemma:DPindex} Let $P$ be the program that computes the mapping $\mathcal{F}$.
Let $\Pi$ be the \system protocol corresponding to the construction of the DP index. The
views and outputs of the \textsf{AS} and the \textsf{CSP} are denoted as $
View_1^{\Pi}(P,\mathcal{D},\epsilon_A)$, $ Output_1^{\Pi}(P,\mathcal{D},\epsilon_A)$  and $
View_2^{\Pi}(P,\mathcal{D},\epsilon_A)$, $ Output_2^{\Pi}(P,\mathcal{D},\epsilon_A)$ respectively.
There exists PPT simulators $Sim_1$
and $Sim_2$ such that:
\squishlist
\item \scalebox{0.9}{$Sim_1 (P^{CDP}_B(\mathcal{D},\epsilon_A)) \equiv_c (View_1^{\Pi}(P,\mathcal{D},\epsilon_A),Output^{\Pi}(\mathcal{D},\epsilon_A))$}, and
\item $Sim_2 (P^{CDP}_B(\mathcal{D},\epsilon))\equiv_c(View_2^{\Pi}(P,\mathcal{D},\epsilon_A),Output^{\Pi}(\mathcal{D},\epsilon_A))$.
  \squishend $Output^{\Pi}(P,\mathcal{D},\epsilon_A))$ is the combined
  output of the two parties \label{lem:DPindex}\vspace{-2mm}
  \end{lemma}
  The proof of the above lemma is presented in \ifpaper the full paper \cite{anom}\else Appendix \ref{app:index-imp}\fi. Here we present the intuition behind it. From the secure sorting algorithm  (steps 1 and 4), it is evident that the servers cannot associate the records of the encrypted sorted dataset, $\encD_s$, with the data owners. The \AS can learn nothing from $\encD_s$ due to the semantic security of the 
  \textsf{LHE} scheme used. This ensures that the DP index construction of \system satisfies the \textsf{SIM-CDP} privacy guarantee.
\squishlist
\item \textbf{Optimized feature}: This optimization speeds up the program execution by reducing the total number of rows to be processed for the program.
\item \textbf{Trade-off}: The trade-off is a possible increase in error as some of the rows that satisfy the filter condition may not be selected due to the noisy index. 
\item \textbf{Privacy Cost}: Assuming the index is constructed with  privacy parameter $\epsilon_A$, the selection of a subset of rows using it will be $\epsilon_A$-bounded DP (Lemma \ref{lem:DPindex}).  If $\epsilon_L$ is the parameter used for the subsequent measurement primitives, then by Theorem 1, the total privacy parameter is $\epsilon_A+\epsilon_L$.
\squishend
\begin{figure}
      \includegraphics[width=0.9\linewidth,height=2cm]{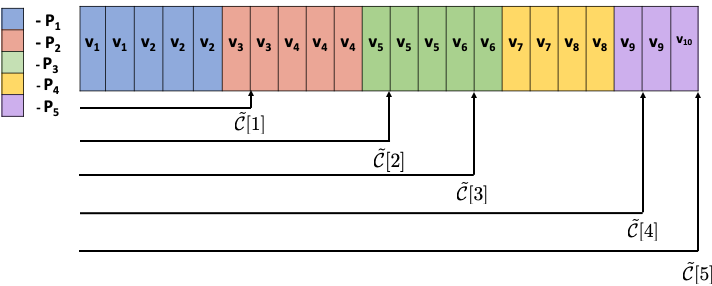}\vspace{-0.5cm}
        \caption{Illustrative example for DP Index}\label{fig:index}\vspace{-8mm}
    \end{figure}
\stitle{Discussion}: Here we discuss the various parameters  in the construction of a DP index. The foremost parameter is the indexing attribute $A$ which can be chosen with the help of the following two heuristics. First, $A$ should be frequently queried so that a large number of queries can benefit from this optimization. Second, choose $A$ such that the selectivity of the popularly queried values of $A$ is high. This would ensure that the first selection performed alone on $A$ will filter out the majority of the rows, reducing the intermediate dataset size to be considered for the subsequent operators.  
The next parameter is the fraction of the program privacy budget, $\rho$ ($\epsilon_A=\rho \cdot \epsilon$ where $\epsilon$ is the total program privacy budget) that should be used towards building the index. The higher the value of $\rho$, the better is the accuracy of the index (hence better speed-up). However, the privacy budget allocated for the rest of the program decreases resulting in increased noise in the final answer. This trade-off is studied in Figures \ref{fig:error:} and \ref{fig:time:} in Section \ref{sec:experiments}. 
Another parameter is the number of bins $k$. Finer binning gives more resolution but leads to more error due to DP noise addition. Coarser binning introduces error in indexing but has lower error due to noise. We explore this trade-off in Figures \ref{fig:error:NumberOfBins} and \ref{fig:time:NumberOfBins}. To increase accuracy we can also consider bins preceding $i_{s}$ and bins succeeding $i_{e}$. This is so because, since the index is noisy, it might miss out on some rows that satisfy the filter condition. For example, in Figure \ref{fig:index}, both the indices $i_{s}=\tilde{\mathcal{C}}[1]+1=6$ and $i_{e}=\tilde{\mathcal{C}}[3]=13$ miss a row satisfying the filter condition $\phi=A \in [v_3,v_6]$; hence including an extra neighboring bin would reduce the error.

Thus, in order to gain in performance, the proposed DP index optimization allows some DP leakage of the data. This is in tune with the works in \cite{Mazloom:2018:SCD,He:2017:CDP,Chan:2019:FDO:3310435.3310585,Groce}. However, our work differs from  earlier work in the fact that we can achieve pure DP (albeit SIM-CDP). In contrast, previous work achieved a weaker version of DP, approximate DP \cite{approxDP}, and added one-sided noise (i.e., only positive noise). One-sided noise requires addition of dummy rows in the data, and hence increases the data size. However, in our \system programs, all the rows in the noisy set are part of the real dataset.%This means that in \system,  the filter can exclude rows that actually satisfy the condition, but we do not need to add dummy rows and increase the size of the table. 
%\vspace{-1.5cm}
\subsection{Crypto-Engineering Optimizations} \label{sec:im_optimization}
\stitle{{\normalfont (1)} DP Range Tree}:
%\am{If we have a DP index on an attribute, why do we need a DP range tree? A DP index can be used to do the same kinds of optimization the DP-Range tree provides. So, move this back to the previous section and maybe say this: Indexes can help speed up the answering of range counting queries. The DP index requires the table be sorted by the attribute A. For other attributes, one can construct index structure like the 1-D range tree to avoid a table scan for answering range counting queries. One can think of these as secondary indexes ... and shorten the rest of the description.}
%A 1-dimensional range tree for an attribute $A$ is an ordered hierarchical data structure that can efficiently answer all possible range queries on $A$.
%such that the leaf nodes correspond to the individual counts for each possible value of $A$, while the parent nodes cover a range of values and store the sum of the counts of its children.
If range queries are common, pre-compu-ted noisy range tree is a useful optimization. For example, building a range tree on $Age$ attribute can improve the accuracy for P1 and P2 in Table~\ref{tab:programexamples}. The sensitivity for  such a noisy range tree is $\log s_A$ where $s_A$ is the domain size of the attribute on which the tree is constructed. Any arbitrary range query requires access to at most $2\log s_A$ nodes on the tree. Thus to answer all possible range queries on $A$, the total squared error accumulated is $O(\frac{s^2(\log s_A)^2 }{\epsilon})$. In contrast for the naive case, we would have incurred error $O(\frac{s_A^3}{\epsilon})$ \cite{cdf}. Note that, if we already have a DP index on $A$, then the DP range tree can be considered to be a secondary index on $A$. %Hence this range tree optimization not only gives us a huge performance boost but also results in better answer accuracy.

%A 1-dimensional range tree for an attribute $A$ is an ordered data structure such that the leaf nodes correspond to the individual counts $ct_{A,i}$, $v_i \in dom(A)$ while each parent node stores the sum of the counts of its children. Hence an useful optimization for \system can be pre-computing the noisy range tree for some of the most popular attributes. This can be useful for programs P1 and P2 in Table 3. %In Crypt$\epsilon$ we construct a noisy range tree for some of the attribute.
%The sensitivity for each such noisy range tree is $\log k$ where $k$ is the domain size of the attribute on which the tree is constructed. For answering any arbitrary range query, we need to access at most $2\log k$ nodes of the range tree. Thus to answer all possible range queries for the given attribute, the total squared error accumulated is $O(\frac{k^2(\log k)^2 }{\epsilon})$. In contrast for the naive case, we would have incurred error $O(\frac{k^3}{\epsilon})$. Hence this range tree optimization not only gives us a huge performance boost but also results in better answer accuracy.

\squishlist
\item \textbf{Optimized Feature}: The optimization  reduces both execution time and expected error when executed over multiple range queries.
\item \textbf{Trade-off}: The trade-off for this optimization is the storage cost of the range tree $(O(2\cdot s_A))$.
\item \textbf{Privacy Cost}: If the range tree is constructed with privacy parameter $\epsilon_R$, then any measurement on it is post-processing. Hence, the privacy cost is $\epsilon_R$-bounded DP.
\squishend

\stitle{{\normalfont (2)} Precomputation}:  
%Recall that the data owners $\textsf{DO}_i$ send per-attribute encrypted one-hot-codings of their data to the \textsf{AS}.
The \textsf{CrossProduct} primitive generates the one-hot-coding of data across two attributes. However, this step is costly due to the intermediate interactions with the \textsf{CSP}. Hence, a  useful optimization is to pre-compute the one-hot-codings for the data across a set of frequently used attributes $\bar{A}$ so that for subsequent program executions, the \textsf{AS} can get the desired representation via simple look-ups. For example, this benefits P3 (Table \ref{tab:programexamples}).
\squishlist
\item \textbf{Optimized Feature}: This reduces the execution time of \system programs. The multi-attribute one-hot-codings can be re-used for all subsequent programs.
%Moreover, once the multi-attribute one-hot-codings are created it can be re-used for all subsequent programs.
\item  \textbf{Trade-off}: The trade-off is the storage cost (O($m\cdot s_{\bar{A}}=m\cdot \prod_{A \in \bar{A}}s_A$), $m =$ the number of data owners) incurred to store the multi-attribute one-hot-codings for $\bar{A}$.
\item \textbf{Privacy Cost}: The computation is carried completely on the encrypted data, no privacy budget is expended.
\squishend

\stitle{{\normalfont (3)} Offline Processing}:
For $\textsf{GroupByCountEncoded}$, the \textsf{CSP} needs to generate the encrypted one-hot-codings for the masked histogram. Note that the one-hot-encoding representation for any such count would simply be a vector of $(|\encD|-1)$ ciphertexts for `0', $labEnc_{pk}(0)$ and 1 ciphertext for `1', $labEnc_{pk}(1)$. Thus one useful optimization is to generate these ciphertexts offline (similar to offline generation of Beaver's multiplication triples \cite{Beaver} used in SMC). Hence, the program execution will not be blocked by encryption.
\squishlist
\item \textbf{Optimized Feature}: This optimization results in a reduction in the run time of \system programs. \item \textbf{Trade-off}: A storage cost of O($m\cdot s_A$) is incurred to store the ciphers for attribute $A$.
\item \textbf{Privacy Cost}: The computation is carried completely on the encrypted data, no privacy  budget is expended.
\squishend

%\paragraph*{\textbf{Optimized Crypt$\epsilon$ programs}}
%Let us reconsider the example programs covered in section 5. Both Program 1 and Program 2 can be optimized by constructing a range tree over attribute $Age$. Program  4 and Program 5  on the other hand can be improved by the differentially private index over attribute $NativeCountry$ while for Program 6 we can create the index over attribute $Gender$.

%\vspace{-5mm}
\section{Experimental Evaluation}\label{sec:experiments}
\begin{figure*}[ht]
    \begin{subfigure}[b]{0.25\linewidth}
        \centering
         \includegraphics[width=1\linewidth]{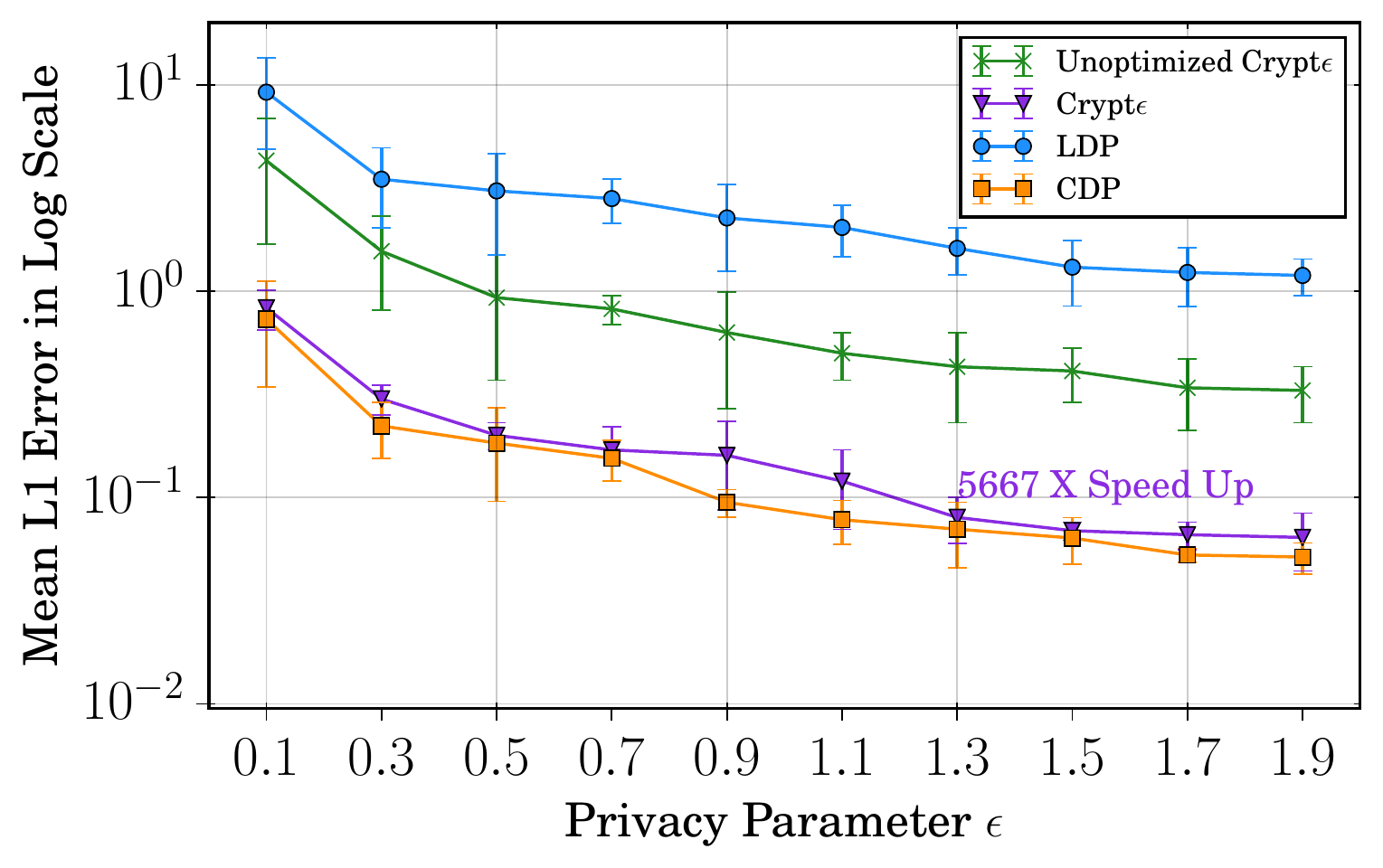}
            
        \caption{ Program 1}
        \label{fig:P1}
    \end{subfigure}%%
    \begin{subfigure}[b]{0.25\linewidth}
    \centering \includegraphics[width=1\linewidth]{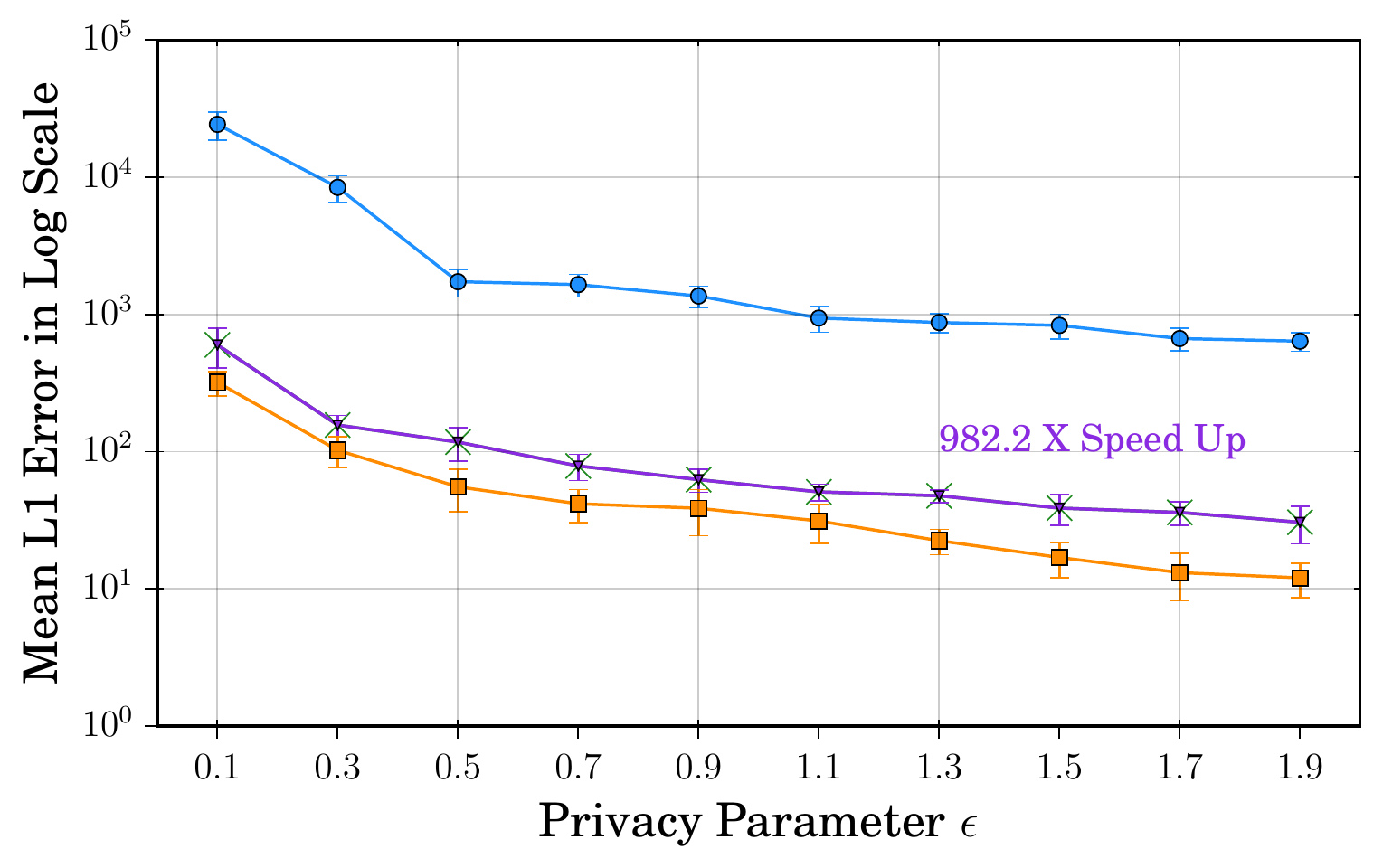}   
 \caption{Program 3}
        \label{fig:P3}\end{subfigure}%%
    \begin{subfigure}[b]{0.25\linewidth}
    \centering    \includegraphics[width=1\linewidth]{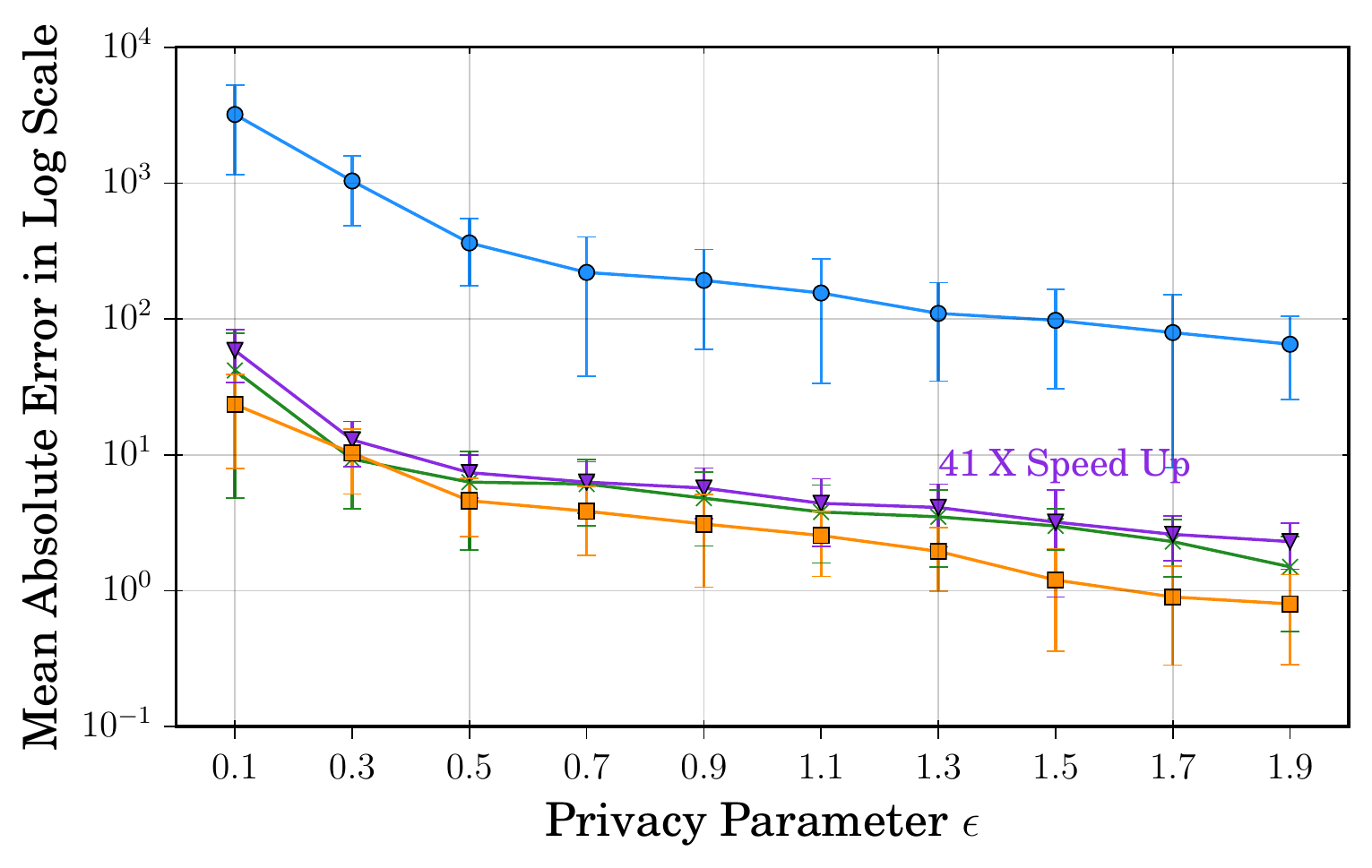}
        \caption{Program 5}
        \label{fig:P5}\end{subfigure}%%
      \begin{subfigure}[b]{0.25\linewidth}
    \centering    \includegraphics[width=1\linewidth]{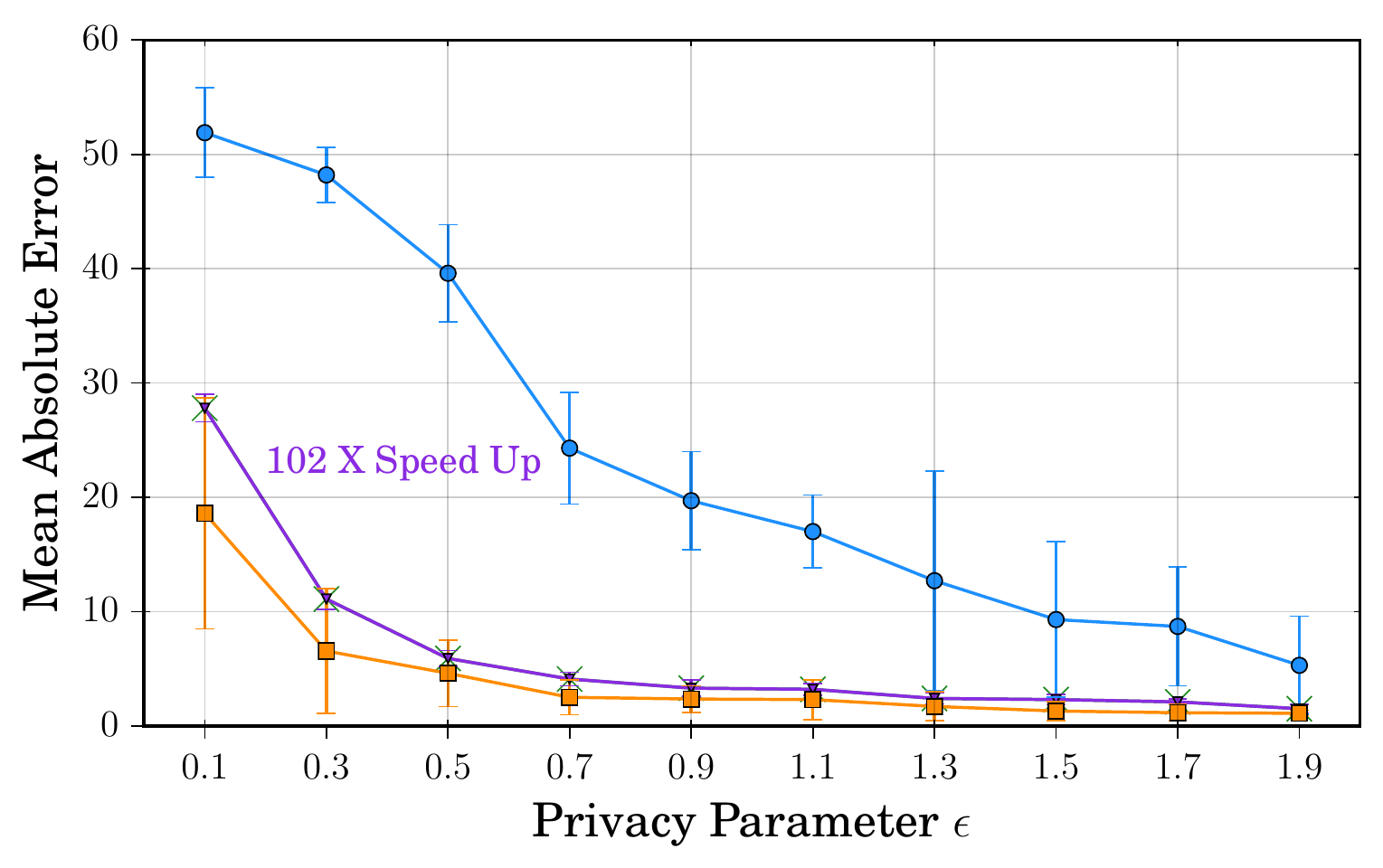}    
        \caption{Program 7}
        \label{fig:P7}
    \end{subfigure}
    \vspace{-8mm}
   \caption{Accuracy Analysis of Crypt$\epsilon$ Programs}
   \label{accuracy}
\end{figure*}
In this section, we describe our evaluation of Crypt$\epsilon$ along two dimensions, accuracy and  performance of Crypt$\epsilon$ programs. Specifically, we address the following questions: \squishlist \item \textbf{Q1}: Do Crypt$\epsilon$ programs have significantly lower errors than that for the corresponding state-of-the-art \textsf{LDP} implementations? Additionally, is the accuracy of \system programs comparable to that of the corresponding \textsf{CDP} implementations? \item \textbf{Q2}: Do the proposed optimizations provide substantial performance improvement over unoptimized Crypt$\epsilon$? \item \textbf{Q3}: Are Crypt$\epsilon$ programs practical in terms of their execution time and do they scale well? \squishend
\textbf{Evaluation Highlights}:
\squishlist \item \system can achieve up to $50\times$ smaller error than the corresponding
\textsf{LDP} implementation on a data of size $\approx 30K$ (Figure \ref{accuracy}). Additionally, \system errors are at most $2\times$ more than that of the corresponding \textsf{CDP} implementation.
\item The optimizations in \system can improve the performance
of unoptimized \system by up to $5667\times$ (Table \ref{perf}).
\item A large class of \system programs execute within
 3.6 hours for a dataset of size $10^6$,  and they scale linearly with the dataset size (Figure \ref{fig:scale}). The \textsf{AS} performs majority of the work for most programs (Table \ref{perf}).
\squishend

\vspace{-0.8em}
\subsection{Methodology} 
\textbf{Programs}:
To answer the aforementioned questions, we ran the experiments on the Crypt$\epsilon$ programs previously outlined in Table \ref{tab:programexamples}. Due to space limitations, we present the results of only four of them in the main paper namely P1, P3, P5 and P7. The rationale behind choosing these four is that they cover all three classes of programs (Section \ref{sec:implementation}) and showcase the advantages for all of the four proposed optimizations. %Each program is compared with the corresponding state-of-the-art \textsf{LDP} implementation \cite{LDP1}.
\\\textbf{Dataset}:
We ran our experiments on the Adult dataset from the UCI repository \cite{UCI}. The dataset is of size $32,651$. For the scaling experiments (Figure \ref{fig:scale}), we create toy datasets of sizes 100K and 1 million by copying over the Adult dataset.
\\\textbf{Accuracy Metrics}:  Programs with scalar outputs (P5, P7) use \textit{absolute error} $|c-\hat{c}|$ where $c$ is the true count and $\hat{c}$ is the noisy output. Programs with vector outputs (P1, P3) use the \textit{L1 error metric} given  by $Error=\sum_{i}|V[i]-\hat{V}[i]|, i \in [|V|]$ where $V$ is the true vector and $\hat{V}$ is the noisy vector.
We report the mean and s.t.d of error values over 10 repetitions.
\\\textbf{Performance Metrics}: We report the mean total execution time in seconds for each program, over 10 repetitions.
\\\textbf{Configuration}: We implemented \system in Python with the garbled circuit implemented via EMP
toolkit \cite{EMP}. We use Paillier encryption scheme \cite{Paillier}. All the experiments have been performed on the Google Cloud Platform \cite{GCP} with the configuration  c2-standard-8. %, CPU: 8 cores(vCPUs), 3.8GHz Cascade Lake, Memory: 32 Gb, Disk: 256Gb SSD, OS: Ubuntu Server 16.04.} %The prototype  includes all four optimizations described in section 6.
For Adult dataset, \system constructs a DP index optimization over
the attribute $NativeCountry$ that benefits programs like P4 and P5. Our experiments assign
20\% of the total program privacy parameter towards constructing the index
and the rest is used for the remaining program execution. \system also constructs a DP range tree over $Age$. This helps programs like P1, P2 and P3. This is our default \system implementation. 
\vspace{-1em}
\subsection{End-to-end Accuracy Comparison}
In this section, we evaluate \textbf{Q1} by performing a comparative analysis between the empirical accuracy of the aforementioned four \system programs (both optimized and unoptimized) and that of the corresponding state-of-the-art \textsf{LDP} \cite{ldp} and \textsf{CDP} (under bounded DP; specifically, using the \textsf{CDP} view \system is computationally indistinguishable from as shown in Section \ref{sec:security}) \cite{Dork} implementations. % $\epsilon \in \{0.1,...,0.9\}$. %All the experiments are evaluated on the full dataset and we report our results for 
%We run the aforementioned four \system programs (both optimized and unoptimized) and their corresponding state-of-the-art \textsf{LDP} ~\cite{ldp} and \textsf{CDP}~\cite{Dork} implementations at varying privacy parameters $\epsilon\in\{0.1,\ldots,1.9\}$. 

The first observation with respect to accuracy is that the mean error for a single frequency count for Crypt$\epsilon$ is at least $50\times$ less than that of the corresponding \textsf{LDP} implementation. For example, Figure \ref{fig:P3} shows that for P3, $\epsilon=0.1$ results in a mean error of $599.7$ as compared to an error of $34301.02$ for the corresponding \textsf{LDP} implementation. Similarly, P5 (Figure \ref{fig:P5}) gives a mean error of only $58.7$ for $\epsilon=0.1$. In contrast, the corresponding \textsf{LDP} implementation has an error of $3199.96$.  For P1 (c.d.f on $Age$), the mean error for \system for $\epsilon=0.1$ is given by $0.82$ while the corresponding \textsf{LDP} implementation has an error of $9.2$.  The accuracy improvement on P7 (Figure \ref{fig:P7}) by \system is less significant as compared to the other programs, because P7 outputs the number of age values ([$1-100$]) having  $200$ records. At $\epsilon=0.1$, at least $52$ age values out of $100$ are
reported incorrectly on whether their counts pass the threshold. \system reduces the error almost by  half. Note that the additive error for a single frequency count query in the \textsf{LDP} setting is at least $\Omega(\sqrt{n}/\epsilon)$, thus the error increases with dataset size. On the other hand, for \system the error is of the order $\Theta(1/\epsilon)$, hence with increasing dataset size the relative the error improvement for \system over that of an equivalent implementation in \textsf{LDP} would increase.

For P1 (Figure \ref{fig:P1}), we observe that the error of \system is around $5\times$ less than that of the unoptimized implementation. %For instance, the mean error for \system is $20\times$ lower than that of unoptimized \system for $\epsilon=0.1$. 
The reason is that P1 constructs the c.d.f over the attribute $Age$ (with domain size $100$) by first executing $100$ range queries. %, where the $i$-th query outputs the noisy count for age values in $[1,i], i \in [100]$. This is followed by an isotonic regression post-processing to get a proper c.d.f.
Thus, if the total privacy budget for the program is $\epsilon$, then for unoptimized \system, each query gets a privacy parameter of just $\frac{\epsilon}{100}$. In contrast, the DP range tree is constructed with the full budget $\epsilon$ and sensitivity $\lceil\log 100\rceil$ thereby resulting in lesser error. For P5 (Figure \ref{fig:P5}) however, the unoptimized implementation has slightly better accuracy (around $1.4\times$) than \system. It is because of two reasons; first, the noisy index on $NativeCountry$ might miss some of the rows satisfying the filter condition ($NativeCountry$=Mexico). Second, since only 0.8\% of the total privacy parameter is budgeted for the \textsf{Laplace} operator in the optimized program execution, this results in a higher error as compared to that of unoptimized \system. However, this is a small  cost to pay for achieving a performance gain of $41\times$. The optimizations for P3 (Figure \ref{fig:P3}) and P7 (Figure \ref{fig:P7}) work completely on the encrypted data and do not expend the privacy budget. Hence they do not hurt the program accuracy in any way.
 
Another observation is that for frequency counts the error of \system is around $2\times$ higher than that of the corresponding \textsf{CDP} implementation. This is intuitive because %for guaranteeing bounded $\epsilon$-DP, we need to add noise from $Lap(\frac{2\cdot\Delta}{\epsilon})$ (Section \ref{sec:security}) in \system. Moreover,
 we add two instances of Laplace noise in \system (Section \ref{sec:operator_implementation}).  For P1, the \textsf{CDP} implementation also uses a range tree. %For instance, \system's error for programs 4 (Figure \ref{fig:P3}), 6 (Figure \ref{fig:P5}) and 8 (Figure \ref{fig:P7})  is respectively $1.49\times$, $1.69 \times$ and $2.1\times$ higher  than that of the corresponding \textsf{CDP} implementation for $\epsilon=0.1$.  
 %For P1 however, \system has at least $15\times$ lower error than that of the \textsf{CDP} implementation due to the DP range tree optimization. %Note that the range tree optimization can also implemented in the \textsf{CDP} setting, but here we present the results for just the standard \textsf{CDP} implementation.
 %\vspace{-2mm}
 \subsection{Performance Gain From Optimizations}
In this section, we evaluate \textbf{Q2} (Table \ref{perf}) by analyzing how much speed-up is brought about by the proposed optimizations in the program execution time. 
 \eat{\begin{table}[ht]
\caption{Execution Time Analysis for Crypt$\epsilon$ Programs}
\small
\centering
\begin{tabular}{c  c c c c c}
\toprule
Program &  \multicolumn{3}{c}{Base} & \multicolumn{2}{c}{Optimized} \\ 
 & AS &  CSP & Total & Total & Speed up  \\ &(s)&(s)&(s)&(s)&$\times$\\ % inserts table %heading
\midrule
1 & 0.49& 0.0027& 0.4927 & 0.0029 & 168.9 \\
2 &  6.12 & 0.3  &6.42 & 0.89 & 7.2\\ %197 the communication rounds
3&  3859.52 & 3661.29 & 7520.81&N/A&N/A \\4  &7765.16&3624.05&11389.21& 910.96 & 12.5 \\5&18.56&16.7&35.26&3.49 & 10.1 \\6&1910.01&571.11&2481.12&429.92 & 5.77\\7&6.35 & 1393.89 & 1400.24 &  N/A & N/A \\ [1ex]
\bottomrule
\end{tabular}
\label{c}
\end{table}}
\begin{table}
\vspace{-2mm}
\caption{Execution Time Analysis for Crypt$\epsilon$ Programs}
\centering
\vspace{-2mm}
\scalebox{0.6}{
\begin{tabular}{|c c|c|c|c|c|}
\hline
\multicolumn{2}{|c}{\textbf{Time in (s)}}   & \multicolumn{4}{|c|}{\textbf{Program}}\\
\cline{3-6}&&\textbf{1} & \bf{3} & \bf{5} &\bf{7}  \\ 
\hline \hline
\multirow{3}{*}{\textbf{Unoptimized \system}}& \multicolumn{1}{|c|}{\textsf{AS}} & 1756.71 & 6888.23  & 650.78&290   \\
\cline{2-6}& \multicolumn{1}{|c|}{\textsf{CSP}} & 0.26 & 6764.64 & 550.34  &30407.73\\\cline{2-6}
& \multicolumn{1}{|c|}{Total} & 1756.97 & 13652.87& 1201.12 &30697.73    \\\hline
%\multirow{4}{*}{\textbf{Optimized}}& \multicolumn{1}{|c|}{\textsf{AS}} & .0004 &16.21 &250.17  &\\ \cline{2-6}& \multicolumn{1}{|c|}{\textsf{CSP}} & .0027 &13.92 &0.54 &\\
\multirow{2}{*}{\textbf{\system}}& \multicolumn{1}{|c|}{Total} & 0.31 &     13.9 & 29.21& 299.5 \\
\cline{2-6}& \multicolumn{1}{|c|}{Speed Up $\times$} &5667.64 & 982.2 & 41.1 &    102.49
\\ [1ex]
\hline
\end{tabular}}
\label{perf}
\vspace{-4mm}
\end{table}
\\\stitle{DP Index}: For P5, we observe from Table \ref{perf} that the unoptimized implementation takes around $20$ minutes to run. However,  a DP index over the attribute $NativeCountry$  reduces the execution time to about $30s$ giving us a $41\times $ speed-up. It is so because, only about 2\% of the data records satisfy $NativeCountry$=Mexico. Thus the index drastically reduces the number of records to be processed for the program. %The is because only the cost for the \textsf{Filter} operator decreases, not all the operators scale 10 drop is due to the
 \begin{figure}[ht]
   \begin{subfigure}[b]{0.45\linewidth}
    \centering \includegraphics[width=1\linewidth]{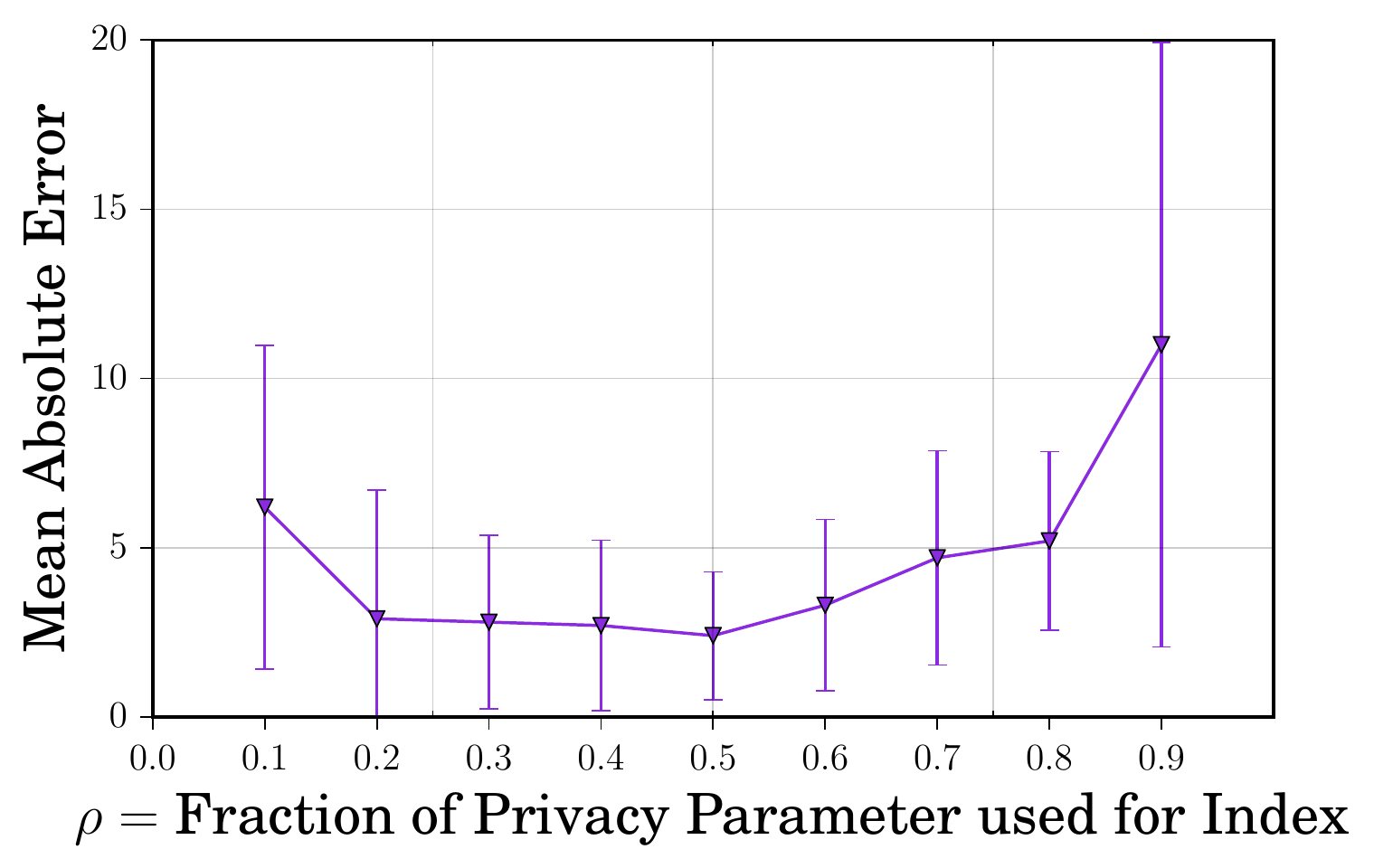}   \caption{}
        \label{fig:error:}\end{subfigure}
        \begin{subfigure}[b]{0.45\linewidth}
        \includegraphics[width=1\linewidth]{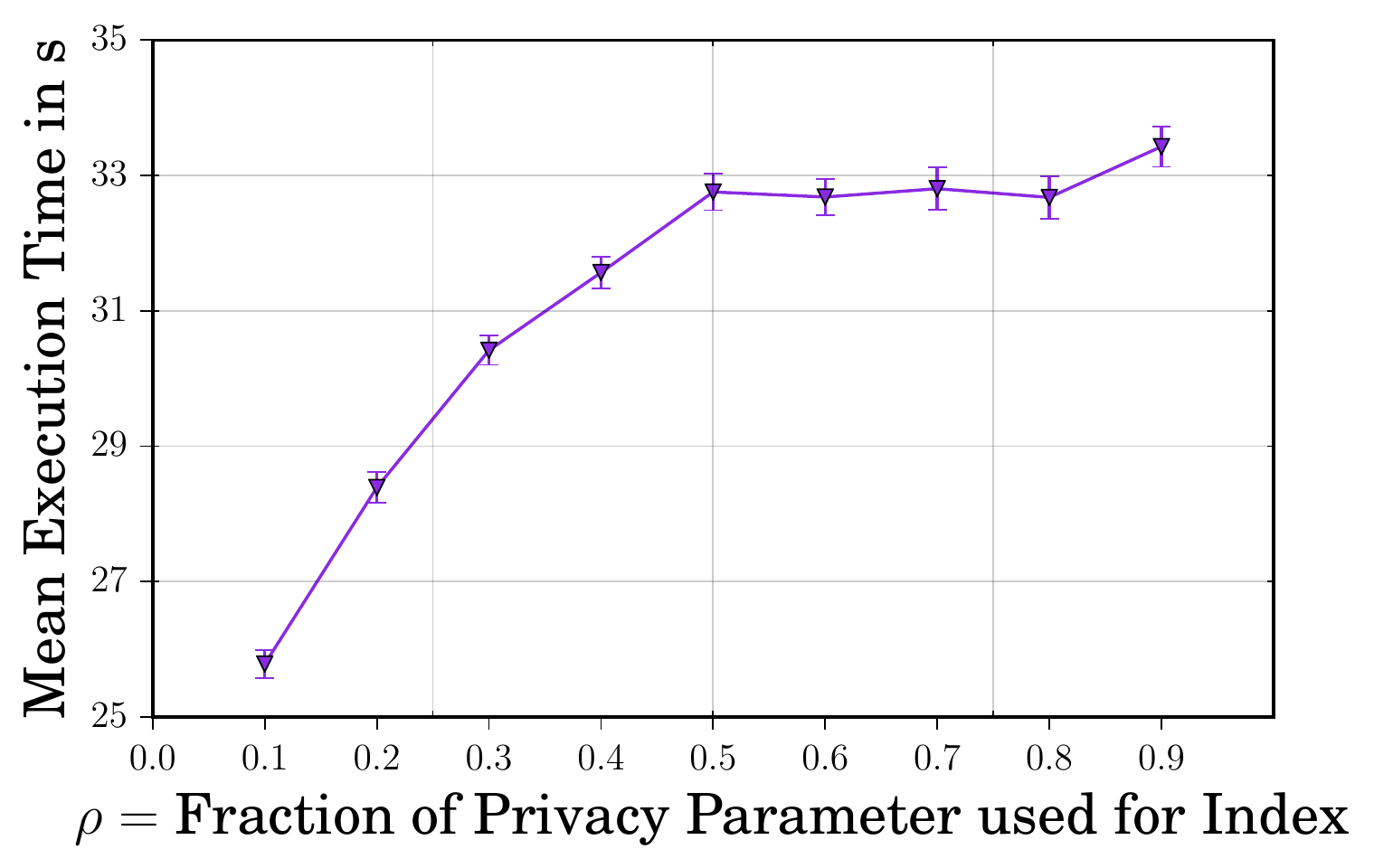}   \caption{}
        \label{fig:time:}
        \end{subfigure}
        \\
         \begin{subfigure}[b]{0.45\linewidth}
    \centering \includegraphics[width=1\linewidth]{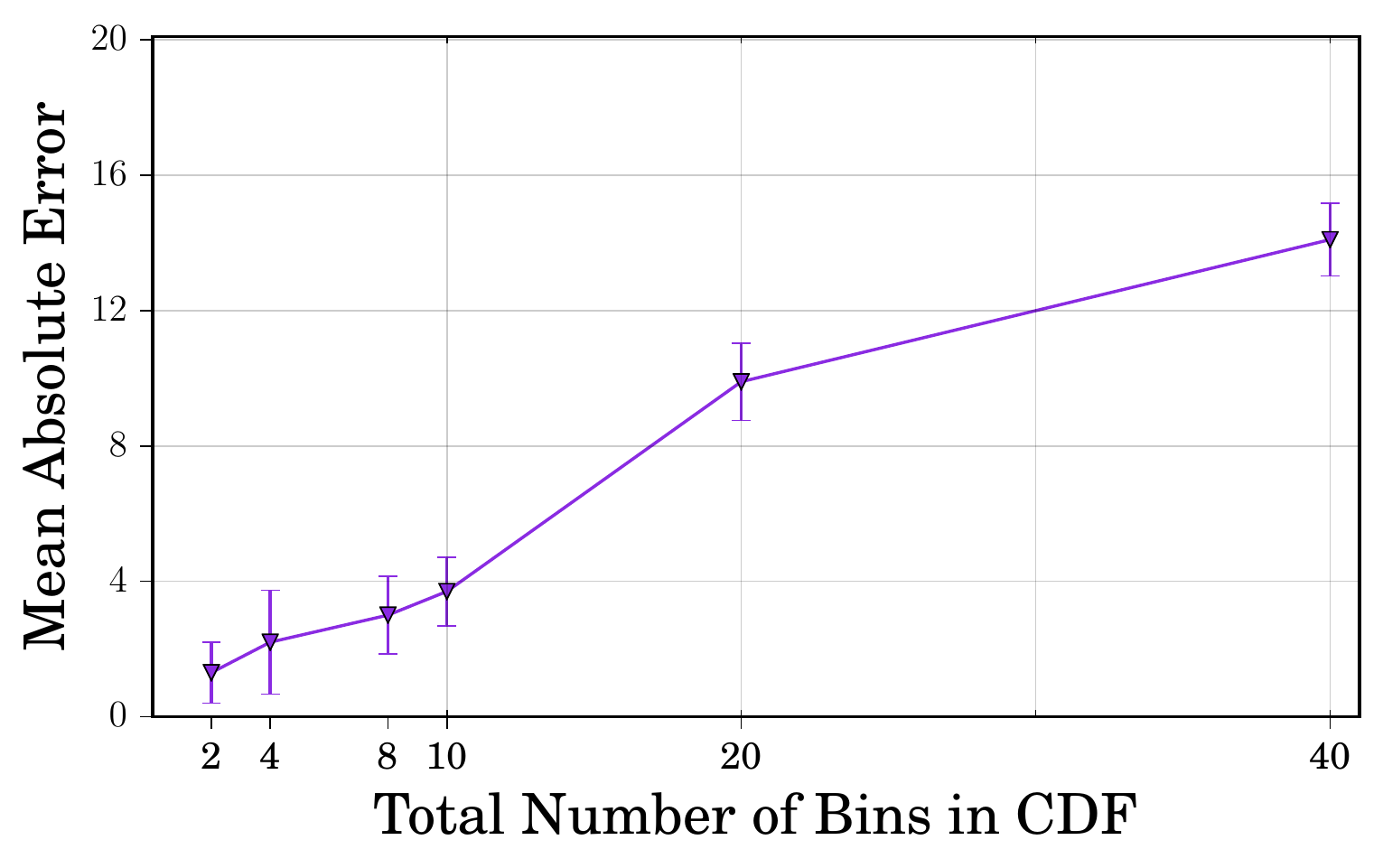}
     %\vspace{-1mm}
        \caption{}
        \label{fig:error:NumberOfBins}\end{subfigure}
        \begin{subfigure}[b]{0.45\linewidth}
        \includegraphics[width=1\linewidth]{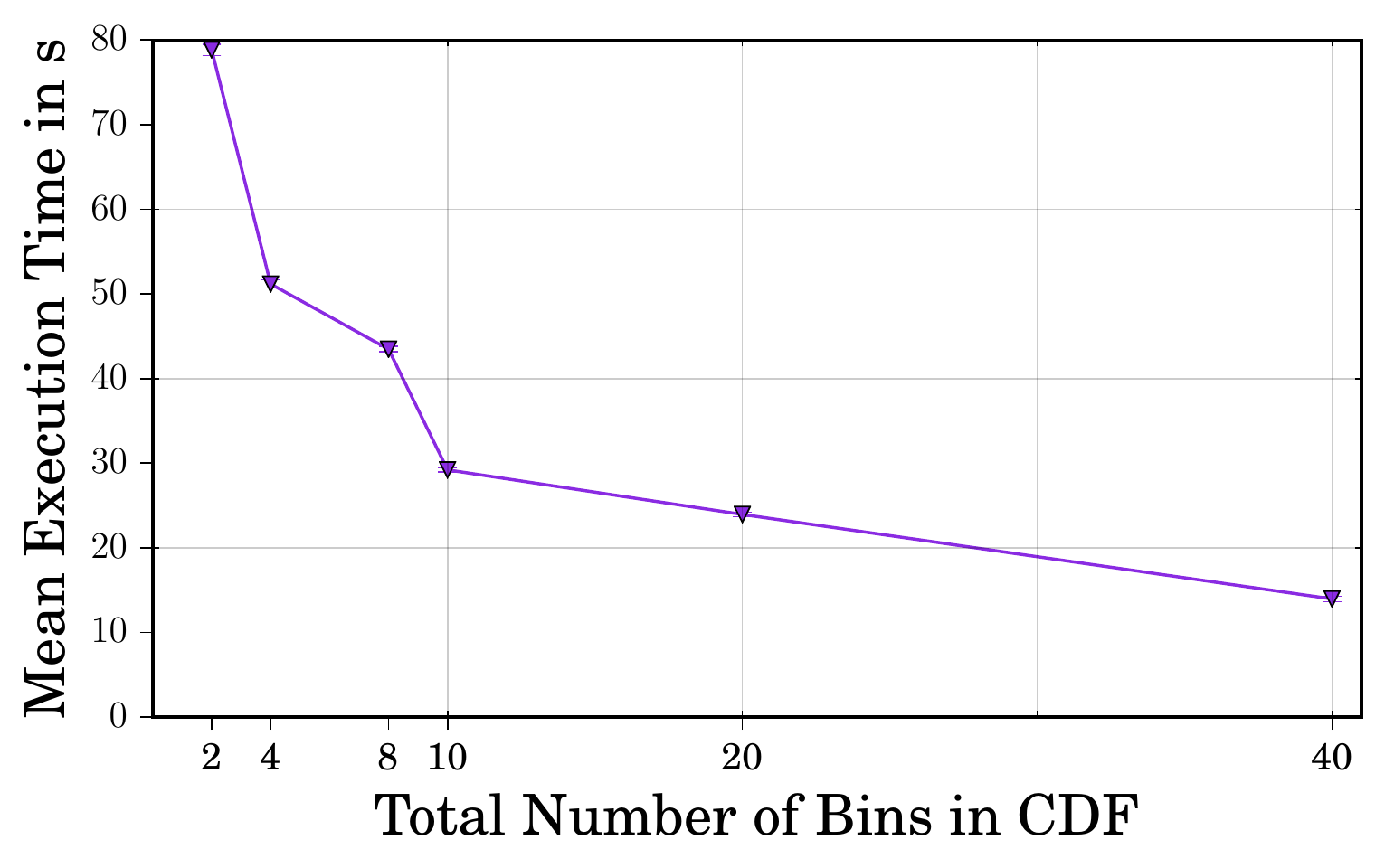}
        % \vspace{-1mm}
        \caption{}
        \label{fig:time:NumberOfBins}
        \end{subfigure}
             \vspace{-1mm}
       \eat{ \\  
        \begin{subfigure}[b]{0.45\linewidth}
    \centering \includegraphics[width=1\linewidth]{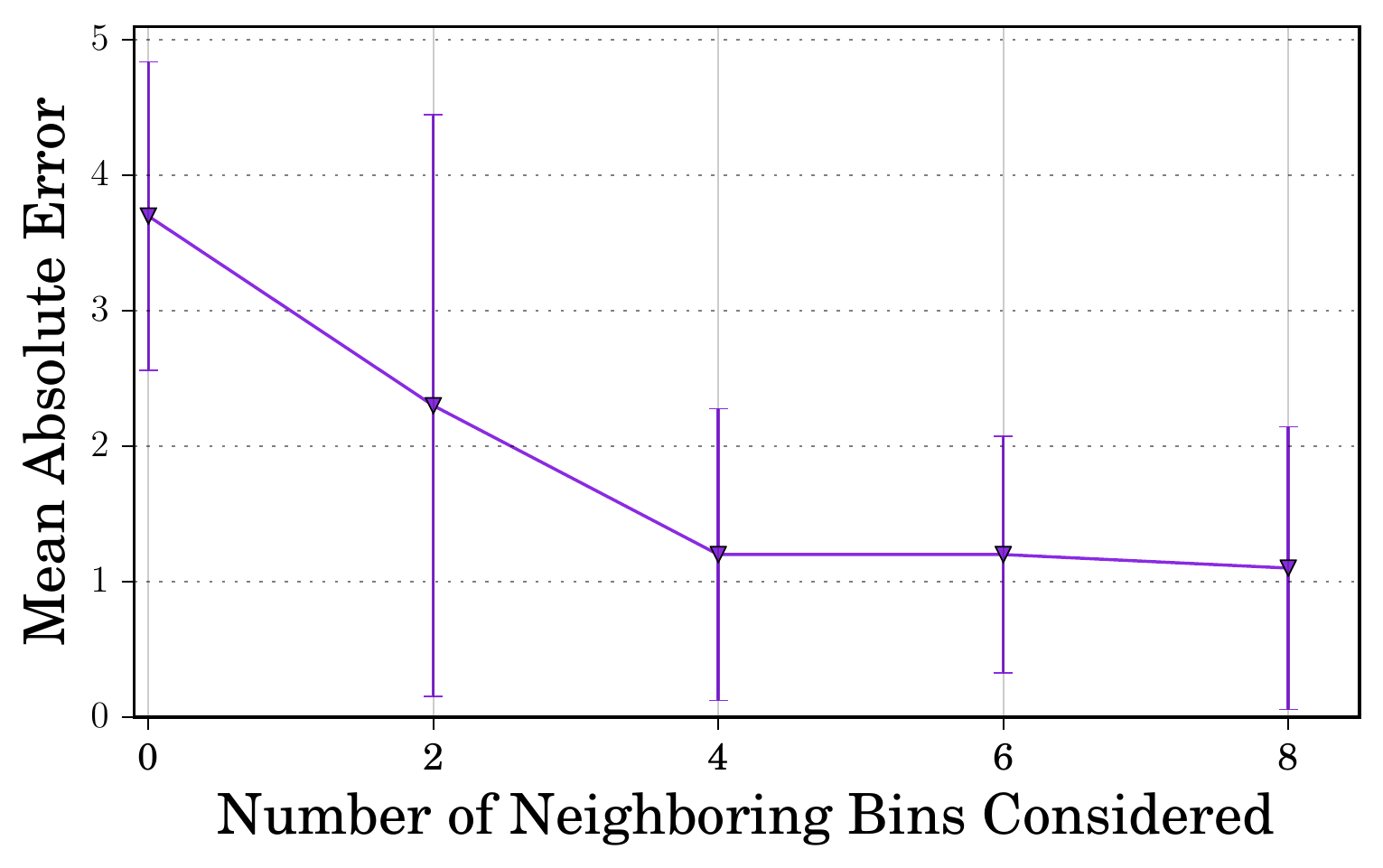}
         \vspace{-2mm}
        \caption{}
        \label{fig:error:NeighboringBins}\end{subfigure}
             \vspace{-1mm}
        \begin{subfigure}[b]{0.45\linewidth}
        \includegraphics[width=1\linewidth]{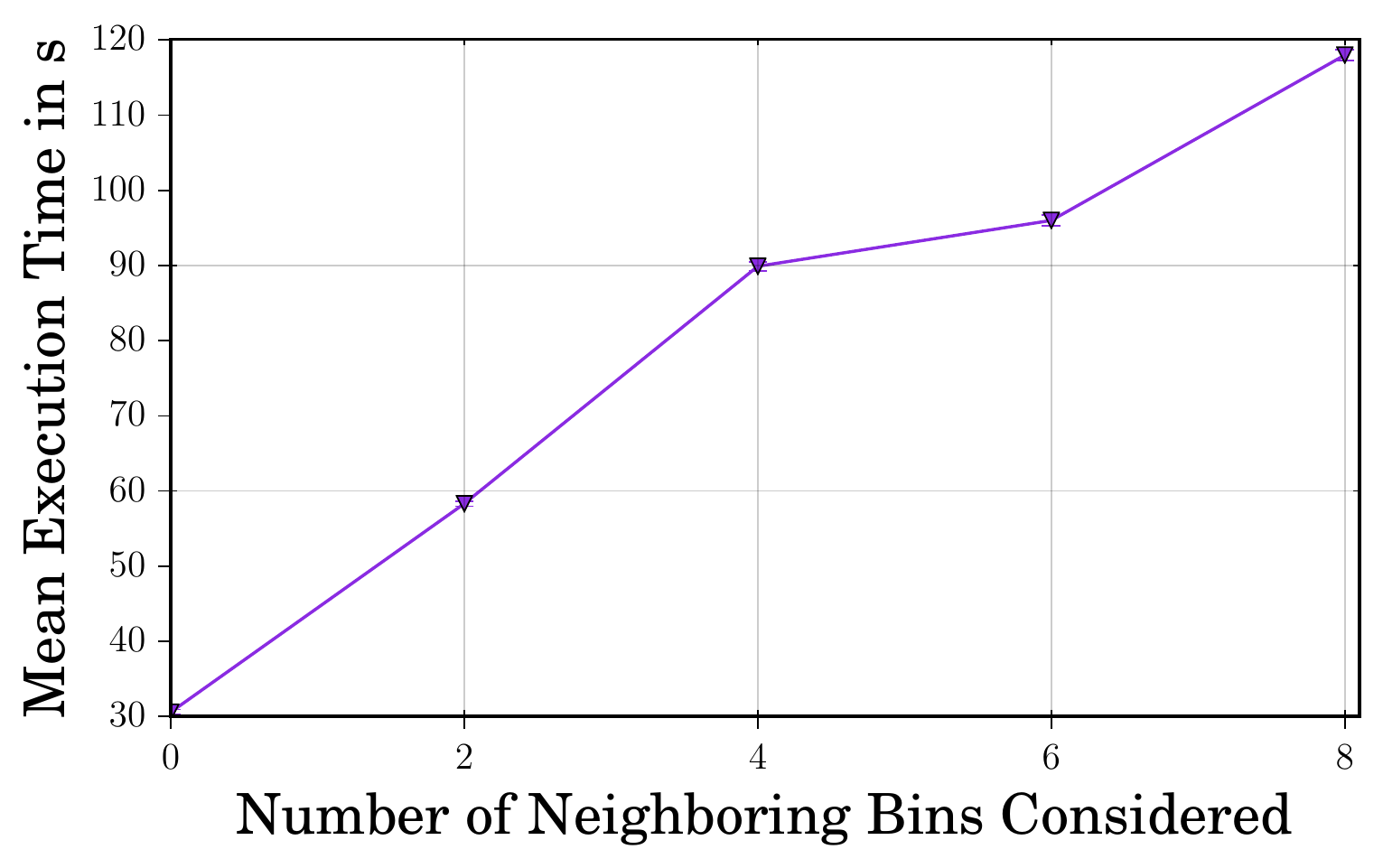}
         \vspace{-2mm}
        \caption{}
        \label{fig:time:NeighboringBins}
        \end{subfigure}}
        \vspace{-3mm}
        \caption{Accuracy and performance of P5 at different settings of the DP index optimization }\label{index}  \vspace{-0.8cm} 
    \end{figure}
    
Additionally, we study the dependency of the accuracy and execution time of P5 implemented with the DP index on three parameters -- (1) fraction of privacy budget $\rho$ used for the index (2) total number of domain partitions (bins) considered (3) number of neighboring bins considered. The default configuration for \system presented in this section uses $\epsilon=2.2$, $\rho=0.2$, total 10 bins and considers no extra neighboring bin. 

%Let $\rho$ represent the fraction of the privacy parameter used towards constructing the DP index. 
In Figure \ref{fig:error:} and \ref{fig:time:} we study how the mean error and execution time of the final result varies with $\rho$ for P5.  From Figure \ref{fig:error:}, we observe that the mean error drops sharply from $\rho=0.1$ to $\rho=0.2$, stabilises till $\rho=0.5$, and starts increasing again. This is because, at $\rho=0.2$, the index correctly identifies almost all the records satisfying the $\textsf{Filter}$ condition. However, as we keep increasing $\rho$, the privacy budget left for the program after $\textsf{Filter}$ (\textsf{Laplace} operator) keeps decreasing resulting in higher error in the final answer. From Figure \ref{fig:time:}, we observe that the execution time increases till $\rho=0.5$ and then stabilizes; the reason is that the number of rows returned after $\rho=0.5$ does not differ by much. %Thus from the two figures we observe that for $\rho=0.2$ we get a high speed up ($40\times$) with relatively low error ($3$). Hence we choose $\rho=0.2$ for our experiments. %A formal accuracy vs speed up trade-off analysis would be very helpful in this regard and is part of our future plan.

%Recall from our discussion in Section \ref{sec:dp_optimization}, for constructing the DP index, first we compute noisy prefix counts on the indexing attribute by partitioning its domain into $k$ bins. 
We plot the mean error and execution time for P5 by varying the total number of bins from $2$ to $40$ (domain size of \textit{NativeCountry} is 40) in Figure \ref{fig:error:NumberOfBins} and \ref{fig:time:NumberOfBins} respectively. %No extra neighboring bins %(i.e., $i_{s}=\hat{C}[s-1]$ and  $i_{e}=\hat{C}[e]$ from Section \ref{sec:dp_optimization}] are considered for the above two experiments. 
From Figure \ref{fig:error:NumberOfBins}, we observe that the error of P5 increases as the number of bins increase. It is so because %as $k$ increases, individual bin size decreases thereby increasing the risk of missing some records that satisfy the requisite \textsf{Filter} condition.  
from the computation of the prefix counts (Section \ref{sec:dp_optimization}), the amount of noise added increases with $k$ (as noise is drawn from $Lap(\frac{k}{\epsilon}))$. Figure \ref{fig:time:NumberOfBins} shows that the execution time decreases with $k$. This is intuitive because increase in $k$ results in smaller bins, hence the number of rows included in $[i_{s}, i_{e}]$ decreases.

To avoid missing relevant rows, more bins that are adjacent to the chosen range $[i_s,i_e]$ can be considered for the subsequent operators. Thus, as the number of bins considered increases, the resulting error decreases at the cost of higher execution time. The experimental results are presented in Figure \ref{fig:error:NeighboringBins} and Figure \ref{fig:time:NeighboringBins} in \ifpaper  full paper \cite{anom}% using paper 
\else 
Appendix \ref{app:evaluation}\fi.%We increase the number of neighbouring bins from $0$ to $8$. As shown in Figure~\ref{fig:error:NeighboringBins}, the error decreases and all the relevant rows are included when $4$ neighbouring bins are considered. However, the execution time naturally increases with extra neighbouring bins as shown in Figure~\ref{fig:time:NeighboringBins}.%For increased accuracy,  bins preceding $s-1$ and following $e$ can be considered for $i_{s}$ and $i_{e}$ respectively. This is showcased in Figure \ref{fig:error:NeighboringBins} with total 10 bins, $\epsilon=1.1$ and $\rho=0.2$. We observe that the error stabilizes after considering 4 extra neighboring bins. It is so because all the records satisfying the \textsf{Filter} condition that were missing in $i_{s}=\hat{C}[s-1]$ and  $i_{e}=\hat{C}[e]$, are captured with just 4 extra neighboring bins. However, as shown in Figure \ref{fig:time:NeighboringBins}, the execution time naturally increases with extra neighboring bins.

\stitle{DP Range Tree}:
For P1, we see from Table \ref{perf} that the total execution time of the unoptimized Crypt$\epsilon$ implementation is about half an hour. However, using the range tree optimization reduces the execution time by $5667\times$. The reason behind this huge speed-up is that the time required by the \textsf{AS} in the optimized implementation becomes almost negligible because it simply needs to do a memory fetch to read off the answer from the pre-computed range tree.

\stitle{Pre-computation}: For P3, the unoptimized execution time on the dataset of $32561$ records is around 4 hours (Table \ref{perf}). This is so because the $\textsf{CrossProduct}$ operator has to perform $10\cdot 32561$ $labMult$ operations which is very time consuming. Hence, pre-computing the one-hot-codings for 2-D attribute over $Race$ and $Gender$ is very useful; the execution time reduces to less than a minute giving us a $982.2\times$ speed up.  %The un-optimized implementation for  Program D takes . It is because of the \textsf{CrossProduct} Primitiev because ot needs to compute . Pre-computaion f thsi saves a lot of time and cuts down the executioj tiem by.

\stitle{Offline Processing}:
The most costly operator for P7 is the \textsf{GroupByCountEncoded} operator since the \textsf{CSP} has to generate $\approx 3300K$ ciphertexts of $0$ and $1$ for the encrypted one-hot-codings. This results in a  total execution time of about 8.5 hours in unoptimized \system. However, by generating the ciphertexts off-line, the execution time can be reduced to just $5$ minutes giving us a speed up of $102.49\times$.

Another important observation from Table \ref{perf} is that the \textsf{AS} performs the major chunk of the work for most program executions. This conforms with our discussion in Section~\ref{sec:design}.\vspace{-0.4cm}
\begin{figure}[ht]
      \includegraphics[width=0.5\linewidth]{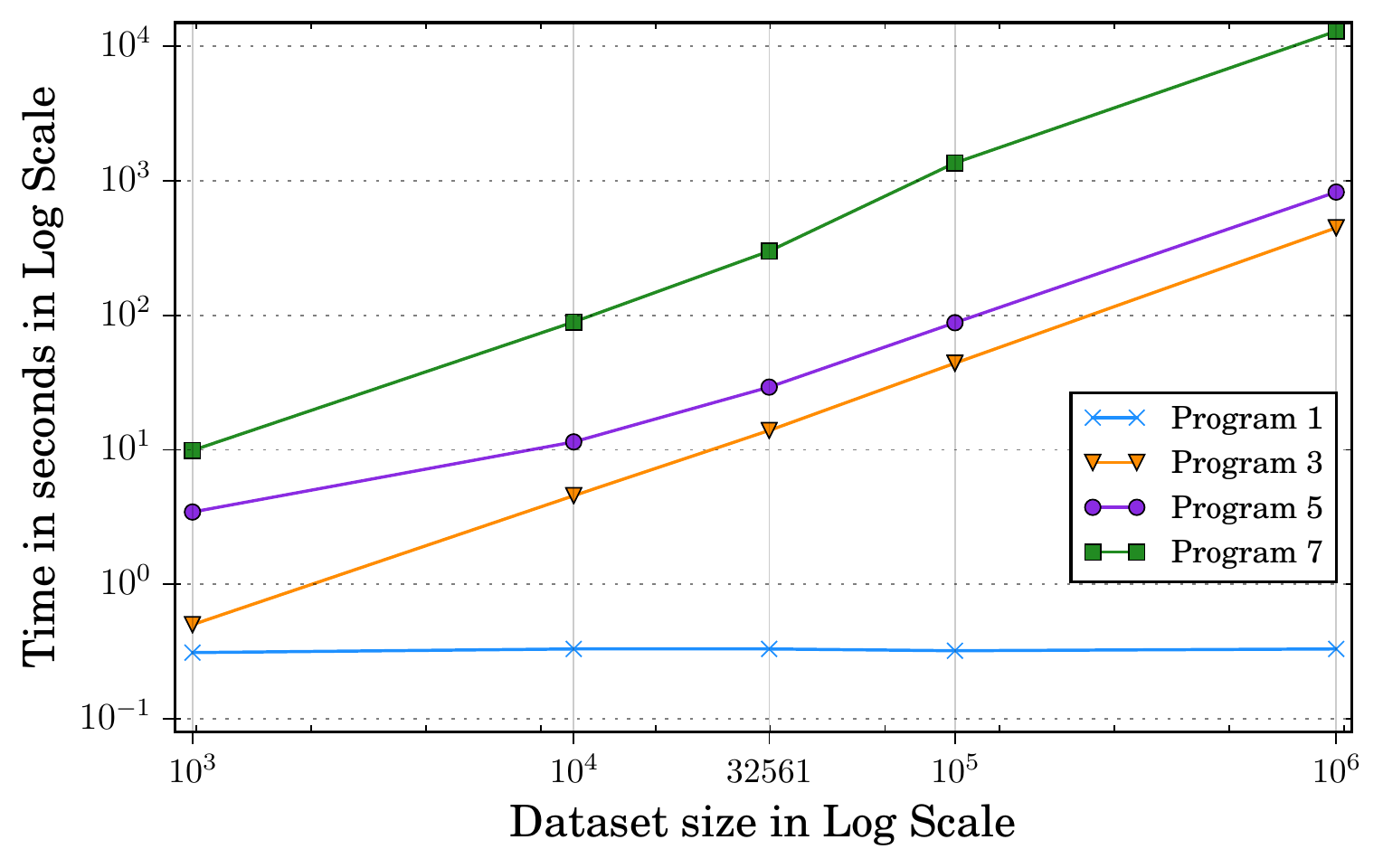}\vspace{-0.3cm}
        \caption{Scalability of \system Programs} \label{fig:scale}%\vspace{-0.5cm}
    \end{figure}
    \vspace{-0.7cm}
    \subsection{Scalability}
%In this section we evaluateFigure \ref{scale} plots the execution times of the aforementioned four \system programs with varying dataset size. 

In this section, we evaluate \textbf{Q3} by observing the execution times of the aforementioned four \system programs for dataset sizes up to 1 million. As seen from Figure \ref{fig:scale}, the longest execution time (P7) for a dataset of 1 million records is $\approx 3.6$ hours; this shows the practical scalability of \system.
All the reported execution times are for default setting. For P1 we see that the the execution time does not change with the dataset size. This is so because once the range tree is constructed, the program execution just involves reading the answer directly from the tree followed by a decryption by the \textsf{CSP}. The execution time for the P3 and P7 is dominated by the $\oplus$ operation for the \textsf{GroupByCount} operator. The cost of $\oplus$ is linear to the data size. Hence, the execution time for P3 and P7 increases linearly with the data size. For P5, the execution time depends on the \% of the records in the dataset that satisfy the condition $NativeCountry=Mexico$ (roughly this many rows are retrieved from the noisy index). 

\vspace{-0.2cm}
\section{Extension of \system to the Malicious Model}\label{sec:malicious}
In this section, we briefly discuss how to extend the current \system system to account for malicious adversaries. We present one approach for the extension here and detail another approach in \ifpaper the full paper \cite{anom}. \else  Appendix \ref{app:sec:malicious}.\fi The first approach implements the \CSP inside a trusted execution environment (TEE) \cite{Boneh2,Prochlo,Aïmeur2008}. This ensures non-collusion (as the \CSP cannot collude with the \AS since its operations are vetted). The measurement operators are implemented as follows (the privacy budget over-expenditure checking remains unchanged from that in Section \ref{sec:operator_implementation} and we skip re-describing it here).
\\
\textbf{\textsf{Laplace}} $\lap_{\epsilon,\Delta}(\encV\textbackslash\encC)$: The new implementation requires only a single instance of noise addition by the \CSP. The \AS sends the ciphertext $\encC$ to the \CSP. The \CSP decrypts the ciphertext, adds a copy of noise, $\eta \sim  Lap(\frac{2\cdot\Delta}{\epsilon})$, and sends it to the \AS. %'The \CSP computes $\tilde{c}=labDec_{sk}(\mathbf{c})+\eta, \eta \sim  Lap(\frac{2\cdot\Delta}{\epsilon})$ and sends it over to the \AS.
\\
\textbf{\textsf{NoisyMax}} $\noisymax^k_{\epsilon,\Delta}(\mathbf{V})$: The new implementation works without the garbled circuit as follows. The \AS sends the vector of ciphertexts, $\encV$, to the \CSP.  The \CSP computes $\tilde{V}[i]=labDecrypt_{sk}(\encV[i]) + \eta[i], i \in [|\encV|]$, where $\eta[i]\sim Lap(2k\Delta/\epsilon)$ and  outputs the indices of the top $k$ values of $\tilde{V}$. \\
\textbf{Malicious AS}: Recall that a \system program, $P$, consists of a series of transformation operators that transform the encrypted database, $\encD$, to a ciphertext, $\encC$ (or an encrypted vector, $\encV$). This is followed by applying a measurement operator on $\encC$ (or $\encV$). %Additionally, as shown in the above discussion, in the very first step of the measurement operators the \AS adds a mask to $\encC$ and sends the masked ciphertext $\hat{\encC} = \encC + M$ to the \CSP. For a given program $P$, 
Let $P_1$ represent the first part of the program $P$  up to the computation of $\encC$ and let $P_2$ represent the subsequent measurement operator (performed by the \CSP inside a TEE).  In the malicious model, the \AS is motivated to misbehave. For example, instead of submitting the correct cipher $\encC = P_1(\encD)$ the \AS could run a different program $P'$ on the record of a single data owner only. %The resulting output could potentially leak or even completely reveal that data owner's record. 
Such malicious behaviour can be prevented by having the \CSP validate the \AS's work via zero knowledge proofs (ZKP) \cite{Oded} as follows (similar proof structure as prior work~\cite{Boneh1}). Specifically, the ZKP statement should prove that the \AS 1) runs the correct program $P_1$ 2) on the correct dataset $\encD$.  For this, the \CSP shares a random one-time MAC key, $mk_{i}, i \in [m]$ with each of the data owners, $DO_i$. Along with the encrypted record $\mathbf{\tilde{D}}_i$, $DO_i$ sends a Pedersen commitment \cite{Pedersen} $Com_i$ to the one-time MAC \cite{CryptoBook} on $\mathbf{\tilde{D}}_i$ and a short ZKP that the opening of this commitment is a valid one-time MAC on $\mathbf{\tilde{D}}_i$. The \AS collects all the ciphertexts and proofs from the data owners and computes $\encC=P_1(\mathbf{\tilde{D}}_1,\cdots,\mathbf{\tilde{D}}_m)$. Additionally, it constructs a ZKP that $\encC$ is indeed the output of executing $P_1$ on $\encD=\{\mathbf{\tilde{D}}_1,\cdots,\mathbf{\tilde{D}}_m\}$ \cite{ZKP1}. Formally, the proof statement is \vspace{-1mm}\begin{gather}\encC=P_1(\mathbf{\tilde{D}}_1,\cdots,\mathbf{\tilde{D}}_m) \wedge \forall i \mbox{ Open}(Com_i)=MAC_{mk_i}(\mathbf{\tilde{D}}_i) \label{ZKP}\end{gather}  The \AS submits the ciphertext $\encC$ along with all the commitments and proofs to the \CSP. By validating the proofs, the \CSP can guarantee  $\encC$ is indeed the desired ciphertext. The one-time MACs ensure that the \AS did not modify or drop any of the records received from the data owners. %These protocols can be made non- interactive using the standard Fiat-Shamir heuristic in the random oracle model \cite{FiatShamir}. \\
\\
\textbf{Efficient proof construction}: Our setting suits that of designated verifier non-interactive zero knowledge (DV NIZK) proofs~\cite{DVNIZK}. In a DV NIZK setting, the proofs can be verified by a single designated entity (as opposed to publicly verifiable proofs \cite{GrothSahai}) who possesses some secret key for the NIZK system. Thus in \system, clearly the \CSP can assume the role of the designated verifier. The framework for efficient DV NIZKs proposed by Chaidos and Couteau~\cite{DVNIZK} can be applied to prove Eq.~\eqref{ZKP}, as this framework enables proving arbitrary relations between cryptographic primitives, such as Pedersen commitment or Paillier encryption. % The authors in \cite{DVNIZK} present a framework for efficient DV NIZKs for a group-dependent language $\mathcal{L}$ where the abelian group $\mathcal{L}$ is initiated on is of order $N$ and $\mathbb{Z}_N$ is the plaintext-space of an homomorphic cryptosystem. Cryptographic primitives like Pedersen commitments and Paillier encryptions are defined over abelian groups. Thus, this framework enables proving arbitrary relations between primitives like Pedersen commitment or Paillier encryption via DV NIZK proofs, in an efficient way.  Since we use Paillier encryption \cite{Paillier} for our prototype \system , this means that eq \eqref{ZKP} can be proven in this framework.  
A detailed construction is given in \ifpaper the full paper \cite{anom} \else Appendix \ref{app:sec:malicious}\fi which shows that all the steps of the proof involve simple arithmetic operations modulo $N^2$ where $N$ is an RSA modulus. To get an idea of the execution overhead for the ZKPs, consider constructing a DV NIZK for proving that a Paillier ciphertext encrypts the products of the plaintexts of two other ciphertexts (this could be useful for proving the validity of our \textsf{Filter} operator, for example). In \cite{DVNIZK}, this involves $4logN$ bits of communication and the operations involve addition and multiplication of group elements. Each such operation takes order of $10^{-5}$ seconds to execute, hence for proving the above statement for $1$ million ciphertexts will take only a few tens of seconds. %\am{This is good!}
 %Note that for transformation operators (\textsf{CrossProduct}, \textsf{GroupByCount}, \textsf{CountDistinct}) need intermediate communications with the \CSP. Thus for programs using them, $P_1$ can be broken into $k$ parts where $P_{11}$ represents the program upto the first communication with the \CSP, $P_{12}$ represents the sub-program starting from obtaining the response from the \CSP for $P_{11}$ up till the second communication to the $\CSP$ and finally $P_{1k}$ corresponds to the sub-program that ends at sending $\encC$ to the \CSP. There has to be $k$ separate ZKP proofs to the effect $$. 
 \\ \textbf{Malicious \CSP}: Recall that our extension implements the \CSP inside a TEE. Hence, this ensures that the validity of each of \CSP's actions in the TEE can be attested to by the data owners. Since the measurement operators ($P_2$) are changed to be implemented completely inside the \CSP, this guarantees the bounded $\epsilon$-DP guarantee of \system programs even under the malicious model. Additionally sending the \CSP the true ciphers $\encC=P_1(\encD)$ also does not cause any privacy violation as it is decrypted inside the TEE. 
 \\ \textbf{Validity of the data owner's records}: The validity of the one-hot-coding of the data records, $\mathbf{\tilde{D}}_i$, submitted by the data owners $DO_i$ can be checked as follows. Let $ \tilde{\mathbf{D}}_{ij}$ represent the encrypted value for attribute $A_j$ in one-hot-coding for $DO_i$. The \AS selects a set of random numbers $R=\{r_k~|~ k \in [|domain(A_j)|] \}$ and computes the set $ P_{ij}=\{labMult(\tilde{\mathbf{D}}_{ij}[k], \\ labEnc_{pk}(r_k))\}$.
Then it sends sets $P_{ij}$ and $R$ to the \CSP who validates the record only if $|P_{ij} \cap R|=1 \thinspace \forall A_j$. Note that since the \CSP does not have access to the index information of $P_{ij}$ and $R$ (since they are sets), it cannot learn the value of $D_{ij}$. Alternatively each data owner can provide a zero knowledge proof for $\forall j,k, \thinspace\thinspace D_{ij}[k] \in \{0,1\} \wedge \sum_k D_{ij}[k] = 1$.

\vspace{-3mm}
\section{Related Work}\label{sec:related-short}
\stitle{Differential Privacy}: Introduced by Dwork et al. in \cite{Dork}, differential privacy has enjoyed immense attention from both academia and industry in the last decade. Interesting work has been done in both the \textsf{CDP} model  \cite{MVG,Blocki,AHP,DAWA,hist1,hist2,hist3,hist4,hist6,hist7,hist8,A1,A2,A3,A4,A5,A6,A7,DPSVTProof,u1,MWEM,DPBench} and the \textsf{LDP} model \cite{HH2, Rappor1,HH,Rappor2,Cormode, CALM,15,itemset,ldp}.
Recently, it has been showed that augmenting the \textsf{LDP} setting by a layer of anonymity  improves the privacy
guarantees \cite{mixnets,Prochlo,amplification}. It is important to note that the power of this new model (known as shuffler/mixnet model) lies strictly between that of \textsf{LDP} and \textsf{CDP}. Crypt$\epsilon$ differs from this line of work in three ways, namely expressibility, precise DP guarantee and trust assumptions (details are in the \ifpaper 
full paper \cite{anom}% using paper 
\else 
 Appendix \ref{app:related}
\fi).  %Firstly, Crypt$\epsilon$ has the same expressibility as that of the \textsf{CDP} model. Secondly, the shuffler/mixnet model gives an approximate DP guarantee $(\epsilon\sqrt{\frac{\log\frac{1}{\delta}}{n}},\delta)$ which incurs an expected error of $O(\epsilon\sqrt{\log\frac{1}{\delta}})$.  In practice, $\delta$ has to be smaller than $\frac{1}{n}$ in order to get meaningful privacy. In contrast Crypt$\epsilon$ achieves the same order of accuracy guarantees as that of \textsf{CDP}. Finally, the shuffler/mixnet model and \system have certain differences in their respective trust assumptions (details in \ifpaper full paper \cite{anom}% using paper \else Appendix \ref{app:sepldp}\fi). %Google's implementation relies on a trusted intermediary shuffler which they implement via trusted hardware enclaves. However truly secure hardware enclaves are notoriously difficult to achieve in practice \cite{Foreshadow}. The mixnet model on the other hand requires a  mix network or mixnet which is a protocol involving several computers that inputs a sequenceof encrypted messages, and outputs a uniformly random permutation of those messages' plaintexts.  Their trust assumption is that at least one of the servers needs to behave honestly. For Crypt$\epsilon$ both the servers are completely untrusted under the constraint that they are non-colluding and follow the protocols semi-honestly.
\\\stitle{Two-Server Model}: The two-server model is popularly used for privacy preserving machine learning approaches where one of the servers manages the cryptographic primitives while the other handles computation~ \cite{Boneh1,Boneh2,Ridge2,Matrix2,secureML,LReg}. \\\stitle{Homomorphic Encryption}: Recently, there has been a surge in  privacy preserving solutions using homomorphic encryptions due to improved primitives. A lot of the aforementioned two-server models employ homomorphic encryption \cite{Boneh1,Boneh2,LReg,Matrix2}.  Additionally, it is used in \cite{CryptoDL,CryptoNet,NN, Irene2, grid}.

%A detailed discussion on related work is presented in the full paper Appendix \ref{app:related} \cite{anom}.

\vspace{-1em}
\section{Conclusions}\label{sec:conclusions}
%There are a number of future work directions for Crypt$\epsilon$. 
In this paper, we have proposed a system and programming framework, \system, for differential privacy that achieves the constant accuracy guarantee and algorithmic expressibility of \textsf{CDP} without any trusted server. This is achieved via two non-colluding servers with the assistance of cryptographic primitives, specifically \textsf{LHE} and garbled circuits. Our proposed system \system can execute a rich class of programs that can run efficiently by virtue of four optimizations.
\par  %One possible future work can be the development of aCrypt$\epsilon$ compiler.
% \system opens several important and interesting research directions. 
Recall that currently the data analyst spells out the explicit Crypt$\epsilon$ program  to the \textsf{AS}. Thus, an interesting future work is constructing a compiler for \system that inputs a user specified query in a high-level-language. The
compiler should next formalize a Crypt$\epsilon$ program expressed in terms of Crypt$\epsilon$ operators with automated sensitivity analysis. 
Another direction is to support a larger class of programs in \system. For example, inclusion of aggregation operators such as sum, median, average is easily achievable. Support for multi-table queries like joins would require protocols for computing sensitivity \cite{elastic} and  data truncation \cite{Kotsogiannis:2019}. 

\eat{can be achieved based on existing works like \emph{elastic sensitivity} \cite{elastic}. Another direction is enabling learning algorithms on \system; linear regression can be based on \cite{LReg} which also uses \textsf{LHE} and a two-server model. %For this, we need to extend \system with a new primitive for matrix multiplications. 
For more involved models like deep learning, DP techniques of \cite{DLDP} could be combined with the homomorphic encryption techniques of CryptoNet \cite{CryptoNet}. %As mentioned in section 3.6, an alternative implementation for \system  can be based on secret shares modulo the assumption that both the servers are benefit from learning the differential privacy output. Hence another useful extension might be re-implementing \system with  secret shares. For this, the functionality of the existing primitives would mostly be the same, only the respective implementations will change. 
%Yet another extension can be removing the second server (\textsf{CSP}) altogether and instead capturing its functionalities within a trusted execution environment (TEE) as mentioned in Section 7.2.
}

\stitle{Acknowledgement}: This work was supported by NSF under grants 1253327, 1408982; and by DARPA and SPAWAR under contract N66001-15-C-4067.
\bibliographystyle{abbrv}
\balance
\bibliography{references.bib}
\newpage\appendix
\section{Background Cntd.} \label{app:background}
The \emph{stability} of a transformation operation is defined as
\begin{definition}\label{def:stability}
A transformation $\mathcal{T}$ is defined to be $t$-stable if for two datasets $D$ and $D'$, we have\begin{gather}|\mathcal{T}(D)\ominus \mathcal{T}(D')| \leq t \cdot |D\ominus D'|  \end{gather} where  (i.e.,  $D \ominus D' = (D-D') \cup (D'-D)$. \end{definition}
Transformations with bounded stability scale the DP guarantee of their outputs, by their stability constant \cite{PINQ}.
\begin{theorem}\label{theorem:stability}
If $\mathcal{T}$ is an arbitrary $t$-stable transformation on dataset $D$ and $\mathcal{A}$ is an $\epsilon$-DP algorithm  which takes output of $\mathcal{T}$ as input, the composite computation $\mathcal{A} \circ \mathcal{T}$ provides $(\epsilon \cdot t)$-DP.\end{theorem}
%\textcolor{blue}{\begin{definition}(Unbounded Differential Privacy). A randomized algorithm $\mathcal{A}$ satisfies unbounded $\epsilon$- differential privacy if for any two databases $D$ and $D'$ such that $D'$ can be obtained from $D$ by either removing or adding a single record 
%{\begin{gather} \forall S \subset Range(\mathcal{A}), Pr \big[\mathcal{A}(D) \in S\big] \leq e^{\epsilon}Pr\big[\mathcal{A}(D') \in S\big]\end{gather}} where $Range(\mathcal{A})$ denotes the set of all possible outputs of $\mathcal{A}$.\label{def:unboundedDP}\end{definition}
%\begin{theorem}If a mechanism $\mathcal{A}$ satisfies unbounded $\epsilon$-DP, then $\mathcal{A}$ also satisfies bounded $2\epsilon$-DP \cite{bookDP}.\label{thm:boundedDP}  \end{theorem}}
\stitle{Labeled Homomorphic Encryption(\textsf{labHE}).}
Let $(Gen,$\\
$Enc,Dec)$ be an \textsf{LHE} scheme with security parameter $\kappa$ and message space $\mathcal{M}$. Assume that a multiplication operation exists in $\mathcal{M}$, i.e., is a finite ring. Let $\mathcal{F}:\{0,1\}^s \times \mathcal{L}\rightarrow \mathcal{M}$ be a pseudo-random function with seed space $\{0,1\}^s$( s= poly($\kappa $)) and the label space $\mathcal{L}$. A \textsf{labHE} scheme is defined as
\squishlist
 \item $\textbf{labGen}(\kappa):$ Runs $Gen(\kappa)$ and outputs $(sk,pk)$.
\item $\textbf{localGen}(pk):$ For each user $i$ and with the public key as input, it samples a random seed $\sigma_i \in \{0,1\}^s$ and computes $pk_i = Enc_{pk}(\underline{\sigma_i})$ where $\underline{\sigma_i}$ is an  encoding of $\sigma_i$ as an  element of $\mathcal{M}$. It outputs $(\sigma_i,pk_i)$.
\item $\textbf{labEnc}_{pk}(\sigma_i, m , \tau):$ On input a message $m \in \mathcal{M} $ with label $\tau \in \mathcal{L}$  from user $i$, it computes $b=\mathcal{F}(\sigma_i, \tau)$ (mask) and outputs the labeled ciphertext $\mathbf{c}=(a,d) \in \mathcal{M} \times \mathcal{C}$ with $ a= m- b$  (hidden message) in $\mathcal{M}$ and $d=Enc_{pk}(b)$. For brevity we just use notation $\textbf{labEnc}_{pk}(m)$ to denote the above functionality, in the rest of paper. 
\item $\textbf{labDec}_{sk}(\mathbf{c}):$ This functions inputs a cipher $\mathbf{c}=(a,d) \in \mathcal{M} \times \mathcal{C}$  and decrypts it as $m=a-Dec_{sk}(d)$.\squishend
%\textsf{LHE} and \textsf{labHE} provides semantic security guarantee \cite{Katz}. In addition to the operations supported by an \textsf{LHE}  scheme, \textsf{labHE} supports multiplication of two \textsf{labHE} ciphers. 
\squishlist
\item $\textbf{labMult}(\mathbf{c}_1,\mathbf{c}_2)$ - On input two \textsf{labHE} ciphers $\mathbf{c}_1=(a_1,d_1)$ and $\mathbf{c}_2=(a_2,d_2)$, it computes a "multiplication" ciphertext  $\mathbf{e}=labMult(\mathbf{c_1,}$ $\mathbf{c_2})=Enc_{pk}(a_1,a_2)\oplus cMult(d_1,a_2) \oplus cMult(d_2,a_1)$. Observe that $Dec_{sk}(\mathbf{e})=m_1\cdot m_2 -b_1 \cdot b_2$.
\item $\textbf{labMultDec}_{sk}(d_1,d_2,\mathbf{e})$ - On input two encrypted masks $d_1,d_2$ of two \textsf{labHE} ciphers $\mathbf{c_1},\mathbf{c_2}$, this algorithm decryts the output $\mathbf{e}$ of $labMult(\mathbf{c_1},\mathbf{c_2})$ as $m_3=Dec_{sk}(\mathbf{e})+Dec_{sk}(d_1)\cdot Dec_{sk}(d_2)$ which is equals to $m_1\cdot m_2$.   
\squishend
\begin{table}[t]
\centering
\caption {Comparative analysis of different DP models}
\scalebox{0.7}{ \begin{tabular}{|c| c c c|}  \toprule
\multicolumn{1}{|c}{\textbf{Features}} & \textbf{LDP}  & \textbf{CDP}  & \textbf{Crypt$\epsilon$}  \\ [0.5ex]
 \hline \hline\# Centralized Servers & 1& 1 & 2\\\hline
Trust Assumption & & & Untrusted \\   for Centralized & Untrusted & Trusted & Semi Honest \\ Server &  &   &  Non-Colluding  \\ \hline
Data Storage & \multirow{2}{*}{N/A} & \multirow{2}{*}{Clear} & \multirow{2}{*}{Encrypted} \\in Server & &  &  \\\hline
\multirow{2}{*}{Adversary} & Information & Information & Computationally \\& Theoretic & Theoretic & Bounded\\\hline
 Error on Statistical Counting Query& $O\Big(\frac{\sqrt(n)}{\epsilon}\Big)$& $O\Big(\frac{1}{\epsilon}\Big)$ & $O\Big(\frac{1}{\epsilon}\Big)$\\
  [1ex]
 \bottomrule
 \end{tabular}}\label{DPCompare}
\end{table}
\section{Security Proof}\label{sec:proof}
In this section we present the formal proof for Theorem~\ref{thm:security}.
\begin{proof}
 We have {\it nine operators} in our paper (see Table~\ref{tab:operators}).
\begin{itemize}
\item \textsf{NoisyMax} and \textsf{CountDistinct} use
  ``standard'' garbled circuit construction and their security proof
  follows from the proof of these schemes.

\item All other operators except \textsf{Laplace} essentially use 
homomorphic properties of our encryption scheme and thus there 
security follows from semantic-security of these scheme.

\item The proof for the \textsf{Laplace} operator is given below.
\end{itemize}
The proof for an entire program $P$ (which is a composition 
of these operators) follows from the composition theorem~\cite[Section 7.3.1]{Oded}

We will prove the theorem for the \textsf{Laplace} operator.
In this case the views are as follows (the outputs of the
two parties can simply computed from the views):
\begin{eqnarray*}
View_1^{\Pi}(P,\mathcal{D},\epsilon) & = & (pk,\encD,\eta_1,P(\mathcal{D})+\eta_2+\eta_1) \\
View_2^{\Pi}(pk,sk,P,\mathcal{D},\epsilon) & = & (\eta_2,labEnc_{pk} (P(\mathcal{D})+\eta_1))
\end{eqnarray*}
The random variables $\eta_1$ and $\eta_2$ are random variables
generated according to the Laplace distribution $Lap(\frac{2\cdot\Delta}{\epsilon})$ where $\Delta$ is the program sensitivity (computed w.r.t Definition \ref{def:dp}). The simulators $Sim_1 (z^B_1)$ (where $z_1 = (y_1, |\mathcal{D}|)$ is the random
variable distributed according to $P^{CDP}_B(\mathcal{D},\epsilon))$, $y_1$ being the random variable distributed as $P^{CDP}(\mathcal{D},\epsilon/2)$) performs the
following steps:
\begin{itemize}
\item Generates a pair of keys $(pk_1,sk_1)$ for the encryption scheme
  and generates random data set $\mathcal{D}_1$ of the same size as $\mathcal{D}$ and
  encrypts it using $pk_1$ to get $\encD_1$.

\item Generates $\eta'_1$ according to the Laplace distribution $Lap(\frac{2\cdot\Delta}{\epsilon})$.
\end{itemize}
The output of $Sim_1(z_1)$ is $(\encD_1,\eta'_1, y_1+\eta'_1)$.
Recall that the view of the \textsf{AS} is
$(\encD,\eta_1,P(\mathcal{D})+\eta_2+\eta_1)$.  The computational
indistinguishability of $\encD_1$ and $\encD$ follows from the
semantic security of the encryption scheme. The tuple
$(\eta'_1,y_1+\eta'_1)$ has the same distribution as
$(\eta_1,P(\mathcal{D})+\eta_2+\eta_1)$ and hence the tuples are
computationally indistinguishable.  Therefore, $Sim_1 (z_1)$ is
computational indistinguishable from $View_1^{\Pi}(P,\mathcal{D},\epsilon)$.

The simulators $Sim_2 (z_2)$ (where $z_2=(y_2,|\mathcal{D}|)$ is the random
variable distributed according to $P^{CDP}_B(\mathcal{D},\epsilon))$, $y_2$ being the random variable distributed as $P^{CDP}(\mathcal{D},\epsilon/2)$) performs the following steps:
\begin{itemize}
\item Generates a pair of keys $(pk_2,sk_2)$ for our encryption scheme.

\item Generates $\eta'_2$ according to the Laplace distribution $Lap(\frac{2\cdot\Delta}{\epsilon})$.
\end{itemize}
The output of $Sim_2 (z_2)$ is $(\eta'_2,labEnc_{pk}(y_2)+\eta'_2)$.
By similar argument as before $Sim_2 (z_2)$ is computationally indistinguishable 
from $View_2^{\Pi}(P,\mathcal{D},\epsilon)$.
\end{proof}

\section{Extension To Malicious Model Cntd.}\label{app:sec:malicious}
\subsection{First Approach Cntd.}
\stitle{Efficient proof construction}: Here we will outline an efficient construction for the aforementioned proof. First note that our setting suits that of designated verifier non-interactive zero knowledge (DV NIZK) proofs. In a DV NIZK setting, the proofs can be verified by a single designated entity (as opposed to publicly verifiable proofs) who possesses some secret key for the NIZK system. Thus in \system, clearly the \CSP can assume the role of the designated verifier.  This relaxation of public verifiability leads to boast in efficiency for the proof system.  \par 
The authors in \cite{DVNIZK} present a framework for efficient DV NIZKs for a group-dependent language $\mathcal{L}$ where the abelian group $\mathcal{L}$ is initiated on is of order $N$ and $\mathbb{Z}_N$ is the plaintext-space of an homomorphic cryptosystem. In other words, this framework enables proving arbitrary relations between cryptographic primitives such as Pedersen commitments or Paillier encryptions via DV NIZK proofs, in an efficient way. 
In what follows, we show that the proof statement given by eq \eqref{ZKP} falls in the language $\mathcal{L}$ and consists of simple arithmetic computations. \par
 The construction of the proof works as follows. First the \AS creates linearly homomorphic commitments (we use Pederson commitments) $Com^e_i$ on the encrypted data records $\tilde{\mathbf{D}_i}$ and proves that $Com_c = P_1(Com^e_1,\ldots, Com^e_m)$ where $Open(Com_c)=\mathbf{c}$. This is possible because of the homomorphic property of the Pederson commitment scheme;  all the operations in $P_1$ can be applied to $\{Com^e_i\}$ instead. We use Paillier encryption scheme \cite{Paillier} for our prototype \system construction and hence base the rest of the discussion on it. Paillier ciphertexts are elements in the group $(\mathbb{Z}/N^2 \mathbb{Z})^*$ where $N$ is an RSA modulus.  Pedersen commitments to such values can be computed as $Com=g^xh^r \in \mathbb{F}^*_p, 0 \leq r < N^2, g,h \in \mathbb{F}^*_p, \mbox{ Order}(g)=\mbox{ Order}(h) = N^2, p \mbox{ is a prime such that } p = 1 \mbox{mod }N^2$. This allows us to prove arithmetic relations on committed values modulo $N^2$. Finally, the \AS just needs to show that $Com_i$ opens to a MAC of the opening of $Com^e_i$. For this, the MACs we use are built from linear hash functions $H(x) = ax + b$ \cite{Boneh1} where the MAC signing key is the pair of random values $(a, b) \in (\mathbb{Z}/N^2 \mathbb{Z})$. Proving to the \CSP that the opening of $Com_i$ is a valid MAC on the opening of $Com^e_i$ is a simple proof of arithmetic relations. Thus, quite evidently an efficient DV NIZK proof for eq \eqref{ZKP} can be supported by the framework in \cite{DVNIZK}. To get an idea of the execution overhead for the ZKPs, consider constructing a DV NIZK for proving that a Paillier ciphertext encrypts the products of the plaintexts of two other ciphertexts requires (this could be useful for proving the validity of our \textsf{Filter} operator). In the framework proposed in in \cite{DVNIZK}, this involves $4logN$ bits of communication and the operations involve addition and multiplication of group elements. Each such operation takes order of $10^{-5}$s execution time, hence for proving the above statement for $1$ mil ciphers will take only a few tens of seconds.   
 \subsection{Second Approach}
 %Note that for transformation operators (\textsf{CrossProduct}, \textsf{GroupByCount}, \textsf{CountDistinct}) need intermediate communications with the \CSP. Thus for programs using them, $P_1$ can be broken into $k$ parts where $P_{11}$ represents the program upto the first communication with the \CSP, $P_{12}$ represents the sub-program starting from obtaining the response from the \CSP for $P_{11}$ up till the second communication to the $\CSP$ and finally $P_{1k}$ corresponds to the sub-program that ends at sending $\encC$ to the \CSP. There has to be $k$ separate ZKP proofs to the effect $$. 
In this section, we describe the second approach to extend \system to account for a malicious adversary.
For this we propose the following changes to the implementation of the measurement operator. \\
\textbf{\textsf{Laplace }}$Lap_{\Delta,\epsilon}(\encC\textbackslash\encV)$: Instead of having  both the servers, \textsf{AS} and \textsf{CSP} add two separate instances of Laplace noise to the true answer,  single instance of the Laplace noise is jointly computed via a SMC protocol ~\cite{Djoin,DworkOurData} as follows. First the \AS adds a random mask $M$ to the encrypted input $\mathbf{c}$ to generate $\hat{\mathbf{c}}$ and sends it to the \CSP. Next the \CSP generates a garbled circuit that 1) inputs two random bit strings $S_1$ and $R_1$ from the \AS 2) inputs another pair of random strings $S_2$ and $R_2$ and a mask $M'$ from the \CSP 3) uses $S_1$ and $S_2$ to generate an instance of random noise, $\eta \sim Lap(\frac{2\cdot\Delta}{\epsilon})$ using the fundamental law of transformation of probabilities 4) uses $R_1 \oplus R_2$ as the randomness for generating a Pedersen commitment for $M', Com(M')$ 4) outputs $\tilde{\mathbf{c}}' = \hat{\mathbf{c}} + labEnc_{pk}(\eta) + labEnc_{pk}(M')$, $Com(
M')$, and $labEnc_{pk}(r)$ ($r$ is the randomness used for generating $Com(M')$). The \CSP sends this circuit to the \AS who evaluates the circuit and sends $\tilde{\mathbf{c}}'$, $Com(
M')$, and $labEnc_{pk}(r)$ back to the \CSP. Now, the \CSP decrypts $\tilde{\mathbf{c}}'$ and subtracts the mask $M'$ to return $\tilde{c}' = labDec_{sk}(\tilde{\mathbf{c}}') - M'$ to the \AS.  Finally the \AS can subtract out $M$ to compute the answer $\tilde{c}=\tilde{c}'-M$. Note that one can create an alternative circuit to the one given above which decrypts $\mathbf{\tilde{c}}$ inside the circuit. However, decrypting Pailler ciphertexts inside the garbled circuit is costly. The circuit design given above hence results in a simpler circuit at the cost of an extra round of communication.
\\
\stitle{\textsf{NoisyMax}} $\noisymax^k_{\epsilon,\Delta}(\cdot)$: The \AS sends a masked encrypted vector, $\hat{\encV}$ to the \CSP $\hat{\encV}[i]=\encV[i]+M[i], i \in [|\encV|]$. The \CSP generates a garbled circuit that
1) inputs the mask vector $M$, a vector of random strings $S_1$, a random number $r$ and its ciphertext $\mathbf{c_r}=labEnc_{pk}(r)$ from the \AS
2) inputs the secret key $sk$ and another vector of random strings $S_2$ the from the \CSP
3) checks if $ labDec_{sk}(\mathbf{c_r}) == r$, proceed to the next steps only if the check succeeds else return $-1$
4) uses $S_1$ and $S_2$ to generate a vector $\eta[i]\sim Lap(\frac{2\cdot k \cdot \Delta}{\epsilon})$
using the fundamental law of transformation of probabilities 
5) computes $V[i] = labDec_{sk}(\hat{\encV}[i])+ \eta[i] - M[i]$
6) finds the indices of the top $k$ highest values of $V$ and outputs them. The \CSP sends this circuit to the \AS who evaluates it to get the answer.
Note that here we are forced to decrypt Paillier ciphertexts inside the circuit because in order to ensure DP in the Noisy-Max algorithm, the noisy intermediate counts cannot be revealed.
\\\stitle{Malicious AS:}
Recall that a \system program $P$ consists of a series of transformation operators that transforms the encrypted database $\encD$ to a ciphertext $\encC$ (or a vector of ciphertexts $\encV$). This is followed by applying a measurement operator on $\encC$ ($\encV$). Additionally, as shown in the above discussion, in the very first step of the measurement operators the \AS adds a mask to $\encC$ and sends the masked ciphertext $\hat{\encC} = \encC + M$ to the \CSP. For a given program $P$, 
let $P_1$ represent first part of the program up till the computation of $\encC$ ($\encV$). The zero knowledge proof structure is very similar to the one discussed in Section 7.2 except for the following changes. Now the \CSP sends a one-time MAC key $k_{AS}$ to the \AS as well and the \AS sends the masked ciphertext $\hat{\encC}$(or $\hat{\encV})$, along with the commitments and zero knowledge proofs from the data owners and an additional commitment to the one-time MAC on the mask $M$, $Com_{AS}$ and a proof for the statement \begin{gather*}\encC=P_1(\boldsymbol{\tilde{\mathcal{D}}}_1,\cdots,\boldsymbol{\tilde{\mathcal{D}}}_m) \wedge \forall i \mbox{ Open}(Com_i)=MAC_{mk_i}(\boldsymbol{\tilde{\mathcal{D}}}_i)\\ \wedge \thinspace \hat{\encC}=\encC+labEnc_{pk}(M) \wedge \mbox{ Open}(Com_{AS})=MAC_{k_{AS}}(M) \end{gather*} The \CSP proceeds with the rest of the computation only if it can validate the above proof. As long as one of the bit strings (or vectors of bit strings) in $\{S_1,S_2\}$ (and $\{R_2,R_2\}$ in case of the \textsf{Laplace} operator) is generated truly at random (in this case the honest \CSP will generate truly random strings), the garbled circuits for the subsequent measurement operators will add the correct Laplace noise. Additionally, the mask $M'$ prevents the \AS from cheating in the last round of communication with the \CSP in the protocol. It is so because, if the \AS does not submit the correct ciphertext to the \CSP in the last round, it will get back garbage values (thereby thwarting any privacy leakage). Hence, this prevents a malicious AS from cheating during any \system program execution. Note that the construction of the ZKP is similar to the one discussed in Section 7.2 and can be done efficiently via the framework in \cite{DVNIZK}.
\\\stitle{Malicious \CSP}: As discussed in Section 3, the \CSP maintains a public ledger with the following information\\
(1) total privacy budget $\epsilon^B$ which is publicly known
\\(2) the privacy budget $\epsilon$ used up every time the \AS submits a ciphertext for decryption\\
Since the ledger is public, the \AS can verify whether the per program reported privacy budget is correct preventing any disparities in the privacy budget allocation.
\par Recall that the \CSP receives a masked cipher $\hat{\encC}$ from the \AS at the beginning of the measurement operators. The mask $M$ protects the value of $\encC$ from the \CSP. We discuss the setting of a malicious \CSP separately for the two measurement operators as follows. \\
\textbf{\textsf{Laplace}}: In case of the \textsf{Laplace} operator, a malicious \CSP can cheat by 1) the generated garbled circuit does not correspond to the correct functionality 2) reports back incorrect decryption results.
The correctness of the garbled circuit can be checked by standard mechanisms \cite{Wang_GC} where the \AS specifically checks  that a) the circuit functionality is correct b) the circuit uses the correct value for $\hat{\mathbf{c}}$. 
For the second case, the \CSP provides the \AS with a zero knowledge proof for the following statement\begin{gather*} Open(Com(M'),r) = labDec_{sk}(\tilde{\encC}') - \tilde{c}'\end{gather*}
\\\textbf{\textsf{NoisyMax}}: The garbled circuit for the \textsf{NoisyMax} operator is validated similarly by standard mechanisms \cite{Wang_GC} where the \AS  checks a) whether the circuit implements the correct functionality b) the correct value of $\hat{\encV}$ is used.  Note that the equality check of step (3) in the circuit validates if the \CSP has provided the correct secret key $sk$ thereby forcing it to decrypt the ciphertexts correctly. \\
Note that certain operators like \textsf{CrossProduct}, \textsf{GroupByCount*} and \textsf{CountDistinct} the involve interactions with the \CSP as well but their validity can also be proven by standard techniques similar to the ones discussed above. Specifically \textsf{CrossProduct} and \textsf{GroupByCount*} can use zero knowledge proof in the framework \cite{DVNIZK} while the garbled circuit in \textsf{CountDistinct}  can use \cite{Wang_GC}.

\section{Additional Implementation Details}\label{app:implement}
\subsection{\textbf{General $n$-way Multiplication for \textsf{labHE}}}\label{genlab_appen}
The $labMult()$ operator of a \textsf{labHE} scheme allows the multiplication of two ciphers.
However, it cannot be used directly for a $n$-way muplication where $n>2$.  It is so because the "multiplication" cipher $\mathbf{e}=labMult(\mathbf{c_1},\mathbf{c_2})$ does not have  a corresponding label, i.e., it is not in the correct \textsf{labHE} cipher representation. Hence we propose Algorithm~\ref{algo:genlabmult} to generate a label $\tau'$ and a seed $b'$ for every intermediary product of two multiplicands so that it we can do a generic $n$-way multiplication on the ciphers. Note that the mask $r$ protects the value of $(m_1\cdot m_2)$ from the \textsf{CSP} (Step 3) and $b'$ hides $(m_1\cdot m_2)$ from the \textsf{AS} (Step 6). 
For example, suppose we want to multiply the respective ciphers of  $4$ messages $\{m_1,m_2,m_3,m_4\} \in \mathcal{M}^4$ and obtain $\mathbf{e}=labEnc_{pk}(m_1\cdot m_2\cdot m_3 \cdot m_4)$. For this, the \textsf{AS} first generates $\mathbf{e_{1,2}}=labEnc_{pk}(m_1\cdot m_2)$ and $\mathbf{e_{3,4}}=labEnc_{pk}(m_3\cdot m_4)$ using Algorithm~\ref{algo:genlabmult}. Both operations can be done in parallel in just one interaction round between the \textsf{AS} and the \textsf{CSP}. In the next round,  the \textsf{AS} can again use Algorithm~\ref{algo:genlabmult} with inputs $\mathbf{e_{1,2}}$ and $\mathbf{e_{3,4}}$ to obtain the final answer $\mathbf{e}$. %consider a case of mu Now with the true \textsf{labHE} cipher $\mathbf{c}=(a',d')$ for the product the \textsf{AS} can compute further multiplications on it. 
Thus for a generic $n-way$ multiplication the order of multiplication can be, in fact, parallelized as  shown in Figure ~\ref{genlab-fig} to require a total of $\lceil \log n\rceil$ rounds of communication with the \textsf{CSP}.
\begin{algorithm}[b]
\caption{$genLabMult$ - generate label for $labMult$}\label{algo:genlabmult}
\small
\begin{algorithmic}[1]
\STATEx
\textbf{Input}: $\mathbf{c_1}=(a_1,d_1)=labEnc_{pk}(m_1)$ and $\mathbf{c_2}=labEnc_{pk}(m_2)$ 
\STATEx where $a_1= m_1-b_1, d_1=Enc_{pk}(b_1)$, $a_2= m_2-b_2, d_2=Enc_{pk}(b_2)$
\STATEx \textbf{Output}: $\mathbf{e}=labEnc_{pk}(m_1\cdot m_2)$
\STATEx \textbf{\textsf{AS}:} 
\STATE Computes $\textbf{e}'=labMult(\mathbf{c_1,c_2}) \oplus Enc_{pk}(r)$ where $r$ is a random mask 
\STATEx  //$e'$ corresponds to $m_1\cdot m_2-b_1\cdot b_2+r$
\STATE Sends $\mathbf{e'},d_1,d_2$ to \textsf{CSP}
\STATEx \textbf{\textsf{CSP}:}
\STATE Computes $e''= Dec_{sk}(\mathbf{e'}) + Dec_{sk}(d_1)\cdot Dec_{sk}(d_2)$
\STATEx //$e''$ corresponds to $m_1\cdot m_2 + r$ 
%\STATE Decrypts $\mathbf{e'}$, to get $Dec_{sk}(\mathbf{e}')=m_1\cdot m_2 -b_2\cdot b_1 + r$
%\STATE Computes $b_1 \cdot b_2$ from $d_1$ and $d_2$.
%\STATE Removes $b_1\cdot b_2$ from $e'$ to compute $e''=m_1\cdot m_2+r$
\STATE Picks a seed $\sigma'$ and label $\tau'$ and computes $b'=\mathcal{F}(\sigma',\tau')$ 
%\STATE Computes $\bar{a}=e''-b'=m_1\cdot m_2 +r -b'$,  and $d'=Enc_{pk}(b')$
\STATE Sends $\bar{e}=(\bar{a},d')$ to \textsf{AS}, where $\bar{a} = e''-b'$ and $d' = Enc_{pk}(b')$
\STATEx //$\bar{a}$ corresponds to $m_1\cdot m_2 + r-b'$.
\STATEx \textbf{\textsf{AS}:}
\STATE Computes true cipher $\mathbf{e}=(a',d')$ where $a'=\bar{a}-r$ %m_1\cdot m_2 - b'$
 \end{algorithmic}
\end{algorithm}

\subsection{Operator Implementation}\label{app:implement_operators}

\stitle{\textsf{CrossProduct}} $\crossproduct_{A_i,A_j\rightarrow A'}(\cdot)$: This operator replaces the two attributes $A_i$ and $A_j$ by a single attribute $A'$. Given the encrypted input table $\encT$, where all attributes are in one-hot-encoding and encrypted, the attributes of $\encT$ except $A_i$ and $A_j$ remain the same. For every row in $\encT$, we denote the encrypted one-hot-encoding for $A_i$ and $A_j$ by $\tilde{\bf{v}}_1$ and $\tilde{\bf{v}}_2$.  Let $s_1$ and $s_2$ be the domain sizes of $A_i$ and $A_j$ respectively. Then the new one-hot-encoding for $A'$, denoted by $\tilde{\bf{v}}$, has a length of $s=s_1\cdot s_2$. For $l\in \{0,1,\ldots, s-1\}$, we have $$\tilde{\bf{v}}[l] = labMult(\tilde{\bf{v}}_1[l/s_2], \tilde{\bf{v}}_2[l\%s_2]).$$
Only one bit in $\tilde{\bf{v}}$ for $A'$ will be encrypted 1 and the others will be encrypted 0s. When merging more than two attributes, \system can use the $genLabMult()$ described in Section~\ref{genlab} to speed up computation.

%Let $D_1$ and $D_2$  be the encrypted one-hot-coding corresponding to two  values $v_1$ and $v_2$ (integral representation) for attributes $A_1$ and $A_2$ respectively. The corresponding encrypted one-hot-encoding for the two-dimensional attribute $A_1\times A_2$ is given by  \begin{gather} D_{1\times 2}[(i-1)\cdot s_{A_2}+j] = labMult(D_1[i], D_2[j])\\ i \in [s_{A_1}], j \in [s_{A_2}]\end{gather} For this particular case, only $D_{1 \times 2}[(v_1-1)\cdot s_{A_2}+v_2]=Enc(1)$ while all other indices will equate to $Enc(0)$. Note that when computing the one-hot-encoding for a t-dimensional attribute $t > 2$,  for the actual implementation, instead of calling $t$ iterative instances of \textsf{CrossProduct}() we use the $genLabMult()$ operator of labeled homomorphic encryption to speed up the computation.

\stitle{\textsf{Project}} $\project_{\bar{A}}(\cdot)$: The implementation of this operator simply drops off all but the attributes in $\bar{A}$ from the input table $\encT$ and returns the truncated table $\encT'$.

\stitle{\textsf{Filter}} $\filter_{\phi}(\cdot)$: The predicate $\phi$ in this operator is a conjunction of range conditions over $\bar{A}$, defined as: for a row $r$ in input table $\encT$,
$\phi(r) = \bigwedge_{A_j\in \bar{A}} ~~(r.{A_j} \in V_{A_j}),$ where $r.A_j$ is the value of attribute $A_j$ in row $r$ and $V_{A_j} \subseteq \{0,1,\ldots,s_{A_j}\}$ (the indices for attribute values of $A_{j}$ with domain size $s_{A_j}$).

First, we will show how to evaluate whether a row $r$ satisfies $r.{A_j} \in V_{A_j}$. Let $\tilde{\bf{v}}_j$ be the encrypted one-hot-encoding of $A_j$, then the indicator function can be computed as
$$I_{r.{A_j}\in V_{A_j}}=\bigoplus_{l\in V_{A_j}}\tilde{\bf{v}}_j[l].$$
If the attribute of $A_j$ in $r$ has a value in $V_{A_j}$, then $I_{r.{A_j}\in V_{A_j}}$ equals $1$; otherwise, $0$.

Next, we can multiply all the indicators using $genLabMult()$ (Section~\ref{genlab}) to check whether all attributes in $A_j\in \bar{A}$ of $r$ satisfy the conditions in $\phi$. Let $\bar{A} = \{A_1,\ldots,A_m\}$, then $$\phi(r) = genLabMult(I_{A_1\in V_{A_1}},\ldots, I_{A_m\in V_{A_m}}).$$

Last, we update the bit of $r$ in $\encB$, i.e., $\encB'[i] = labMult(\encB[i], \phi(r))$, given $r$ is the $i$th row in the input table. This step zeros out some additional records which were found to be extraneous by some preceding filter conditions.

Note that when the \textsf{Filter} transformation is applied for the very first time in a \system program and the input predicate is conditioned on a single attribute $A \in V_A$, we can directly compute the new bit vector using $I_{r.A\in V_{A}}$, i.e., for the $i$th record $r$ in input table $\encT$, we have $\encB'[i] =\bigoplus_{l\in V_A} \tilde{\bf{v}}_j[l]$.  This avoids the unnecessary multiplication $labMult(\encB[i],\phi(r))$.

\stitle{\textsf{Count}} $\countagg(\cdot)$: To evaluate this operator on its input table $\encT$, \system simply  adds up the bits in the corresponding $\encB$, i.e., $\bigoplus_{i}^m \encB[i]$.

%The \textsf{Count} operator takes the associated bit vector $\mathbf{B}$ of its input table $\encT$  and simply adds up its entries to return  \begin{gather}\mathbf{c}=\bigoplus_{i=1}^m\mathbf{B}[i]\end{gather}%\item GroupBy*($\mathbf{V},sk$)- This operator is an extension of the previous GroupBy transformation.

\stitle{\textsf{GroupByCount}} $\groupbystar_{A}(\cdot)$: The implementation steps for \textsf{Project}, \textsf{Filter} and \textsf{Count} are reused here. First, \system projects the input table $\encT$ on attribute $A$, i.e. $\encT_1 = \project_A(\encT)$. Then, \system loops each possible value of $A$. For each value $v$, \system initializes a temporary $\encB_v=\encB$ and filters $\encT'$ on $A=v$ to get an updated $\encB'_v$. Last, \system counts the number of 1s in $\encB'_v$ and release the counts.

\eat{
\begin{enumerate}[label=\alph*)] \item $\mathbf{\tilde{T}}_1$=\textsf{Project}($\mathbf{\tilde{T}}$, $A$) \item $\mathbf{B}$ =  current indicator bit vector \item  for $i = 1:s_A $ \\Intialize bit vector to $\mathbf{B}$  \\$\phi_i= (A==v_{i,A}) $ \\$\hat{\mathbf{T_2}}$ = \textsf{Filter}($\mathbf{\tilde{T}}_1, \phi_i$)\\ $\mathbf{C}[i]$ = \textsf{Count}($\hat{\mathbf{T_2}}$) \\ end for \item Output $\mathbf{C}$ 
\end{enumerate}
}

\stitle{\textsf{GroupByCountEncoded}} $\groupbystar_A(\cdot)$: The implementation detail of this operator is given by Algorithm 2.

\stitle{\textsf{CountDistinct}}  $\countdistinct(\cdot)$: The implementation of this operator involves both \AS and \CPS. Given the input encrypted vector of counts $\encV$ of length $s$, the AS first masks $\encV$ to form a new encrypted vector ${\bf \mathcal{V}}$ with a vector of random numbers $M$, i.e., for $i\in \{0,1,\ldots, s-1\}$,
${\bf \mathcal{V}}[i] = {\encV}[i] \oplus labEnc_{pk}(M[i]).$
This masked encrypted vector is then sent to \CPS and decrypted by \CPS to a plaintext vector $\mathcal{V}$ using the secret key.

Next, \CPS generates a garbled circuit which takes (i) the mask $M$ from the \AS, and (ii) the plaintext  masked vector $\mathcal{V}$ and a random number $r$  from the \CPS as the input. This circuit first removes the mask $M$ from $\mathcal{V}$ to get $V$ and then counts the number of non-zero entries in $V$, denoted by $c$. A masked count $c'=c+r$ is outputted by this circuit. \CPS send both the circuit and the encrypted random number $labEnc_{pk}(r)$  to \AS.

Last, the \AS evaluates this circuit to the masked count $c'$ and obtains the final output to this operator: ${\bf c} = labEnc_{pk}(c') - labEnc_{pk}(r)$.

\eat{
   ($\mathbf{V},\epsilon$) - The \textsf{CountDistinct} operator is implemented as follows \begin{enumerate}[label=\alph*)]\item Firstly the \textsf{AS} creates a mask vector drawn uniformly at random from $[m]^{s_A}$, i.e.,  \begin{gather*} M[i] \in_R [m], i \in [|V|]\end{gather*} \item \textsf{AS} masks the encrypted true count vector $\mathbf{V}$  as follows \begin{gather*}\boldsymbol{\mathcal{V}}[i]= \mathbf{V}[i] \oplus labEnc_{pk}(M[i])\end{gather*} and sends it to the \textsf{CSP} \item \textsf{CSP} decrypts the masked encrypted vector as \begin{gather*}\mathcal{V}[i]=labDec_{sk}(\mathbf{V}[i]), i \in [|V|]\end{gather*} \item Next the \textsf{CSP} generates the following garbled circuit that\begin{enumerate}[label=\roman*)]  \item takes the mask $M$ as an input from the \textsf{AS} \item takes a random number $r$  as an input from the \textsf{CSP}\item takes the decrypted masked vector $\mathcal{V}$ as an input from the \textsf{CSP} \item removes the mask $M$ from $\mathcal{V}$ as \begin{gather*}V[i]=\mathcal{V}[i]-M[i], i \in [|V|]\end{gather*}\item  counts the number of non-zero entries of $V$ as C \item adds the laplace noises \begin{gather*}\mathcal{C}=C+r\end{gather*} and returns $\mathcal{C}$ \end{enumerate} \item The \textsf{AS} evaluates the above circuit and gets output $\mathcal{C}$ \item The \textsf{AS} gets $labEnc_{pk}(r)$ from the \textsf{CSP} and generates $labEnc_{pk}(\mathcal{C})$ to compute\begin{gather*}\mathbf{C}=labEnc_{pk}(\mathcal{C})-labEnc_{pk}(r)\end{gather*} \end{enumerate}
}

\stitle{\textsf{Laplace}}  $\lap_{\epsilon,\Delta}(\mathbf{V})$:  The implementation of this operator is presented in the main paper in sec 5.2.\eat{Given an encrypted vector counts $\encV$ of size $s$, both \AS and \CPS have to add Laplace noise in this operator. Hence, this implementation involves two steps.

First, the \AS adds encrypted Laplace noise vector to $\encV$, i.e., for $i\in \{0,1,\ldots,s\}$, $\hat{\encV}[i]  = \encV[i] \oplus labEnc_{pk}(\eta_i),$ where $\eta_i\sim Lap(\Delta/\epsilon)$. This encrypted noisy vector $\hat{\encV}$ is then sent to the \CPS.

Next, the \CPS decrypts $\hat{\encV}$ using the secret key and add another Laplace noise vector, i.e., for  $i\in \{0,1,\ldots,s\}$, $\hat{V}[i] = labDec_{sk}(\hat{\encV}[i]) +\eta'_i$, where $\eta'~\sim Lap(\Delta/\epsilon)$. This plaintext noisy vector is returned as the final output of this operator.}

\eat{
($\mathbf{V},\epsilon$)- Recall that both \textsf{AS} and \textsf{CSP} have to add Laplace noise to the output in Crypt$\epsilon$. Hence the \textsf{Laplace} operator has two components. The first component is executed by the \textsf{AS} wherein,
\begin{enumerate} \item \textsf{AS} generates a noisy vector $\eta$ such that $\eta \in [Lap(\frac{1}{\epsilon})]^{|V|}$ \item encrypts $\eta$ and adds it to the input vector as \begin{gather*}\boldsymbol{\eta}=labEnc_{pk}(\eta)\\\mathbf{\hat{V}}[i]=\mathbf{V}[i]\oplus \boldsymbol{\eta}[i], i \in [|V|]\end{gather*} \end{enumerate} This encrypted noisy vector $\mathbf{\hat{V}}$ is the input for the second phase of the \textsf{Laplace} operator which is executed by the \textsf{CSP} as follows \begin{enumerate}\item Decrypts $\mathbf{\hat{V}}$ \begin{gather*}\hat{V}=labDec_{sk}(\mathbf{\hat{V}})\end{gather*}  \item Generates a noisy vector $\eta'$ such that $\eta' \in [Lap(\frac{1}{\epsilon})]^{|\hat{V}|}$ \item Finally adds the noise $\eta'$ to $\hat{V}$ \begin{gather*}\hat{\mathcal{V}}[i]=\hat{V}[i]+\eta'[i], i \in [|\hat{V}|]\end{gather*} \item Returns $\hat{\mathcal{V}}$ to \textsf{AS} \end{enumerate} 
% Note that in the Crypt$\epsilon$ implementation we need to add two instances of the Laplace noise as opposed to a single instance in the standard central differential privacy setting. After the addition of the first instance of the laplace noise, $\eta$ (by the AS),  the encrypted answer is sent to the CSP. becuse of CSP has only a differential private view Hence the addition of the second instance of the laplace noise can be looked upon as a post-processing step  However and differential privacy is immune to post processing 
}

\stitle{\textsf{NoisyMax}} $\noisymax^k_{\epsilon,\Delta}(\cdot)$: The input to this operator is an encrypted vector of counts $\encV$ of size $s$. Similar to \textsf{Laplace} operator, both \AS and \CPS are involved. First,  the \AS adds to $\encV$
an encrypted Laplace noise vector and a mask $M$, i.e., for $i\in \{0,1,\ldots,s\}$,
$\hat{\encV}[i]  = \encV[i] \oplus labEnc_{pk}(\eta_i) \oplus M[i],$
where $\eta_i\sim Lap(2\cdot k \cdot\Delta/\epsilon)$. This encrypted noisy, masked vector $\hat{\encV}$ is then sent to the \CPS.

 The \CSP first checks whether $\sum_{i=1}^t \epsilon_i +\epsilon \leq \epsilon^B$ where $\epsilon_i$ represents the privacy budget used for a previously executed program $P_i$ (we presume that a total of $t \in \mathbb{N}$ programs have been executed hitherto the details of which are logged into the \CSP's public ledger). Only in the event the above check is satisfied, the \CSP proceeds to decrypt $\hat{\encV}$ using the secret key, i.e., for  $i\in \{0,1,\ldots,s\}$, $\hat{V}[i] = labDec_{sk}(\hat{\encV}[i])$. Next the \CSP records $\epsilon$ and the current program details in the public ledger. This is followed by the \CSP adding another round of Laplace noise to generate 
$\hat{\encV'}[i]  = \hat{\encV}[i] \oplus labEnc_{pk}(\eta'_i),$
where $\eta'_i\sim Lap(2 \cdot k \cdot \Delta/\epsilon), i\in \{0,1,\ldots,s\}$. (This is to ensure that as long as one of the parties is semi-honest, the output does not violate DP.) Finally, the \CPS generates a garbled circuit which takes  (i) the noisy, masked vector $\hat{V}$ from the \CPS, and (ii) the mask $M$ from the \AS as the input. This circuit will remove the mask from $\hat{V}$ to get the noisy counts $\hat{V}'$ and find the indices of the top-$k$ values in $\hat{V}'$.

Finally, the \AS evaluates the circuit above and returns the indices as the output of this operator.

\eat{
($\mathbf{V},\epsilon,k$)- The input to the NoisyMax operator is an encrypted vector $\mathbf{V}$ where each entry $V[i]$ is a count. The operator is implemented via the following steps.  \begin{enumerate}
\item First the \textsf{AS} adds noise to the input encrypted vector as follows \begin{gather*} \eta \in [Lap(\frac{1}{\epsilon})]^{|V|}\\\boldsymbol{\eta}=labEnc_{pk}(\eta)\\\mathbf{\hat{{V}}}[i]=\mathbf{V}[i]+ \boldsymbol{\eta}[i], i \in [|V|] \end{gather*} \item Next the \textsf{AS} creates a mask vector $M$ drawn uniformly at random from $[m]^{s_A}$, i.e.,  \begin{gather*} M[i] \in_R [m], i \in [|V|]\end{gather*} \item \textsf{AS} masks the encrypted noisy vector $\mathbf{\hat{V}}$  as follows \begin{gather*}\boldsymbol{\mathcal{V}}[i]= \mathbf{\hat{V}}[i] \oplus labEnc_{pk}(M[i]), i \in [|V|]\end{gather*} and sends it to the \textsf{CSP} \item \textsf{CSP} decrypts the masked encrypted noisy vector as \begin{gather*}\mathcal{V}[i]=labDec_{sk}(\mathbf{\hat{V}}[i]), i \in [|V|]\end{gather*} \item Next, the following garbled circuit is evaluated which
    \begin{enumerate}[label=\roman*]\item takes noisy masked  vector $\mathcal{V}$ as an input from the \textsf{CSP} \item takes mask $M$ as the input from \textsf{AS}  \item removes the mask from  $\mathcal{V}$  as \begin{gather*} \hat{V}[i]=\mathcal{V}[i]-M[i], i \in [|V|]\end{gather*}  \item computes the top $k$ element over  $\hat{V}$ and returns $arg_{\textit{top k}}\max{\hat{V}[i])}$
    \end{enumerate}
    \end{enumerate}
}

\subsection{DP Index Optimization Cntd.}\label{app:index-imp}

\eat{The DP index optimization can be implemented via a garbled circuit which aims to compute two pieces of information: a sorted encrypted database based on $A$ and a $\epsilon$-differentially private index on $A$. This circuit takes (i) the secret key $sk$ from the \CPS, and (ii) the entire database $\encD$ and the attribute $A$ from the \AS. The attribute $A$ has a domain of size $s_A$ and is uniformly partitioned into $k$ ranges $\{R_1,\ldots, R_i, \ldots, R_{s_A/k}\}$, where $R_i = [\frac{s_A}{k}i, \frac{s_A}{k}(i+1))$.

This circuit first decrypts $\encD$ using the secret key and sorts the decrypted database on attribute $A$ in ascending order. The sorted database is then encrypted again. Then the circuit computes a histogram $V$ on the $k$ ranges of $A$, denoted by $[c_1,\ldots,c_k]$  and perturbs each count with Laplace noise, i.e., $\hat{c}_i = c_i + \eta_i$, where $\eta_i\sim Lap(1/\epsilon)$. The noisy counts $\encV$ are then used to construct a cumulative histogram over $A$, where $\hat{cdf}[i] = \sum_{j=0}^i \hat{c}_i$ and post-processed such that they are in non-decreasing order and non-negative, and $\hat{cdf}[k] = |\encD|$~\cite{cdf}.

Given the differentially private cdf and the sorted database, when a program would like to select rows with $A\in [v_i, v_j]$, we find the ranges that contain $v_i$ and $v_j$ respectively, denoted by $R_i$ and $R_j$. Then we return all records in the sorted database which cover all ranges from $R_i$ and $R_j$. This corresponds to row $(\hat{cdf}[i-1]+1)$ to row $(\hat{cdf}[j])$.
}
\begin{customlemma}{5}\rm Let $P$ be the program that computes the mapping $\mathcal{F}$.
Let $\Pi$ be the protocol corresponding to the construction of the DP index in \system. The
views and outputs of \textsf{AS} and \textsf{CSP} are denoted follows:
\[
\begin{array}{cc}
View_1^{\Pi}(P,\mathcal{D},\epsilon_A) & Output_1^{\Pi}(P,\mathcal{D},\epsilon_A) \\
View_2^{\Pi}(P,\mathcal{D},\epsilon_A) & Output_2^{\Pi}(P,\mathcal{D},\epsilon_A) \\
\end{array}
\]
There exists Probabilistic Polynomial Time (PPT) simulators $Sim_1$
and $Sim_2$ such that:
\squishlist
\item $Sim_1 (P^{CDP}_B(\mathcal{D},\epsilon_A))$ is computationally indistinguishable ($\equiv_c$)
from $(View_1^{\Pi}(P,\mathcal{D},\epsilon_A),Output^{\Pi}(\mathcal{D},\epsilon_A))$, and
\item $Sim_2 (P^{CDP}_B(\mathcal{D},\epsilon_A))$ is $\equiv_c$ to
  $(View_2^{\Pi}(P,\mathcal{D},\epsilon_A),Output^{\Pi}(\mathcal{D},\epsilon))$.
  \squishend $Output^{\Pi}(P,\mathcal{D},\epsilon_A))$ is the combined
  output of the two parties\end{customlemma}
\begin{proof} Recall that protocol $\Pi$ consists of two parts; in the first part $\Pi_1$, the \AS obtains the sorted encrypted database $\encD_s$ via a garbled circuit. Next $\Pi_2$ computes $\mathcal{F}$ via a \system program. The security of the garbled circuit in $\Pi_1$ follows from standard approaches \cite{GC_Security}. Hence in this section we concentrate on $\Pi_2$. The proof of the entire protocol $\Pi$ follows from the composition theorem [88,Section 7.3.1]. The  views of the servers for $\Pi_2$ are as follows: 
\begin{gather*}View_1^{\Pi_2}(P,\mathcal{D},\epsilon_A) = (pk, \encD, \encD_s, \mathcal{F})\\View_2^{\Pi_2}(pk, sk, P,\mathcal{D},\epsilon_A) = 
(\mathcal{F})\end{gather*}
The simulators $Sim_1(z_1)$ (where $z_1=(y_1,|\mathcal{D}|)$ is the random variable distributed as $P^{CDP}_B(\mathcal{D},\epsilon_A)$, $y_1$ being the random variable distributed as $P^{CDP}(\mathcal{D},\epsilon_A/2)$) performs the following steps:
\begin{enumerate}\item Generates a pair of keys $(pk_1,sk_1)$ for the encryption scheme and generates random data set $\mathcal{D}_1$ of the same size as $\mathcal{D}$ and encrypts it using $pk_1$ to get $\mathbf{\tilde{\mathcal{D}_1}}$
\item Generates another random dataset $\mathcal{D}_2$ of the same size and encrypts it with $pk$ to get $\mathbf{\tilde{\mathcal{D}_2}}$. 
\end{enumerate} The computational indistinguishability of $\mathbf{\tilde{\mathcal{D}_1}}$ and $\encD$  follows directly from the semantic security of the encryption scheme. From the construction of the secure sorting algorithm, it is evident that the records in $\encD_s$ cannot be associated back with the data owners by the \AS. This along with the semantic security of the encryption scheme ensures that $\mathbf{\tilde{\mathcal{D}_2}}$ and $\encD_s$ are computationally indistinguishable as well. The tuples $(pk_1, \mathbf{\tilde{\mathcal{D}_1}},\mathbf{\tilde{\mathcal{D}_2}},y_1)$ has the same distribution as $(pk, \encD,\encD_s,\mathcal{F})$ and hence are computationally indistinguishable. Therefore, $Sim_1(z_1)$ is computational indistinguishable from $V iew^{\Pi_2}_1(P,\mathcal{D},\epsilon_A)$.\\ For the simulator $Sim_2(z_2)$ (where $z_2=(y_2,|\mathcal{D}|)$ is the random variable distributed according to $P^{CDP}_B(\mathcal{D},\epsilon_A)$, $y_2$ being the random variable distributed as $P^{CDP}(\mathcal{D},\epsilon_A/2)$), clearly tuples $(y_2)$ and $(\mathcal{F})$ have identical distribution. Thus, $Sim_2(z_2)$ is also computationally indistinguishable from $\\View_2^{\Pi_2}(P,\mathcal{D},\epsilon_A)$ thereby concluding our proof.
\end{proof}

\eat{
\begin{enumerate}\item takes the entire database $\boldsymbol{\mathcal{\tilde{D}}}$ as an input and the attribute $A$ as an input from the \textsf{AS}.
\item takes the secret key $sk$ as an input from  the \textsf{CSP} \item Decrypts $\boldsymbol{\mathcal{\tilde{D}}}$ \item Sorts the decrypted database on $A$, i.e., the first $ct_{A,v_1}$ rows are the ones with value $v_1$ for attribute $A$, the next $ct_{A,v_2}$ are  the records with value $v_2$ for attribute $A$ and so on. \item  re-encrypts the sorted database \item Divide the domain of $A$ into $k$ bins such that each bin contains $s_A/k$ consecutive domain values. \item Construct a $k$ lengthed vector $\hat{V}$ such that $\hat{V}[i]=\sum_jct_{A,j}+\eta_i, i \in [k], j \in [\frac{s_A}{k}(i-1)+1,\frac{s_A}{k}i]$ where $\eta_i$ is a random laplace noise drawn from the distribution $Lap(\frac{k}{\epsilon})$ \item Return $\hat{V}$ and sorted $\boldsymbol{\mathcal{\tilde{D}}}_{sort}$\end{enumerate}
}

%\subsection{Illustration genLabMult}\label{app:genlabhe}
%Figure~\ref{genlab-fig} illustrates how a $n$-way multiplication described in Section~\ref{genlab} can be parallelized. This approach requires a total of $\lceil \log n\rceil$ rounds of communication between \AS and \CPS.

\begin{figure}\includegraphics[width=\columnwidth]{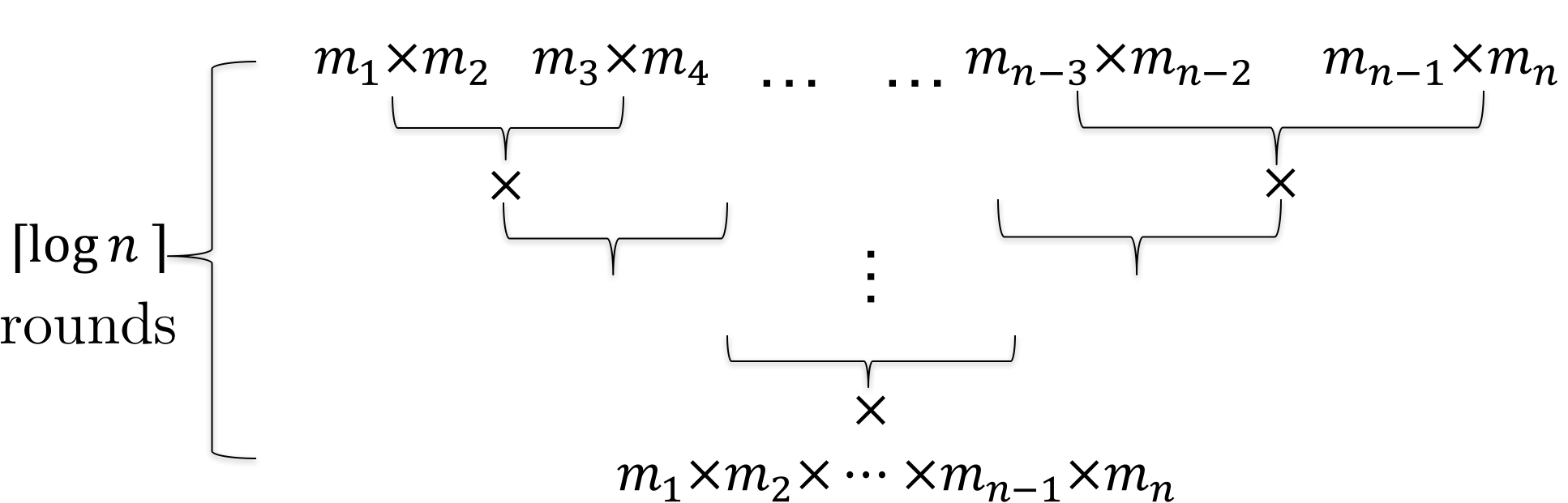} \caption{ $genLabMult()$ - Batching of multiplicands for \textsf{labHE}} \label{genlab-fig}\end{figure}

\begin{algorithm}[H]
\caption{\textsf{GroupByCountEncoded }$\groupby_A(\mathbf{\tilde{T}})$}
\begin{algorithmic}[1]
\STATEx
\textbf{Input}: $\mathbf{\tilde{T}}$
\STATEx \textbf{Output}: $\tilde{\encV}$
\STATEx \textbf{\textsf{AS}:} \STATE Computes $\mathbf{V}=\groupbystar_{A}(\encT)$.
\STATE Masks the encrypted histogram $\mathbf{V}$ for attribute $A$ as follows \begin{gather*}\boldsymbol{\mathcal{V}}[i]= \mathbf{V}[i] \oplus labEnc_{pk}(M[i])\\M[i] \in_R [m], i \in [|V|]\end{gather*}
\STATE Sends $\boldsymbol{\mathcal{V}}$ to \textsf{CSP}.
\STATEx \textbf{\textsf{CSP}:}
\STATE Decrypts  $\boldsymbol{\mathcal{V}}$ as $\mathcal{V}[i]=labDec_{sk}(\boldsymbol{\mathcal{V}}), i \in [|V|]$.\STATE Converts each entry of $\mathcal{V}$ to its corresponding one-hot-coding and encrypts it, $\boldsymbol{\tilde{\mathcal{V}}}[i]=labEnc_{pk}(\tilde{\mathcal{V}[i]}), i \in [|V|]$
\STATE Sends $\boldsymbol{\tilde{\mathcal{V}}}$ to \textsf{AS}.
\STATEx \textbf{\textsf{AS}}:
\STATE  Rotates every entry by its corresponding mask value to obtain the desired  encrypted one-hot-coding $\boldsymbol{\tilde{V}}[i]$. \begin{gather*}\boldsymbol{\tilde{V}}[i]=RightRotate(\boldsymbol{\tilde{\mathcal{V}}},M[i]), i \in [|V|]\end{gather*} 
 \end{algorithmic} \label{groupby-imp}
\end{algorithm} 
 
\section{Classification of \system Programs}\label{app:sec:classification}
\system programs are grouped into three classes based on the number and type of interaction between the \textsf{AS} and the \textsf{CSP}.

\stitle{Class I: Single Decrypt Interaction Programs}\\
%Recall that the transformation operators output encrypted data.  Since  the \textsf{CSP} has exclusive access to the secret key, it is the only party in \system capable of decryption. Thus f
For releasing any result (noisy) in the clear, the \textsf{AS} needs to interact at least once with the \textsf{CSP} (via the two measurement operators) as the latter has exclusive access to the secret key. \system programs like P1, P2 and P3 (Table~\ref{tab:programexamples}) that require only a single interaction of this type fall in this class. %Typically these programs filter the database on a single attribute. 
%Examples of this type of programs are . %The post-processing step in P1 is done in the clear and hence requires no more interactions with the \textsf{CSP}.

\stitle{Class II: \textsf{LabHE} Multiplication Interaction Programs}\\
\system supports a $n$-way multiplication of ciphers for $n > 2$ as described in Section~\ref{genlab} which requires intermediate interactions with the \textsf{CSP}. Thus all \system programs that require multiplication of more than two ciphers need interaction with the \textsf{CSP}. Examples include %All programs that filter the database on more than three attributes, such as 
P4 and P5 (Table~\ref{tab:programexamples}).

\stitle{Class III: Other Interaction Programs}\\
 The \textsf{GroupByCountEncoded} operator requires an intermediate interaction with the \textsf{CSP}. %(for generating the encrypted one-hot-coding for the new attribute). 
 The \textsf{CountDistinct} operator also uses a garbled circuit (\ifpaper 
details in full paper \cite{anom}% using paper 
\else 
details in Appendix~\ref{app:implement_operators}
\fi) and hence requires interactions with the \textsf{CSP}. %This is in addition to the interaction required for decrypting the noisy answer (as explained in Class I above). 
Therefore, any program with the above two operators, like P6 and P7 (Table~\ref{tab:programexamples}), requires at least two rounds of interaction. 
\section{Additional Evaluation}\label{app:evaluation}
In this section we present some additional evaluation results for \system programs.

%P1, P3, P5 and P7  w.r.t their  corresponding \textsf{CDP} implementations under unbounded DP (Figure \ref{fig:unboundedDP}. The only difference in observation here as compared to that of Figure \ref{accuracy} is that the error of \system is around $4\times$ higher than that of the corresponding \textsf{CDP} implementation. Recall that for guaranteeing bounded $\epsilon$-DP, we need to add noise from $Lap(\frac{2\cdot\Delta}{\epsilon})$ (Section \ref{sec:security}) in \system. This in tandem with the fact that we add two instances of Laplace noise in \system (Section \ref{sec:operator_implementation}) contributes to the $4\times$ reduction in accuracy.
\eat{We present additional accuracy study for \system programs P2, P4, and P6 from Table~\ref{tab:programexamples}. Figure~\ref{accuracy-appendix} shows the empirical accuracy of these three programs (both optimized and unoptimized) and that of the corresponding state-of-the-art \textsf{LDP} \cite{LDP1} and \textsf{CDP} \cite{Dork} implementations  with varying privacy parameter $\epsilon \in \{0.1,...,0.9\}$. We observe similar results for P2, P4, and P6 as that of the other programs (P1, P3, P5, P7 in Figure~\ref{accuracy}). The error values of \system  and unoptimized \system  are comparable to that of \textsf{CDP} and are much smaller than that of \textsf{LDP}.}
\subsection{DP Index Analysis Cntd.}
Here we discuss the effect of including neighboring bins
in the program execution. To avoid missing relevant rows, more bins that are adjacent to the chosen range $[i_s,i_e]$ can be considered for the subsequent operators. We increase the number of neighbouring bins from $0$ to $8$. As shown in Figure~\ref{fig:error:NeighboringBins}, the error decreases and all the relevant rows are included when $4$ neighbouring bins are considered. However, the execution time naturally increases with extra neighbouring bins as shown in Figure~\ref{fig:time:NeighboringBins}.
\begin{figure}
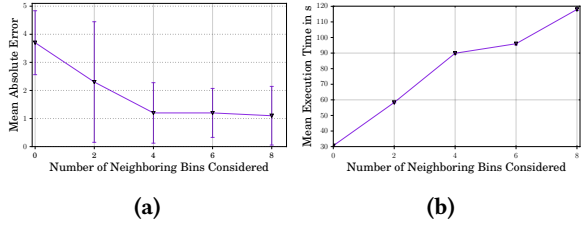

  \begin{subfigure}[b]{0.45\linewidth}
    \centering \includegraphics[width=1\linewidth]{index_NeighboringBins_error.pdf}
        \caption{}
        \label{fig:error:NeighboringBins}\end{subfigure}
        \begin{subfigure}[b]{0.45\linewidth}
        \includegraphics[width=1\linewidth]{index_NeigboringBins_time_new.pdf}
        \caption{}
        \label{fig:time:NeighboringBins}
        \end{subfigure}
        %\vspace{-3mm}
        \caption{Accuracy and performance of P5 with varying number of neighboring bins considered for the DP index optimization }  
    \end{figure}
\subsection{Communication Costs}
We use Paillier encryption scheme \cite{Paillier} in our prototype \system (Section 9.1). This means that each ciphertext is a random number in the group $(\mathbb{Z}/N^2\mathbb{Z})^*$ where $N$ is a RSA moduli. Thus sending an encrypted data record entails in each data owner sending $\sum_j |domain(A_j)|$, where $A_j$ is an attribute of the database schema, such numbers to the \AS. Communication is also needed for the measurement operators and \textsf{GroupByCountEncoded}  where the \AS needs to send a ciphertext (or a vector of ciphertexts) to the \CSP. Additionally operators like \textsf{NoisyMax} and \textsf{CountDistinct} need a round of communication for the garbled circuit however these circuits are simple and dataset size independent. The most communication intensive operator is the \textsf{CrossProduct} which requires $log_2m$ (Appendix D.1) where $m$ is the dataset size rounds of interactions. However, this can be done as a part of pre-processing (Section  and hence does not affect the actual program execution time. Hence overall, \system programs are not communication intensive. 

\section{Related Work}\label{app:related}
\subsection{Differential Privacy}\label{app:dp}
Introduced by Dwork et al.~\cite{Dork}, differential privacy has enjoyed immense attention from both academia and industry in the last decade. We will  discuss the recent directions in two models of differential privacy: the \textit{centralized differential privacy} (CDP), and \textit{local differential privacy} (LDP).

The \textsf{CDP} model assumes the presence of a trusted server which can aggregate all users' data  before perturb the query answers. This allows the design of a complex algorithm that releases more accurate query answers than the basic DP mechanisms. For example, an important line of work in the \textsf{CDP} model has been towards proposing "derived" mechanisms'' \cite{MVG} or  "revised algorithms" \cite{Blocki} from basic DP mechanisms (like exponential mechanism, Laplace mechanism, etc.). The design of these mechanisms leverages on specific properties of the query and the data, resulting in a better utility than the basic mechanisms.  One such technique is based on data partition and aggregation \cite{AHP,hist1,hist2,hist3,hist4,hist6,hist7,hist8} and is helpful in answering histogram queries. The privacy guarantees of these mechanisms can be ensured via the composition theorems and the post-processing property of differential privacy \cite{Dork}. We would like to build \system that can support many of these algorithms. %Another technique involves non-uniform data weighting where each data sample is weighed based on their query contribution. Research in this line of work include \cite{u1,u2,MWEM}. Yet another popular method is to utilize past/auxiliary information to improve the utility of the query answers. Examples are \cite{A1,A2,A3,A4,A6,A7,A8}.

%Another interesting line of work has been towards developing programming frameworks to enable non-experts to write easy differentially private programs. This line of work was started by the PINQ platform \cite{PINQ} and there has been a series of follow up work  \cite{FWPINQ,p2, airavat}. The most recent one is the Ektelo \cite{ektelo} framework where all existing algorithms for answering linear counting queries can be expressed as a composition of its operators. 

The notion of \textsf{LDP} and related ideas has been around for a while ~\cite{Kasivi,Evfimievski:2003:LPB:773153.773174,RR}. Randomized response proposed by Warner in 1960s~\cite{RR} is one of the simplest \textsf{LDP} techniques. The recent \textsf{LDP} research techniques~\cite{LDP1, HH2, Rappor1} focus on constructing a frequency oracle that estimates the frequency of any value in the domain. However, when the domain size is large, it might be computationally infeasible to construct the histogram over the entire domain. To tackle this challenge, specialized and efficient algorithms have been proposed to compute heavy hitters~\cite{HH,Rappor2}, frequent itemsets~\cite{15,itemset}, and marginal tables~\cite{Cormode, CALM}. As the \textsf{LDP} model does not require a trusted data curator, it enjoyed significant industrial adoption, such as Google~\cite{Rappor1, Rappor2}, Apple~\cite{Apple}, and Samsung~\cite{Samsung}.

%To tackle this challenge, specialized algorithms to compute the most frequently occurring values, also known as the heavy hitters, have been proposed \cite{HH,Rappor2,HH2}. Another practical setting can be when the user's data is a set of items and the aggregator is interested in  the $k$ most frequent item sets~\cite{15,itemset}. In \cite{Cormode, CALM} the authors propose efficient constructions of marginal tables in the local differential privacy setting. Due to their attractive trust model, \textsf{LDP} has also enjoyed significant industrial adoption.  Google has integrated RAPPOR \cite{Rappor1, Rappor2} with Chrome. It is primarily tasked with collecting user statistics like default browser homepage, default search engine et al in order to monitor malicious hijacking of user settings. Apple \cite{Apple} has also deployed differential privacy to collect of data like most frequent emojis, help with auto-completion of spellings etc.  Samsung \cite{Samsung} proposed a similar system which enables the collection of both categorical  (like screen resolution) as well as numerical data (like time of usage, battery volume), although it is not clear whether they went ahead with the actual deployment. \par

Recently it has been showed that augmenting randomized response mechanism with an additional layer of anonymity in the communication channel can improve the privacy guarantees. The first work to study this was PROCHLO~\cite{Prochlo} implementation by Google.
%In \cite{Prochlo} the authors propose a  Encode, Shuffle, Analyze (ESA) architecture which relies on an explicit intermediate shuffler that processes the randomized LDP reports from users to ensure their anonymity.
PROCHLO necessitates this intermediary to be trusted, this is implemented via trusted hardware enclaves (Intel's SGX). However, as showcased by recent attacks \cite{Foreshadow}, it is notoriously difficult to design a  truly secure hardware in practice. Motivated by PROCHLO, the authors in \cite{amplification}, present a tight upper-bound on the worst-case privacy loss. Formally, they show that  any permutation invariant algorithm satisfying $\epsilon$-\textsf{LDP} will satisfy $O(\epsilon\sqrt{\frac{\log(\frac{1}{\delta})}{n}},\delta)$-\textsf{CDP}, where $n$ is the data size. Cheu et al.~\cite{mixnets} demonstrate privacy amplification by the same factor for 1-bit randomized response by using a mixnet architecture to provide the anonymity. This work also proves another important result that the power of the mixnet model lies strictly between those of the central and local models.

A parallel line of work involves efficient use of cryptographic primitives for differentially private
functionalities.  Agarwal et al.~\cite{kamara} proposed an algorithm for computing histogram over encrypted data. Rastogi et al.~\cite{Rastogi} and Shi et al.~\cite{Shi} proposed algorithms that allow an untrusted aggregator to periodically estimate the sum of $n$ users' values in a  privacy preserving fashion.However, both schemes are irresilient to user failures. Chan et al.~\cite{Shi2} tackled this issue by constructing binary interval trees over the users.

\subsection{Two-Server Model}\label{app:2servermodel}
The two-server model is a popular choice for privacy preserving machine learning techniques. Researchers have proposed privacy preserving ridge regression systems with the help of a cryptographic service provider~\cite{Boneh1,LReg,Ridge2}. %\xh{The following discussion should (i) address how cryptographic service provider works in their models; (ii) why this is different/similar to our work. We need to draw relation of this work with ours instead of laundry listing their contributions.} 
While the authors in \cite{Ridge2}  use a hybrid multi-party computation scheme with a secure inner product technique, Nikolaenko et al.  propose a hybrid approach in \cite{Boneh1} by combining homomorphic encryptions and Yao's garbled circuits. Gascon et al.~\cite{Ver} extended the results in \cite{Boneh1} to include vertically partitioned data and  the authors in \cite{LReg} solve the problem using just linear homomorphic encryption.  Zhang et al. in \cite{secureML} also propose secure machine learning protocols using a privacy-preserving stochastic gradient descent method. Their main contribution includes developing efficient algorithms for secure arithmetic operations on shared decimal numbers and proposing alternatives to non-linear functions such as sigmoid and softmax tailored for MPC computations.  In \cite{Boneh2} and \cite{Matrix2} the authors solve the problem of privacy-preserving matrix factorization. In both the papers, use a hybrid approach combining homomorphic encryptions and Yao's garbled circuits for their solutions.

\subsection{Homomorphic Encryption}\label{app:he}
%\xh{Need to describe how these techniques relate to our work. Discuss whether we can use them as extension?}
With improvements made in implementation efficiency and new constructions developed in the recent past, there has been a surge in practicable privacy preserving solutions employing homomorphic encryptions. A lot of the aforementioned two-server models employ homomorphic encryption \cite{Boneh1,Boneh2,LReg,Matrix2}.   In \cite{CryptoDL,CryptoNet,NN} the authors enable neural networks to be applied to homomorphic-ally encrypted data. Linear homomorphic encryption is used in \cite{Irene2} to enable privacy-preserving machine learning for ensemble methods while %\cite{FHEReg}
uses  fully-homomorphic encryption
to approximate the coefficients of a logistic-regression model.
\cite{grid} uses somewhat-
homomorphic encryption scheme to compute the forecast
prediction of consumer usage for smart grids. 
%Privacy preserving multi-party machine learning with homomorphic encryption

%\input{discussion}

\end{document}